\documentclass[twoside,11pt]{article}

\usepackage{amsmath,bm}
\usepackage{float}
\usepackage{graphicx}
\usepackage{xcolor}
\usepackage{appendix}
\usepackage{comment}
\usepackage{array,multirow}
\usepackage{xr}
\usepackage{tabularx,ragged2e,booktabs,caption}
\usepackage{jmlr2e}

\newcolumntype{C}[1]{>{\Centering}m{#1}}

\newtheorem{assumption}{Assumption}

\newcommand{\br}{\bm{r}}

\newcommand{\bv}{\bm{v}}
\newcommand{\bw}{\bm{w}}

\newcommand{\by}{\bm{y}}

\newcommand{\Ib}{\mathbf{I}}

\newcommand{\bA}{\bm{A}}

\newcommand{\bE}{\bm{E}}

\newcommand{\bG}{\bm{G}}
\newcommand{\bH}{\bm{H}}
\newcommand{\bI}{\bm{I}}

\newcommand{\bM}{\bm{M}}

\newcommand{\bO}{\bm{O}}

\newcommand{\bQ}{\bm{Q}}

\newcommand{\bS}{\bm{S}}

\newcommand{\bU}{\bm{U}}

\newcommand{\bW}{\bm{W}}
\newcommand{\bX}{\bm{X}}

\newcommand{\bZ}{\bm{Z}}

\newcommand{\cS}{{\mathcal{S}}}

\newcommand{\EE}{\mathbb{E}}

\newcommand{\II}{\mathbb{I}}

\newcommand{\RR}{\mathbb{R}}

\newcommand{\bbeta}{\bm{\beta}}
\newcommand{\bgamma}{\bm{\gamma}}
\newcommand{\bdelta}{\bm{\delta}}

\newcommand{\bzeta}{\bm{\zeta}}

\newcommand{\brho}{\bm{\varrho}}

\newcommand{\btau}{\bm{\tau}}

\newcommand{\bphi}{\bm{\phi}}

\newcommand{\bGamma}{\bm{\Gamma}}

\newcommand{\bLambda}{\bm{\Lambda}}

\newcommand{\bSigma}{\bm{\Sigma}}

\newcommand{\bOmega}{\bm{\Omega}}

\newcommand{\diag}{{\rm diag}}

\def\T{{ \mathrm{\scriptscriptstyle T} }}

\usepackage{lastpage}

\begin{document}

\jmlrheading{24}{2023}{1-\pageref{LastPage}}{7/22; Revised
3/23}{8/23}{22-0834}{Jing Ouyang, Kean Ming Tan and Gongjun Xu}
\ShortHeadings{High-dimensional Inference for GLMs with Hidden Confounding}{Ouyang, Tan and Xu}


\title{High-Dimensional Inference for Generalized Linear Models  \\
with Hidden Confounding}

\author{\name Jing Ouyang \email jingoy@umich.edu \\
\name Kean Ming Tan \email keanming@umich.edu \\
\name Gongjun Xu \email gongjun@umich.edu \\
       \addr Department of Statistics\\
       University of Michigan\\
       Ann Arbor, MI 48109, USA
     }

\editor{Mladen Kolar}

\maketitle

\begin{abstract}
Statistical inferences for high-dimensional regression models have been extensively studied for their wide applications ranging from  genomics, neuroscience, to economics. However, in practice, there are often potential unmeasured confounders  associated with both the response and covariates, which can lead to invalidity of standard debiasing methods. This paper focuses on a generalized linear regression framework with hidden confounding  and proposes a debiasing approach to address this high-dimensional problem, by  adjusting for the effects induced by the unmeasured confounders. We establish consistency and asymptotic normality for the proposed debiased estimator. The finite sample performance of the proposed method is demonstrated through extensive  numerical studies and an application to a genetic data set. 
\end{abstract}

\begin{keywords}
High-dimensional inference; Generalized linear model; Latent variable; Unmeasured confounder.
\end{keywords}

\section{Introduction}

Statistical inferences for high-dimensional regression models have received a growing interest due to the increasing number of complex data sets with high-dimensional covariates that are collected across many different scientific disciplines ranging from genomics to social science to econometrics \citep{peng2010,belloni2012,fan2014nsr, buhlmann2014}.   
One of the most popularly used high-dimensional regression methods is the lasso linear regression, which assumes that the underlying regression coefficients are sparse \citep{tibshirani1996}.     
However, the lasso penalty introduces non-negligible bias that renders high-dimensional statistical inference challenging \citep{wainwright2019}.     
To address this challenge, 
\citet{zhang2014} and \citet{van2014} proposed the debiasing method to construct confidence intervals for the lasso regression coefficients: the main idea is to first obtain the lasso estimator, and then correct for the bias of the lasso estimator using a low-dimensional projection method. We refer the reader to  \cite{javanmard2014confidence}, \cite{belloni2014}, \cite{ning2017}, \cite{Chernozhukov2018}, among others, for detailed discussions of the debiasing approach, and also \cite{wainwright2019} for an overview of high-dimensional statistical inference.

The aforementioned studies were established under the assumption that there are no unmeasured confounders that are associated with both the response and covariates. However, this assumption is often violated in observational studies.
For instance, in genetic studies, the effect of certain segments of DNA on the gene expression may be confounded by population structure and microarray expression artifacts \citep{Jennifer2010}. 
Another example is in healthcare studies where the effect of nutrients intake on the risk of cancer may be confounded by  physical wellness, social class, and behavioral factors \citep{Fewell2007}.
Without adjusting for the unmeasured confounders, the resulting inferences from the standard debiasing methods could be biased and consequently, lead to spurious scientific discoveries.

Various methods have been proposed to perform valid statistical inferences for regression parameters in the presence of hidden confounders. One commonly used approach is instrumental variables regression, which typically requires domain knowledge to identify valid instrumental variables, making it challenging for high-dimensional applications   \citep{kang2016, burgess2017review, guo2018iv, Windmeijer2019}.  
Recently, under a relaxed linear structural equation
modeling framework,  \cite{guo2021doubly}  proposed a deconfounding approach for statistical inference for individual regression coefficients. \cite{fan2022latent} developed a factor-adjusted debiasing method and conducted statistical estimation and inference for the coefficient vector. However, these two methods rely on the linearity assumption between the response and covariates, which may not hold when the response is not continuous but categorical (binary, count, etc.). Such  categorical data commonly occurs in genetic, biomedical, and social science applications, and thus alternative methods are needed to perform valid statistical inferences in these cases.

To bridge the gap in the existing literature, we propose a novel framework to perform  statistical inference in the context of high-dimensional generalized linear models with hidden confounding. The main idea of our proposed method involves estimating the unmeasured confounders using high-dimensional factor analysis techniques and then applying the debiasing method on the regression coefficient of interest, with the estimated unmeasured confounders treated as surrogate variables. This method does not rely on the linear model assumption or any specific model form and it is applicable to more general models beyond the generalized linear models, such as graphical models~\citep{zhao2015, zhu2020} and additive hazards models~\citep{lin1994semiparametric, lin2013}.

Theoretically, we show that under some mild scaling conditions, the estimation errors of proposed estimators achieve comparable rates as that of the $\ell_1$-penalized generalized linear model without unmeasured confounders. 
We further show that the debiased estimator for the coefficient of interest is asymptotically Gaussian after adjusting for the unmeasured confounders, which results in a valid statistical inference for high-dimensional generalized linear models with hidden confounding. It is worth highlighting that when using a factor model to relate covariates and unmeasured confounders,
we make more general assumptions on the random noise compared to existing works. Specifically, we allow the random noise to be non-identically distributed. This represents a significant improvement over the majority of previous works, which assume that the random noise follows an independent and identical distribution~\citep{guo2021doubly, fan2022latent}. Furthermore, unlike existing methods under the linear framework \citep{guo2021doubly, fan2022latent}, generalized linear models pose new challenges as many transformation and decomposition techniques commonly used for the linear models are not directly applicable. Consequently, more general and refined intermediate results are needed as ``building blocks" in the proofs  (see Remark~\ref{remark:estimate U} in Appendix D.1 for details).

Our paper is organized and structured as follows. In Section~\ref{model setup}, we introduce the model setup and provide a comprehensive discussion of related literature. Section~\ref{Sec:methodology} presents our two-step approach for estimating the parameter of interest while adjusting for the effect of unmeasured confounders. Section~\ref{Sec:theoretical results} establishes the theoretical properties of the parameter including the estimation consistency as well as the asymptotic normality for the estimator. In Section~\ref{sec:numerical}, we demonstrate the performance of our proposed method and the validity of theoretical results via extensive simulation studies. We also provide an application to genetic data containing gene expression quantifications and stimulation statuses in mouse bone marrow-derived dendritic cells, where we identify significant gene expressions under stimulations, which are consistent with the experimental findings in genetic studies~\citep{IL1B, SAA3, rsad2, IL6}. 
Lastly, in Section~\ref{Sec:discussion}, we provide concluding remarks and outline potential future directions for this research.

\section{Generalized Linear Models with Hidden Confounding}
\label{model setup}
In this section, we first set up a generalized linear model with hidden confounding and introduce a scientific application of our model framework. Then we will discuss related high-dimensional models in the existing literature.

\subsection{Problem Setup}
Consider a high-dimensional regression problem with a unidimensional response $y$ and a $p$-dimensional observed covariates $\bX$.  
In addition, assume that there is a $K$-dimensional unmeasured confounders $\bU$ that are related to both $y$ and $\bX$. 
Without loss of generality, we write $\bX = (D, \bQ^\T)^\T$, where $D \in \RR $ is the covariate of interest and $\bQ \in \RR^{p-1}$ is a $(p-1)$-dimensional vector of nuisance covariates. 
Furthermore, we denote $\theta \in \RR$ as the univariate parameter of interest, $\bv \in \RR^{p-1}$ as parameters for the nuisance covariates, and $\bbeta \in\RR^K$ as parameters that quantify the effects induced by the unmeasured confounders. 
The goal is to perform statistical inference on the parameter $\theta$. 

We assume that $y$ given $D$, $\bQ$, and the unmeasured confounders  $\bU$ follows a generalized linear model with probability density (mass) function:
 \begin{equation}
 \label{eq:full y model}
 	f(y) = \exp \left[ \{y ( \theta {D}  +\bv^\T \bQ + \bbeta^\T\bU )-b( \theta {D}  +\bv^\T \bQ + \bbeta^\T\bU )\}/{a(\phi)} +c(y, \phi)\right],
 \end{equation}
where $\phi$ is the scale parameter, and $a(\cdot)$, $b(\cdot)$, and $c(\cdot)$ are some known functions.  As the distribution of $y$ belongs to the exponential family, we have $E(y) =b^{\prime}(\theta {D}  +\bv^\T \bQ + \bbeta^\T\bU )$ and $\operatorname{var}(y) =b^{\prime \prime}(\theta {D}  +\bv^\T \bQ + \bbeta^\T \bU )a(\phi)$. 
For simplicity, we take $a(\phi) = 1$. 
For notational convenience, let  $\bZ = (D, \bQ^\T, \bU^\T)^\T$ be a vector that includes all observed  covariates and the unmeasured confounders, and let $\bm{\eta} = (\theta, \bv^\T, \bbeta^\T)^\T$ be the corresponding parameters. 
We now provide three commonly used examples.

\begin{example}[Logistic Regression] Let $y \in \{0,1\}$ be a binary variable. Given covariates $D$, $\bQ$, and unmeasured confounders $\bU$, the response $y$ follows the logistic regression model with $\phi = 1$, $a(\phi) = 1$, $b(t) = \log \{1+\exp(t)\}$, and  $c(y, \phi) =0$.
\end{example}

\begin{example}[Poisson Regression] Let $y \in \{0, 1, 2, \ldots \}$ be a discrete variable. Given covariates $D$, $\bQ$, and unmeasured confounders $\bU$, the response $y$ follows the Poisson regression model with $\phi = 1$, $a(\phi) = 1$, $b(t) = \exp(t)$, and $c(y, \phi) = -\log(y!)$.
\end{example}

\begin{example}[Linear Regression] Let $y \in \RR$ be a real-valued response variable. Given covariates $D$, $\bQ$, and unmeasured confounders $\bU$, the response $y$ follows the linear regression model $y =  \theta {D}  +\bv^\T \bQ + \bbeta^\T\bU + \varepsilon$ with $E(\varepsilon) = 0$ and $\text{var}(\varepsilon) = \sigma^2$. The model parameters are $\phi = \sigma^2$, $a(\sigma^2) = \sigma^2$, $b(t) = t^2/2$ and $c(y,\sigma^2) = -{y^{2}}/(2 \sigma^{2})-\log (2\pi\sigma^{2})/2$.
\end{example}

In addition, we assume that the relationship between covariates $\bX$ and unmeasured confounders $\bU$ is captured by the following model:
\begin{equation}
\label{eq:XUmodel}
 \bX=  \bW^{\T}\bU + \bE, 
\end{equation}
where $\bW \in \RR^{K\times p}$ is the loading matrix that describes the linear effect of unmeasured confounders $\bU$ on covariates $\bX$, and $\bE$ is the random noise independent of $\bU$. While similar model as in~\eqref{eq:XUmodel} is  considered in \cite{guo2021doubly} and \cite{fan2022latent}, here we assume model~\eqref{eq:XUmodel} is an approximate factor model which allows for weak correlation and non-identical distribution of the random noise. This is a general setting compared to many existing works assuming $\bE$ to be identically and independently distributed~\citep{guo2021doubly, fan2022latent}. We will elaborate on this model setup and expound the assumptions for weakly correlated random noise in Sections~\ref{Sec:methodology} and~\ref{Sec:theoretical results}.

The aforementioned structural equation modeling framework can be applied to many scientific applications. In the following, we provide one motivating example in genetic studies.  
Various authors have found that the effect of gene expression in response to the environmental conditions, e.g., viral or bacterial stimulation, might be confounded by unmeasured factors \citep{price2006, lazar2013batch}. 
One interesting scientific problem is to assess the effect of gene expression responding to the viral stimulation to cells, while adjusting for the confounding effects from the unmeasured variables. The viral stimulation status, a binary variable, is considered as the response variable $y$. Therefore, a generalized linear model for modeling $y$ is preferred.  In this example, $y$ is the viral stimulation status, $\bX$ is a vector of high-dimensional gene expression, and $\bU$ is a vector of possible unmeasured confounders. More details of the data application will be provided in Section~\ref{Sec:real data application}.

 \subsection{Related Models in Existing Literature}
 \label{sec:related models}
In high-dimensional regression, the adjustment for the hidden confounding effects is a challenging and intriguing problem. There are various related models in the literature. For instance, to address the difficulty of model selection when $p \gg n$,  \cite{paul2008} assumed that $y$ and $\bX$ are connected via a low-dimensional latent variable model: $y = \bbeta^T\bU + \varepsilon$ and $\bX = \bW^T\bU+ \bE$, where the latent factors are associated with response and covariates are only used to infer the latent factors. However, the response and covariates are not directly associated.
Different from the latent variable model in~\cite{paul2008}, \cite{bai_ng2006} considered an additional low-dimensional covariate $\bG$ to the latent variable model, that is expressed as $y = \bbeta^T\bU + \brho^T\bG + \varepsilon$ and $\bX = \bW^T\bU+ \bE$.
Besides, there are other factor-adjusted models that can be extended to adjust hidden confounding. For example, \cite{fan2020} studied the high-dimensional model selection problem when covariates are highly correlated. As most commonly used model selection methods may fail with highly-correlated covariates, they used a factor model to reduce the dependency among covariates and proposed a factor-adjusted regularized model section method. \cite{fan2020} considered a generalized linear model between response and covariate, which together with the factor model forms a similar model framework as ours. However, the problem they studied is fundamentally different than our problem. They did not assume hidden confounding and the factor model is only used to identify a low rank part of highly-correlated covariates whereas in our problem, we focus on the regression problem with unmeasured confounders associated with response and covariates.  Similar factor-adjusted methods have also been studied in other settings  \citep[e.g.,][]{johann2011, wang2019}; however, as noted in \cite{Domagoj2020}, related theoretical justifications are still underdeveloped in the literature.

Recently, many researchers studied the following linear hidden confounding model,
 which can be viewed as a special case of our framework,
\begin{equation}
    y = \theta {D}  +\bv^\T \bQ  +  \bbeta^\T \bU + \varepsilon, \quad \bX = \bW^\T\bU + \bE. \label{eq:FARM}
\end{equation}
Under this model,~\cite{alois2011factor} used the principal component method to estimate the unmeasured confounders and then applied a selection procedure on a projected model. 
\cite{Domagoj2020} proposed a method that first performs the spectral transformation pre-processing step and then applies the lasso regression on the transformed response and covariates. However, these two works focused on estimation consistency and did not address inference issues. Several works have investigated statistical inference on the covariate coefficients. For example,  \cite{guo2021doubly} proposed a doubly debiased lasso method to perform statistical inference for $\theta$.  Different from the approach of   \cite{guo2021doubly} that implicitly adjusts for the hidden confounding effects, 
\cite{fan2022latent} proposed to first use the principal component method to estimate unmeasured confounders and then construct the bias-corrected estimator for $\|({\theta}, {\bv}^{\T})\|_{\infty}$,
which involves the decomposition of the estimation error relying on the linear form of the response and uses the projection of the response onto the factor space. 
 The motivation of \cite{fan2022latent} originated from \cite{fan2020}. When covariates are highly correlated, the leading factors are likely to have extra impacts on the response. So they augmented the factor into the sparse linear regression model between response and covariates, which is written as~\eqref{eq:FARM}. However, these techniques are designed for linear models and may not be directly applicable to generalized linear model settings.

\section{Estimation Method}
\label{Sec:methodology}
In this section, we propose a novel framework to perform statistical inference for a parameter of interest in the context of high-dimensional generalized linear models with hidden confounding. In the proposed framework, we first estimate the unmeasured confounders using a factor analysis approach.
Subsequently, the estimated unmeasured confounders are treated as surrogate variables for fitting a high-dimensional generalized linear model, and a debiased estimator is constructed to perform statistical inference.

Throughout this section, we assume that the observed data $\{y_i, \bX_i\}_{i=1,\dots, n}$ and the unmeasured confounders $\{\bU_i\}_{i= 1,\dots, n}$ are realizations of~\eqref{eq:full y model} and~\eqref{eq:XUmodel}.
Moreover, the random noise $\bE_i = (E_{i1},\ldots,E_{ip})^{\T}$ has mean zero and variance $\bOmega_i = \EE(\bE_i\bE_i^{\T}).$ Let $\bSigma_{e} = \diag(n^{-1}\sum_{i=1}^n \bOmega_i)$, where $\diag(\bA)$ denotes a diagonal matrix by setting off-diagonal entries in $\bA$ to zero. In the $p \times p$ diagonal matrix $\bSigma_{e}$, we denote the $j$-th diagonal element to be $\sigma_j^2 = n^{-1}\sum_{i=1}^n \tau_{i, jj}$, where $\tau_{i, jj}$ is the $(j, j)$ element of $\bOmega_i$. The model assumption on the random noise is general as it does not assume that the random noise $\bE_i$ is identical nor does it require the covariance matrix $\bOmega_i$ to be diagonal.
The detailed theoretical assumptions regarding the random noise will be presented in Section~\ref{Sec:theoretical results}.

{\em Estimation of the Unmeasured Confounders:}  In this work, we consider the dimension $K$ as pre-specified. In practice, there are various methods to estimate the dimension of  unmeasured confounders such as scree plot~\citep{cattell1966scree}, cross-validation method~\citep{owen2016bicv}, information criteria method~\citep{bai2002determining}, the eigenvalue ratio method~\citep{lam2012factor, ahn2013eigenvalue}, among others. In the implementation of our proposed method, we recommend the  parallel analysis approach because of its good finite sample performance, easy implementation, and popularity in scientific applications \citep{hayton2004factor, costello2005best, brown2015confirmatory} -- it has shown superior performances compared to many other methods in various empirical studies~\citep{zwick1986comparison, pedro2005stopping}.
Detailed discussions on the estimation of the dimension of unmeasured confounders are presented in Appendix C.
 
 We employ the maximum likelihood estimation to estimate the unmeasured confounders under~\eqref{eq:XUmodel}.
  Without loss of generality, let $\bar{\bU} = n^{-1}\sum_{i=1}^n \bU_i = {0}$ and let  $\bS_u = n^{-1}\sum_{i=1}^n \bU_i\bU_i^\T$ be the sample variance of $\bU$.
 Similarly, let $\bS_{x}=n^{-1} \sum_{i=1}^{n}(\bX_{i}-\bar{\bX})(\bX_{i}-\bar{\bX})^{\T}$ be the sample variance of $\bX$, where $\bar{\bX} = n^{-1}\sum_{i=1}^n \bX_i$. 
Given unmeasured confounders $\bU_1, \dots, \bU_n$, define an approximation of population variance of $\bX$ to be $\bSigma_{x}=\bW^\T \bS_{u} \bW+\bSigma_{e}$. Note that $\bSigma_{x}$ is not exactly the covariance matrix of $\bX$ because we do not restrict $\bOmega_i$ to be diagonal and define $\bSigma_{e}$ to be diagonal by setting the off-diagonal of $n^{-1}\sum_{i=1}^{n} \bOmega_i$ to be zero.  
Based on the factor model in~\eqref{eq:XUmodel}, the maximum likelihood estimators of $\bW$, $\bS_u$ and ${\bSigma}_e$ are obtained as follows:
\begin{equation}
    (\hat{\bW}, \hat{\bS}_u, \hat{\bSigma}_{e}) =  \underset{{\bW}, {\bS}_u, {\bSigma}_{e}}{\operatorname{argmax}} \left\{ -(2p)^{-1} \log |\bSigma_{x}|-(2 p)^{-1} \operatorname{tr}(\bS_{x} \bSigma_{x}^{-1})\right\}. \nonumber
\end{equation}

 Computationally, we employ the Expectation-Maximization (EM) algorithm to obtain the maximum likelihood estimators as suggested in \citet{bai2012} and \citet{bai2016maximum},  where the authors proved the EM solutions are the stationary points for the likelihood function. Specifically, in this iterated EM algorithm, we use the principal components estimator as the initial estimator. Because principal component estimators are shown to be consistent estimators under similar model assumptions as ours~\citep{fan2013large, wang2017asymptotics}, using the principal component estimators in initialization instead of using random initialization helps to improve algorithm efficiency and find more refined estimation results.  At the $t$-th iteration, denote the estimators at this step to be $\bW^{(t)}$ and $\bSigma_e^{(t)}$. The EM algorithm updates the estimators to be 
\begin{eqnarray}
\{\bW^{(t+1)}\}^{\T} &=& \left[ n^{-1} \sum_{i=1}^n\EE(\bX_i \bU_i^{\T} \mid \bX, \bW^{(t)}, \bSigma_e^{(t)} )\right]  \left[ n^{-1} \sum_{i=1}^n\EE(\bU_i \bU_i^{\T} \mid \bX, \bW^{(t)}, \bSigma_e^{(t)} )\right]^{-1} \nonumber \\
\bSigma_e^{(t+1)} &=& \diag (\bS_x - \bW^{(t+1)} \{\bW^{(t)}\}^{\T} \{\bSigma_x^{(t)}\}^{-1} \bS_x) \nonumber 
\end{eqnarray}
where $\bSigma_x^{t} = \{\bW^{(t)}\}^{\T}\bW^{(t)} + \bSigma_e^{(t)}$,
\begin{eqnarray}
 n^{-1} \sum_{i=1}^n\EE(\bX_i \bU_i^{\T} \mid \bX, \bW^{(t)}, \bSigma_e^{(t)} ) &=& \bS_x \{\bSigma_e^{(t)}\}^{-1} \{\bW^{(t)}\}^{\T}, \nonumber \\
 n^{-1} \sum_{i=1}^n\EE(\bU_i \bU_i^{\T} \mid \bX, \bW^{(t)}, \bSigma_e^{(t)} ) &=&  \bW^{(t)} \{\bSigma_e^{(t)}\}^{-1} \bS_x \{\bSigma_e^{(t)}\}^{-1} \{\bW^{(t)}\}^{\T}  \nonumber \\
&& + \II_K - \bW^{(t)} \{\bSigma_e^{(t)}\}^{-1}\{\bW^{(t)}\}^{\T}.\nonumber
\end{eqnarray}
The iterative steps stop when $\|\bW^{(t+1)} -  \bW^{(t)}\|_F$ and $\|\bSigma_e^{(t+1)} -  \bSigma_e^{(t)}\|_F$ are less than certain tolerance value. Let $\bW^{\dag}$ and $\bSigma_e^{\dag}$ to be the estimators at the last step, and $\mathcal{V}$ to be the matrix containing eigenvectors of $p^{-1} \bW^{\dag}(\bSigma_e^{\dag})^{-1} (\bW^{\dag})^{\T}$ corresponding to the descending eigenvalues. The maximum likelihood estimators are $\hat{\bW} = \mathcal{V}^{\T}\bW^{\dag}$ and $\hat{\bSigma}_e = \bSigma_e^{\dag}$. The estimator $\hat{\bS}_u$ is obtained after estimating $\hat{\bW}$ and $\hat{\bSigma}_e$. As $\hat{\bS}_u$ is not our focus in this method, we omit its derivation details in this paper.

 With $\hat{\bW}$ and $\hat{\bSigma}_e$, we then estimate $\bU_i$ using the generalized least squares estimator:
\begin{equation}
\hat{\bU}_{i} = (\hat{\bW} \hat{\bSigma}_{e}^{-1} \hat{\bW}^\T)^{-1} \hat{\bW} \hat{\bSigma}_{e}^{-1}(\bX_{i}-\bar{\bX}).  \label{Ui hat} 
\end{equation}
 
Next, we treat the estimators $\hat{\bU}_{1},\ldots,\hat{\bU}_n$ as surrogate variables for the underlying unmeasured confounders to fit a high-dimensional generalized linear model.  
We then construct a debiased estimator for the parameter of interest by generalizing the decorrelated score method proposed by \citet{ning2017}. Other debiasing approaches such as that of \cite{van2014} could also be similarly developed.

\medskip
{\em Initial $\ell_1$-Penalized Estimator:} Recall that $\bm{\eta} = (\theta,\bv^{\T},\bbeta^{\T})^{\T}$.  
For vector $\br = (r_1, \dots, r_l)^{\T}$, define  
 $\|\br\|_1 = \sum_{j=1}^l |r_j|$. 
To fit a high-dimensional generalized linear model, we solve the $\ell_1$-penalized optimization problem  
\begin{equation}
\hat{\bm{\eta}} =\underset{\bm{\eta} \in \RR^{(p+K)}}{\operatorname{argmin}} \left[ -\frac{1}{n} \sum_{i=1}^{n} \{y_{i} (\theta D_{i}+\bv^\T \bQ_i + \bbeta^\T\hat{\bU}_i)- b (\theta D_{i}+\bv^\T \bQ_i + \bbeta^\T \hat{\bU}_i) \} + \lambda (|\theta| +     \|\bv\|_{1})\right], \nonumber
\end{equation}
where $\hat{\bm{\eta}}=(\hat{\theta}, \hat{\bv}^{\T}, \hat{\bbeta}^{\T})^\T$ and $b(t)$ is a known function, given a specific generalized linear model. 
 Throughout the manuscript, for notational convenience, let ${\bzeta} = (\bv^\T, \bbeta^\T)^\T$ be regression coefficients for the nuisance covariates and unmeasured confounders. Accordingly, we have $\bm{\eta} = (\theta, \bzeta^\T)^\T$, with the goal of performing statistical inference on $\theta$. In the following, we construct a debiased estimator for $\theta$, generalizing the approach in \cite{ning2017} to situations involving unmeasured confounders.

\bigskip
{\em A Debiased Estimator:}
Before we unfold the details of constructing a debiased estimator, we start with introducing some notations.  Let $\bI = E\{b^{\prime\prime}(\theta D_{i}+\bv^\T \bQ_i + \bbeta^\T\bU_i)\bZ_i \bZ_i^\T \}$ be the Fisher information matrix, and let $\bw^\T =\bI_{\theta \bzeta}\bI_{\bzeta \bzeta}^{-1}$, where $\bI_{\theta \bzeta}$ and $\bI_{\bzeta \bzeta}$ are corresponding block matrices of $\bI$.
In addition, let ${I}_{\theta\mid \bzeta} = E\{b^{\prime\prime}(\theta D_{i}+\bv^\T \bQ_i + \bbeta^\T \bU_i) D_i(D_i - \bw^\T \bM_i )\}$ be the partial  Fisher information matrix,
where $\bM_i = (\bQ_i^\T, \bU_i^\T)^\T$ is a vector of nuisance covariates and unmeasured confounders.
Finally, let $l(\theta, \bzeta) = -n^{-1} \sum_{i=1}^n \{y_i (\theta D_i + \bzeta^T \dot{\bM}_i) - b(\theta D_i + \bzeta^T \dot{\bM}_i) \}$ with $\dot{\bM}_i = (\bQ_i^\T, \hat{\bU}_i^\T)^\T$ be the loss function. 

We define $S(\theta, \bzeta) = \nabla_{\theta} l(\theta, \bzeta) - \bw^\T \nabla_{\bzeta} l(\theta, \bzeta)$ as the generalized decorrelated score function, where $\nabla_{\theta} l(\theta, \bzeta)$ and $\nabla_{\bzeta} l(\theta, \bzeta)$ are the partial derivatives of the loss function with respect to $\theta$ and $\bzeta$, respectively.  
Different from the existing definition of the decorrelated score function in \citet{ning2017}, the generalized decorrelated score function takes into account the effects induced by the unmeasured confounders. 
Specifically, in the presence of unmeasured confounders, the generalized decorrelated score function is uncorrelated with the score function corresponding to the nuisance covariates as well as the unmeasured confounders, i.e.,  $E\{S(\theta, \bzeta)^\T \nabla_{\bzeta} l(\theta, \bzeta)\}= \bI_{\theta \bzeta} - \bI_{\theta \bzeta} \bI_{\bzeta \bzeta}^{-1} \bI_{\bzeta \bzeta} = \bm{0}$. 
The debiased estimator of $\theta$ is constructed by solving for $\tilde{\theta}$ from the first-order approximation of the generalized decorrelated score function
$\hat{S}(\hat{\theta}, \hat{\bzeta}) + \hat{I}_{\theta\mid \bzeta} (\tilde{\theta} -\hat{\theta}) = 0$.   
From the first-order approximation equation, we see that to establish the debiased estimator $\tilde{\theta}$, we need to construct two estimators $\hat{S}(\hat{\theta}, \hat{\bzeta})$ and $\hat{I}_{\theta \mid \bzeta}$, and the key is to estimate $\bw$.

We estimate $\bw$ by solving the following convex optimization problem:
\begin{equation}
	\hat{\bw} = \underset{\bw \in \mathbb{R}^{(p+K-1)}}{\operatorname{argmin}}~\frac{1}{2n} \sum_{i=1}^{n} \{\bw^\T \nabla_{\bzeta\bzeta} l_{i}(\hat{\theta}, \hat{\bzeta}) \bw - 2 \bw^{\T} \nabla_{\bzeta\theta} l_{i}(\hat{\theta}, \hat{\bzeta})\}+\lambda^{\prime} \|\bw\|_1, \nonumber
\end{equation}
where $l_i(\theta, \bzeta) = -y_i (\theta D_i + \bzeta^T \dot{\bM}_i) + b(\theta D_i + \bzeta^T \dot{\bM}_i)$ is the $i$th component of the loss function and $\lambda'>0 $ is a sparsity tuning parameter for $\bw$. Equivalently, the estimator $\hat{\bw}$ is obtained by
\begin{equation}
\hat{\bw} = \underset{\bw \in \mathbb{R}^{(p+K-1)}}{\operatorname{argmin}}~\frac{1}{2n} \sum_{i=1}^{n}b^{\prime\prime} (\hat{\theta}D_i + \hat{\bzeta}^{\T} \dot{\bM}_i)(D_i -  \bw^{\T}\dot{\bM}_i )^2+\lambda^{\prime} \|\bw\|_1.  \label{eq:w estimation}
\end{equation}The estimator $\hat{\bw}$ is constructed with the intuition of finding a sparse vector such that the generalized decorrelated score function is approximately zero. This coincides with the intuition to solve for $\tilde{\theta}$ from the first-order approximation of the generalized decorrelated score function.
Under the null hypothesis $H_0: \theta^* = \theta^0$, we estimate the generalized decorrelated score function and the partial Fisher information matrix by
\begin{eqnarray}
	\hat{S}(\theta^0, \hat{\bzeta}) &=& -\frac{1}{n}\sum_{i=1}^n\{y_i - b^{\prime}(\theta^0 D_i + \hat{\bv}^\T \bQ_i + \hat{\bbeta}^\T\hat{\bU}_i)\}(D_i - \hat{\bw}^\T\dot{\bM}_i); \nonumber \\
	\hat{{I}}_{\theta\mid \bzeta} &=& \frac{1}{n} \sum_{i=1}^n b^{\prime\prime}(\hat{\theta} D_{i}+\hat{\bv}^\T \bQ_i + \hat{\bbeta}^\T\hat{\bU}_i) D_i(D_i - \hat{\bw}^\T\dot{\bM_i}). \nonumber
\end{eqnarray}
The debiased estimator can then be constructed as $\tilde{\theta} = \hat{\theta} - (\hat{I}_{\theta \mid \bzeta} )^{-1} \hat{S}(\theta^0, \hat{\bzeta})$.
We will show in Section~\ref{Sec:theoretical results} that the debiased estimator $\tilde{\theta}$ is asymptotically normal.
Subsequently, the  $(1-\alpha) \times 100 \% $ confidence interval for $\theta$ 
 can be constructed as 
 \begin{equation}
     \{\tilde{\theta} - (n\hat{I}_{\theta \mid \bzeta})^{-1/2}{\Phi}^{-1}(1-{\alpha}/2),\  \tilde{\theta} + (n\hat{I}_{\theta \mid \bzeta})^{-1/2}{\Phi}^{-1}(1-{\alpha}/2)\}, \label{eq:confidence interval}
 \end{equation}
where $\Phi(t)$ is the cumulative distribution function for the standard normal random variable.  

\section{Theoretical Results}
\label{Sec:theoretical results}
Recall that our proposed method yields estimators  $\hat{\bm{\eta}}$, $\hat{\bw}$, and the debiased estimator $\tilde{\theta}$. 
In this section, we first establish upper bounds for the estimation errors of $\hat{\bm{\eta}}$ and $\hat{\bw}$ under the $\ell_1$ norm.  Subsequently, we show that the debiased estimator $\tilde{\theta}$ is asymptotically normal.

For a vector $\br = (r_1, \dots, r_l)^{\T}$, let $\|\br\|_q = (\sum_{j=1}^l |r_j|^q)^{1/q}$ for $q \geq 1$ and let $\|\br\|_{\infty}= \max_{j=1,\ldots, l} |r_j|$. 
For any matrix $\bA$, let  $\lambda_{\max}(\bA)$ and $\lambda_{\min}(\bA)$ represent the largest and smallest eigenvalues of $\bA$, respectively.
Moreover, for sequences $\{a_n\}$ and $\{b_n\}$, we write $a_n \lesssim b_n $ if there exists a constant $C >0$ such that $a_n \leq C b_n$ for all $n$, and $a_n \asymp b_n$ if $a_n \lesssim b_n $ and $b_n \lesssim a_n$. 
For a sub-exponential random variable $Y_1$, we write  $\|Y_1\|_{\varphi_1}=\inf [s>0 : E \{\exp(Y_1/s)\} \leq 2]$ as the sub-exponential norm. 
For a sub-Gaussian random variable $Y_2$, we write $\|Y_2\|_{\varphi_2} = \inf [s>0 : E \{\exp{(Y_2^2/s^2)}\} \leq 2]$ as the sub-Gaussian norm.
Throughout the manuscript, we will use an asterisk on the upper subscript to indicate the population parameters.   
In addition, we define $s_{\eta} = \text{card}\{(j: \eta^*_j \neq 0)\}$ and $s_{w} = \text{card}\{(j: \bw^*_j \neq 0)\}$ as the cardinalities of $\bm{\eta}^*$ and $\bw^*$, respectively.
All of our theoretical analysis are performed under the regime in which $n$, $p$, $s_{\eta}$, and $s_{w}$ are allowed to increase, and the number of unmeasured confounders $K$ is fixed. 

We start with some conditions on the factor model in~\eqref{eq:XUmodel}. Similar conditions were also considered in \cite{bai2016maximum} in the context of high-dimensional approximate factor model.

\begin{assumption}
\label{assumption1}
For some large constant $C > 0$,

 (a)  $\EE(E_{ij}) = 0$, $\EE({E}_{ij}^{8}) \leq C$.
 
 (b) $\EE(E_{ih}E_{ij}) = \tau_{i, hj}$ with $| \tau_{i, hj}| \leq  \tau_{hj}$ for some $\tau_{hj} > 0$ and all $i = 1, \dots, n$, and $\sum_{h=1}^p\tau_{hj}\leq C$ for all $j=1,\dots,p$.

 (c) $\EE(E_{ij}E_{sj}) = \rho_{is, j}$ with $| \rho_{is, j}| \leq \rho_{is}$ for some $\rho_{is} >0$ and all $j=1,\dots, p$, and $n^{-1}\sum_{i=1}^n \sum_{s=1}^n \rho_{is} $ $  \leq C$.

 (d) For all $j,q = 1, \dots, p$,
 \begin{equation}
     \EE \left\{\left|\frac{1}{\sqrt{n}} \sum_{i=1}^n [E_{ij} E_{iq} - \EE (E_{ij} E_{iq})]\right|^4  \right\} \leq C. \nonumber
 \end{equation}

 (e) $\|\bW_{j}^*\|_2 \leq C$ and  $\sigma_j^2$ are estimated within the set $[C^{-2}, C^2]$ for all $j$. For positive definite matrices $\bGamma^*$ and $\Upsilon^*$,  $\lim _{p \rightarrow \infty} p^{-1} \bW^* (\bSigma_{e}^*)^{-1} (\bW^*)^\T=\bGamma^*$ and $\lim _{p \rightarrow \infty}p^{-1} \sum_{j=1}^{p} ({\sigma}_{j}^*)^{-4}$ $\{(\bW_{j}^*)^\T \otimes(\bW_{j}^*)^\T \}(\bW_{j}^* \otimes \bW_{j}^*)= \Upsilon^*$, where $\otimes$ is the Kronecker product. 
\end{assumption}

 Assumption~\ref{assumption1} is more general than assumptions in classical factor analysis~\citep{anderson1988asymptotic, bai2003, fan2013large, bai2016maximum}. 
Instead of constraining all the $\bE_i$ to have a diagonal covariance matrix, we now only require the higher-order moment of $\bE_i$ to be bounded, the diagonal entries $\sigma_j^*$'s to be bounded, as well as the magnitudes of the correlations among entries of $\bE_i$ to be controlled in Assumption~\ref{assumption1}.  The conditions are mild as we only control the magnitude of correlations rather than assuming zero correlations as in classical factor analysis. 
 Assumption~\ref{assumption1}(e) is a regularity condition for restricting the parameters. Overall, Assumption~\ref{assumption1} follows standard conditions for the approximate factor model in~\cite{bai2016maximum} and is required for the estimation consistency of the unmeasured confounders.

     Comparing Assumption~\ref{assumption1} to the dense confounding assumption~(A2) imposed on the linear framework in~\cite{guo2021doubly}, our assumption is mild as it holds for a broad regime of $n$ and $p$. Specifically, the dense confounding assumption (A2) in \cite{guo2021doubly} is related to our assumption 1(e) that $\lim _{p \rightarrow \infty} p^{-1} \bW^* (\bSigma_{e}^*)^{-1} (\bW^*)^\T=\bGamma^*$ where $\bGamma^*$ is positive definite. Our assumption follows from classical factor analysis literature and as a result, we have $\lambda_q(\bW^*) \asymp \sqrt{p}$, where $\lambda_q(\bW^*)$ is the $q$-th singular value of the factor loading matrix $\bW^*$. With $\bgamma^{\T} = (\theta, \bv^{\T})$ being the coefficient for all covariates and $\gamma_j$ being the individual coefficient of interest, the dense confounding assumption mainly requires that 
\begin{equation}
    \lambda_q(\bW_{-j}^*) \gg \max \left\{M \sqrt{Kpn^{-1}}  (\log p)^{3/4}, \sqrt{MK} p^{1/4} (\log p)^{3/8}, \sqrt{Kn\log p}\right\}, \nonumber
\end{equation}
where $\bW_{-j}^*$ denotes the factor loading matrix $\bW$ with $j$-th column removed. \cite{guo2021doubly} focuses on the setting $p \gg n$ and they point out that in the high-dimensional regime and under certain settings, the dense confounding assumption holds with high probability. Specifically, when the entries of $\bW$ are i.i.d. Sub-Gaussian with zero mean and variance $\sigma_w^2$, it holds that $\lambda_q(\bW_{-j}) \asymp \sqrt{p} \sigma_w$ and the dense confounding assumption requires $p \gg Kn \log p$ and $\min \{n, p \} \gg K^3 (\log p)^{3/2} M^2$ to make $\sigma_w$ diminish to zero. However, this condition is restricted to the high-dimensional regime and may not hold when $p$ is  of relatively lower order.

Moreover, without loss of generality, we assume a working identifiability condition: $\bS_u = \bI_K$ and $p^{-1}\bW^*(\bSigma_e^*)^{-1}(\bW^*)^\T$ is a diagonal matrix with distinct entries. 
Note that the aforementioned working identifiability condition is for presentation purpose only and not an assumption on the model structure. 
As presented in Appendix B, when such a working identifiability condition is not satisfied, our theoretical results in Theorems~\ref{Initial estimator consistency} and~\ref{normality theorem} are still valid. 
We also illustrate this via simulation in Section~\ref{Sec:simulation}. 
Next, we impose some assumptions on the generalized linear model with unmeasured confounders in~\eqref{eq:full y model}.

\begin{assumption} 
\label{glm assumption} (a) The Fisher information $\bI^* = E[b^{\prime\prime}\{(\bm{\eta}^*)^\T \bZ_{i}\}{\bZ}_i{\bZ}_i^\T ]$ satisfies  $\lambda_{\min}(\bI^*) \geq \kappa$, where $\kappa >0$ is some constant.

 (b) For some constant $M >0$,   $\|\bX_i\|_{\infty} \leq M$,  $\|\bU_i\|_{\infty} \leq M$, $\| \bm{\eta}^*\|_{\infty} \leq M$ and $|(\bw_q^*)^\T\bQ_i| \leq M$, where $\bw_q^* =(w_2^*,\ldots, w_p^*)^\T$.
 
  (c) The term $|y_i - b^{\prime}\{(\bm{\eta}^*)^{\T}{\bZ}_i\} |$ is sub-exponential with $\|y_i - b^{\prime}\{(\bm{\eta}^*)^{\T}{\bZ}_i\}\|_{\varphi_1} \leq M$. 
  
  (d) Assume that  $a_1 \leq (\bm{\eta}^*)^{\T}{\bZ}_i \leq a_2$ and $0 \leq |b^{\prime}(t)| \leq B$ with $|b^{\prime}(t_{1})-b^{\prime}(t)| \leq B |(t_1 - t)b^{\prime}(t)| $ and $0 \leq b^{\prime\prime}(t) \leq B$ with $|b^{\prime\prime}(t_{1})-b^{\prime\prime}(t)| \leq B |t_1 - t| b^{\prime\prime}(t)$ for constants $a_1$, $a_2$ and $B$, where $t \in [a_1 - \epsilon, a_2 + \epsilon]$ for $\epsilon > 0$ and sequence $t_1$ satisfies $|t_1 - t| = o(1)$.
\end{assumption}

In the absence of unmeasured confounders, similar conditions  in Assumption~\ref{glm assumption} can be implied from the conditions in Theorem 3.3 in \cite{van2014} and Assumption E.1 in \cite{ning2017}. 
When $\bU_i$ and $\bX_i$ are binary or categorical, Assumption \ref{glm assumption}(b) holds with a constant $M>0$. 
When $\bU_i$ and $\bX_i$ are sub-exponential random vectors, Assumption \ref{glm assumption}(b) holds with  $M = c n^{-1/2} (\log p)^{1/2}$ for some constant $c > 0$, with probability at least $1 - p^{-1}$.
Assumption~\ref{glm assumption}(d) imposes mild regularity conditions on the function $b(t)$, and is commonly used in analyzing high-dimensional generalized linear models without unmeasured confounders. We require the function $b(t)$ to be at least twice differentiable and $b^{\prime}(t)$ and $b^{\prime\prime}(t)$ to be bounded. 
Specifically, $|b^{\prime\prime}(t_1)- b^{\prime\prime}(t)| \leq B|t_{1}-t| b^{\prime \prime}(t)$ can be implied by $|b^{\prime\prime\prime}(t)| \leq B b^{\prime\prime}(t)$ when $b^{\prime\prime\prime}(t)$ exists, which is a weaker condition than the self-concordance property \citep{bach2010}. 
This boundary assumption is important for the concentration of the Hessian matrix of the loss function. 
Assumption~\ref{glm assumption}(d) holds for commonly used generalized linear models.  
For example, for logistic regression where  $b(t) = \log \{1+\exp (t)\}$, Assumption~\ref{glm assumption}(d) holds with $B = 1$, and $a_1, a_2$  extended to infinity and it can be similarly verified to hold at $B = \max \{1, \exp(a_2+\epsilon) \}$ for Poisson regression and $B = \max \{1/(a_2 +\epsilon)^2, 2/|a_2  + \epsilon| \}$ for exponential regression with $a_2 < 0$. 
For linear model, Assumption~\ref{glm assumption} can be relaxed as stated in Remark~\ref{remark:linear model}.

The theoretical analysis on the unmeasured confounders estimator is important in establishing the theoretical guarantee for our debiased method. As the decomposition techniques commonly used in linear models may not be applicable in generalized linear model settings, it is necessary to establish more general and refined intermediate results as the foundation of our theoretical analysis (see Remark~\ref{remark:estimate U} in Appendix D.1 for more details). We first present a uniform convergence result for the estimators of the unmeasured confounders.
\begin{proposition}
\label{prop:uniform U}
Under Assumptions~\ref{assumption1}--\ref{glm assumption}, if $n, p \rightarrow \infty$, we have
\begin{equation}
\max_{i \in \{1, \dots, n\}} \| \hat{\bU}_i - \bU_i^*\|_{\infty} = O_p \left( \frac{1}{\sqrt{n}} + \sqrt{\frac{\log n}{p}} \right). \nonumber
\end{equation}
\end{proposition}

From Proposition~\ref{prop:uniform U}, the estimator of unmeasured confounders $\hat{\bU}_i$ uniformly converges to $\bU_i$ at $p \gg \log n$, which holds naturally under the high-dimensional regime $p \gg n$. Moreover, the convergence rate
$O_p(n^{-1/2} + p^{-1/2} $ $ (\log n)^{1/2})$ is of a similar order to the convergence rates of principal component estimators~\citep{fan2013large, wang2017asymptotics}. The estimation results of $\hat{\bm{\eta}}$ and $\hat{\bw}$ are dependent on the accuracy of the estimator of unmeasured confounders and we next establish upper bounds on the estimation errors for $\hat{\bm{\eta}}$ and $\hat{\bw}$.

\begin{theorem}[Estimation consistency]
\label{Initial estimator consistency}
 	 Suppose that Assumptions~\ref{assumption1}--\ref{glm assumption} hold. With $\lambda \asymp \lambda^{\prime} \asymp n^{-1/2}(\log p)^{1/2} + p^{-1/2} (\log n)^{1/2}$, if the scaling condition $n,p \rightarrow \infty$ and $(s_{w} \vee s_{\eta})$ $ \{ n^{-1/2}(\log p)^{1/2} + p^{-1/2}(\log n)^{1/2}\}$ $ = o_p(1)$ hold, then we have   
	\begin{eqnarray}
		&& \|\hat{\bm{\eta}}-\bm{\eta}^*\|_1  = O_p\left\{s_{\eta} \left( \sqrt{\frac{\log p}{n}} + \sqrt{\frac{\log n}{p}} \right) \right\}; \nonumber \\
  && \|\hat{\bw}-\bw^*\|_1 = O_p\left\{(s_{w} \vee s_{\eta}) \left( \sqrt{\frac{\log p}{n}} + \sqrt{\frac{\log n}{p}} \right)  \right\}. \nonumber\label{estimation error bound}
	\end{eqnarray}
\end{theorem}

  The effect of unmeasured confounders enters the rate of the estimation error in our results through the term $p^{-1/2} (\log n)^{1/2}$. Under the high-dimensional setting with $p \gg n$, the unmeasured confounders effect is dominated by $ n^{-1/2} (\log p)^{1/2}$ and as a result, the convergence rates in Theorem~\ref{Initial estimator consistency} are of the same order as in the oracle case when the unmeasured confounders are assumed to be known~\citep{van2014, ning2017}. With the estimation consistency established, we show that the debiased estimator from the proposed estimation method is asymptotically normal, and thus valid confidence intervals can be constructed.  

\begin{theorem}[Asymptotic normality]
\label{normality theorem}
Suppose that Assumptions~\ref{assumption1}--\ref{glm assumption} hold. With $\lambda \asymp \lambda^{\prime} \asymp n^{-1/2}(\log p)^{1/2} + p^{-1/2} (\log n)^{1/2}$, if the scaling condition $n,p \rightarrow \infty$ and $(s_{w} \vee s_{\eta})$ $(n^{-1/2}\log p + p^{-1}n^{1/2}\log n)= o_p(1)$ hold, then 
\begin{equation}
	{{n}^{1/2} ({I}_{\theta \mid \bzeta}^*})^{1/2}(\tilde{\theta} - \theta^*) \rightarrow_d N(0,1). \label{asy normal}
	\end{equation}
\end{theorem}

The result in Theorem~\ref{normality theorem} is similar to existing inference results for high-dimensional generalized linear models without unmeasured confounders \citep{van2014, ning2017, ma2021, shi2021, cai2021glm}. Theorem~\ref{normality theorem} is established under the condition that the estimation errors of $\hat{\bm{\eta}}$ and $\hat{\bw}$ in Theorem~\ref{Initial estimator consistency} are $o_p(1)$.
As a consequence of the asymptotic normality result in Theorem~\ref{normality theorem} and that $\hat{{I}}_{\theta\mid \bzeta}$ is consistent estimator for ${I}_{\theta \mid \bzeta}^*$, the construction of the confidence interval in~\eqref{eq:confidence interval}
is valid. 
 
\begin{remark}
\label{remark:linear model}
	For the linear model,  estimation consistency and asymptotic normality results in Theorems~\ref{Initial estimator consistency} and \ref{normality theorem}  hold with less stringent conditions than that in Assumption \ref{glm assumption}.
	Suppose Assumption~\ref{assumption1} and Assumption~\ref{glm assumption}(a) hold and assume that the random noise $\varepsilon_i$ is independent sub-Gaussian random variable and $\bZ_i$ is sub-Gaussian vector such that $\|\varepsilon_i\|_{\varphi_2} \leq M$ and $\|Z_{ij}\|_{\varphi_2} \leq M$ for some constant $M > 0$.
	 By choosing  $\lambda \asymp \lambda^{\prime} \asymp n^{-1/2}(\log p)^{1/2} + p^{-1/2} (\log n)^{1/2}$, if $n,p \rightarrow \infty$ and $(s_{w} \vee s_{\eta})(n^{-1/2}\log p + p^{-1}n^{1/2}\log n)= o_p(1)$, the estimation error bounds in Theorem~\ref{Initial estimator consistency} hold and the debiased estimator $\tilde{\theta}$ is asymptotically normal with limiting distribution \eqref{asy normal}. 
Similar assumptions are commonly imposed in many existing inference methods for high-dimensional linear models without unmeasured confounders \citep{zhang2014, javanmard2014confidence}  as well as in the existence of  confounders \citep{Domagoj2020, guo2021doubly}.
\end{remark}

\begin{remark}
\label{remark:sparsity}
 The sparsity assumption on $\bw$ is a standard assumption in high-dimensional regression models without unmeasured confounders. It may be more suitable and even weaker in our proposed framework with unmeasured confounders. 
In high-dimensional regression models without unmeasured confounders, the sparsity assumption on $\bw$ is implied by the sparse inverse population Hessian condition~\citep{van2014, belloni2016post, ning2017, jankova2018semiparametric}.
Specifically, the sparse inverse population Hessian condition implies the sparsity of the coefficient parameter $\bw$ in a weighted node-wise lasso regression $D_i \sim \bQ_i$, which is similar to~\eqref{eq:w estimation} but regresses the covariate of interest $D_i$ on the rest of the covariates $\bQ_i$.

In our proposed setting with unmeasured confounders, as shown from~\eqref{eq:w estimation}, we assume that in the weighted node-wise lasso regression $D_i \sim \bQ_i + \bU_i$, the coefficient of $\bQ_i$, denoted as $\bw_q$, to be sparse. 
That is,  $\bw_q$  is sparse, conditional on the unmeasured confounders $\bU_i$.
The sparsity assumption is mild and can be satisfied in many settings.
For example, under the assumptions in classical factor analysis in which the covariance matrix of the random error $\bE_i$ is diagonal, $D_i$ is uncorrelated with the other covariates $\bQ_i$, conditioned on the unmeasured confounders $\bU_i$. In this case, we have $\bw_q = \bm{0}$, conditioned on $\bU_i$ and thus the sparsity assumption holds naturally. 
We also want to point out that the imposed sparsity assumption may be weaker than that of existing work on high-dimensional inference without unmeasured confounders, where the coefficients of regression $D_i \sim \bQ_i$ are assumed to be sparse.
 For instance, we allow $D_i$ to be densely correlated with $\bQ_i$ marginally.  In other words, we only require $\bw_q$, the coefficient for $\bQ_i$ in $D_i \sim \bQ_i + \bU_i$, to be sparse, conditional on the confounders $\bU_i$ while marginally the coefficients of $D_i \sim \bQ_i$ could be dense. 
\end{remark}

\begin{remark}
For the simplicity of theoretical analysis, we assume the dimension of unmeasured confounders $K$ to be fixed and known, which is also assumed in \cite{wang2017asymptotics}, \cite{fan2022latent}, \cite{wang2022maximum} and many other works. Nevertheless, our theoretical results hold as long as $K$ is consistently estimated.
As introduced in Section~\ref{Sec:methodology}, we use parallel analysis to estimate the dimension of unmeasured confounders in practice. 
Theoretically, it has been shown that parallel analysis consistently selects the dimension of unmeasured confounders in factor analysis~\citep{dobriban2020permutation}. Specifically, when the dimension $p$ is relatively large compared to the sample size $n$, each factor loads on not too few variables, and the signal size of unmeasured confounders is not too large, parallel analysis selects the number of factors with probability tending to one~\citep{dobriban2020permutation}. These conditions can be satisfied under our framework, implying that the dimension of unmeasured confounders is consistently determined. 
Moreover, empirically we find that the proposed method still provides satisfying inference results  under some overestimation of $K$. Intuitively, as long as the corresponding linear combinations of the true underlying factors $\bU$ in the considered models can be well approximated by those of the estimated $\hat \bU$, the developed inference results for $\theta^*$ would still hold.  To further illustrate this,   we perform simulation studies in Appendix~C to  show that the overestimation of $K$ may not affect the asymptotic properties of the debiased estimator.
\end{remark}

\section{Numerical Studies}
\label{sec:numerical}

To illustrate the performance of our proposed method, we conduct numerical experiments including simulation studies and an application of our method to a genetic data set.

\subsection{Simulations}
\label{Sec:simulation}
We consider two different models in~\eqref{eq:full y model}, a linear regression model and a logistic regression model. 
We compare our method with two alternative approaches for performing high-dimensional inference: the oracle method where we perform the debiasing approach assuming that the true values of unmeasured confounders are known; and the naive method in which the unmeasured confounders are neglected in the estimation and we perform the debiasing approach with the observed covariates only. 
In addition, for the linear regression setting, we compare the proposed method with the doubly debiased lasso method proposed by \cite{guo2021doubly}. 
The sparsity tuning parameters for all of the aforementioned methods are selected using 10-fold cross-validation.
Our proposed method involves estimating the dimension of the unmeasured confounders, which we estimate using the parallel analysis \citep{horn1965,dinno2009}.
 To evaluate the performance across the different methods, we construct the 95\% confidence intervals for the parameter of interest, and compute the average confidence length and the coverage probability of the true parameter over 300 independent replications.

First, we generate each entry of the unmeasured confounders $\bU_1,\ldots, \bU_n\in\mathbb{R}^3$ and each entry of the random error $\bE_1,\ldots, \bE_n\in\mathbb{R}^p$ from a standard normal distribution.   
We set \begin{equation}
(\bW^*)^T = 	\left(  \begin{array}{ccc}
\bm{0.5}_{p/3} & \bm{0} & \bm{0}\\
\bm{0} & \bm{1}_{p/3} & \bm{0} \\
\bm{0} & \bm{0} & \bm{1.5}_{p/3} \\
\end{array}\right)_{p\times 3}, \nonumber
\end{equation}
where $\bm{0.5}_{p/3}$, $\bm{1}_{p/3}$, and $\bm{1.5}_{p/3}$ are vectors of dimension $p/3$ with all entries equal 0.5, 1, and 1.5, respectively (in our simulation, $p/3$ is set as integer values). The $\bm{0}$ in the matrix denotes a vector of dimension $p/3$ with all entries equal 0. We then generate the covariates  $\bX_i = (\bW^*)^\T\bU_i +\bE_i$ with $D_i = \bX_{i,1}$ and $\bQ_i = \bX_{i, -1}$. 
It can be verified that the above setting satisfies the working identifiability condition described in Section~\ref{Sec:theoretical results}.  
Later in this section, we also perform additional simulation studies as to illustrate the validity of the theory
when the working identifiability condition does not hold. 

We consider both the linear regression model and the logistic regression model: for linear regression, we generate the response $y_i$ according to $y_i = \theta^* D_i + (\bv^*)^\T \bQ_i + (\bbeta^*)^\T \bU_i + \varepsilon_i $, where the random noise $\varepsilon_i$ is generated from a standard normal distribution; for logistic regression, the response $y_i$ is generated from a Bernoulli distribution with probability $p_i(\theta^*, \bv^*, \bbeta^*) = {1}/(1+\exp [-\{\theta^* D_i + (\bv^*)^\T \bQ_i + (\bbeta^*)^\T \bU_i\}]) $.
The regression coefficients for the unmeasured confounders, parameter of interest, and nuisance parameters are set to be $\bbeta^* = (1, 1, 1)^\T$, $\theta^* = 0$, and $\bv^* = (1, \bm{0}_{p-2})^\T$, respectively.

 \begin{figure}[h!]
 \centering
\includegraphics[scale=0.7]{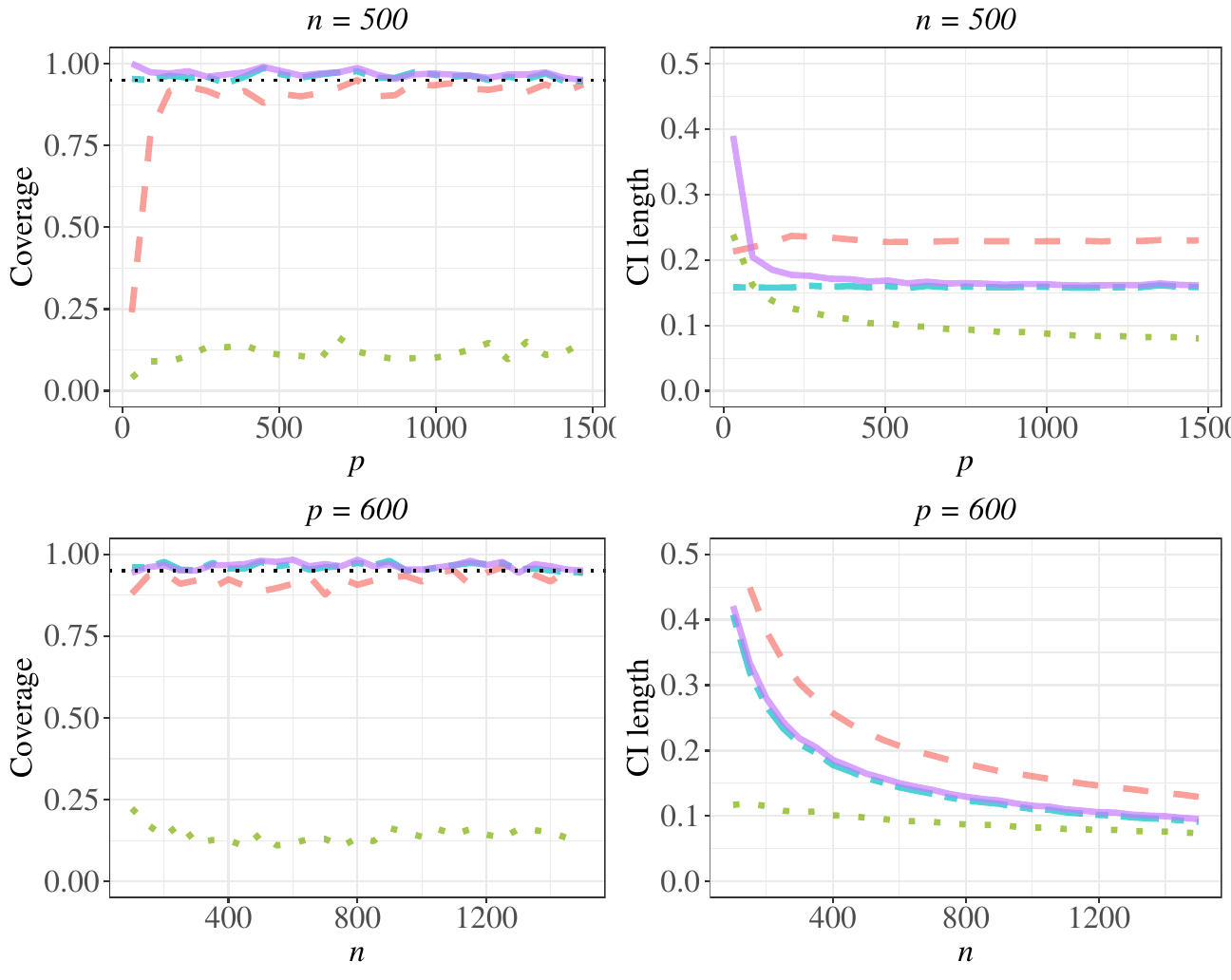}
\caption{Coverage and length of the confidence interval under linear regression, averaged over 300 replications, with varying $p$ and fixed $n = 500$ (top), and with varying $n$ and fixed $p= 600$ (bottom). Black dashlines indicate the 0.95 level. Purple solid lines (\protect\includegraphics[height=0.5em]{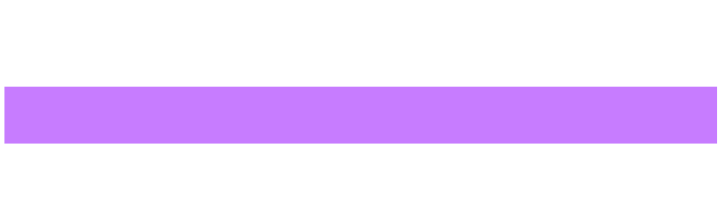}) are our proposed method. 
		Blue two-dashed lines (\protect\includegraphics[height=0.5em]{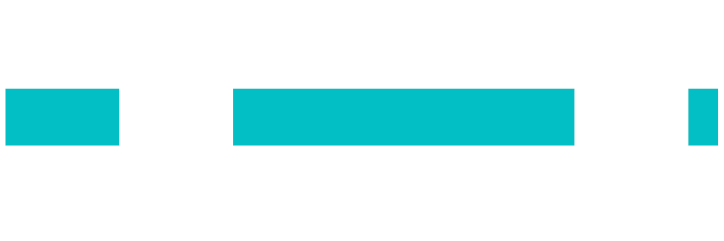})  represent the oracle case. 
		Green dotted lines (\protect\includegraphics[height=0.5em]{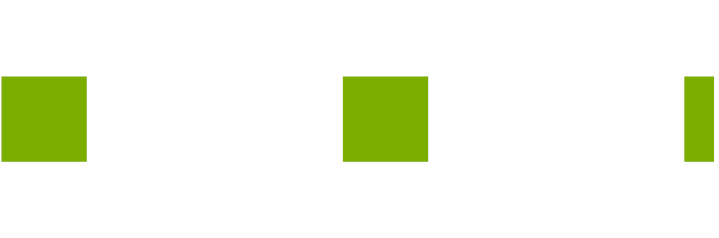}) indicate the naive method. 
		Orange dashed lines
		(\protect\includegraphics[height=0.5em]{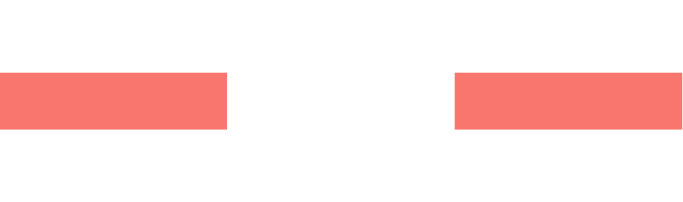}) represent the doubly debiased lasso method.}
	\label{fig:linear_sparse}
\end{figure}

\begin{figure}[h!]
	\centering
	\includegraphics[scale=0.7]{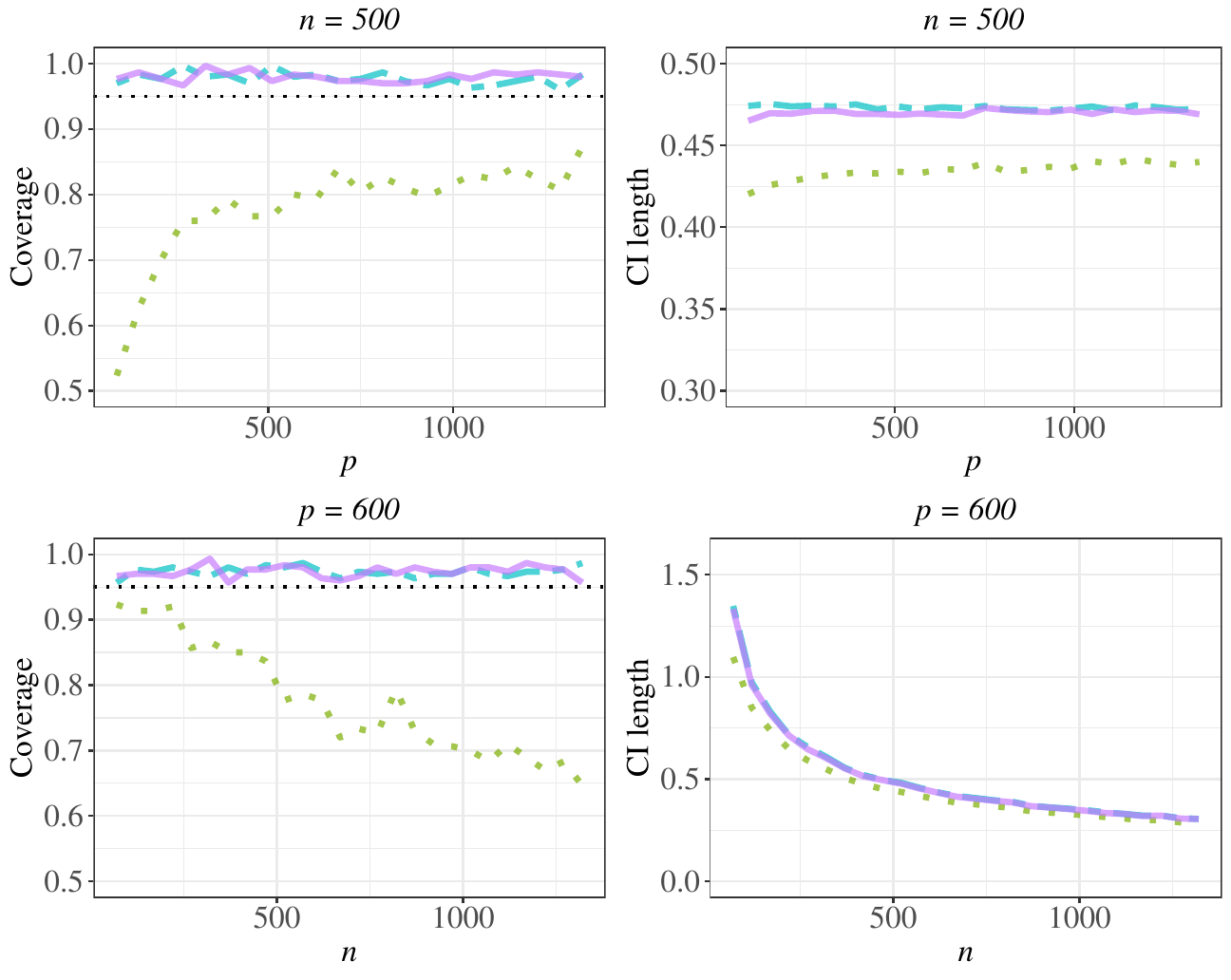}
	\caption{Coverage and length of the confidence interval under logistic regression, averaged over 300 replications, with varying $p$ and fixed $n = 500$ (top), and with varying $n$ and fixed $p= 600$ (bottom). Black dashlines indicate the 0.95 level. 
		Purple solid lines (\protect\includegraphics[height=0.5em]{Figures/purple_solid.png}) are our proposed method.
		Blue two-dashed lines (\protect\includegraphics[height=0.5em]{Figures/blue_twodahsed.png})  represent the oracle case. 
		Green dotted lines (\protect\includegraphics[height=0.5em]{Figures/green_dotted.png}) indicate the naive method.}
	\label{fig:logit_sparse}
\end{figure}

  Results for the cases when $n=500$ and $p$ varies from 100 to 1500 and when $p=600$ and $n$ varies from 100 to 1500 under both linear regression model and logistic regression model are presented in Fig.~\ref{fig:linear_sparse} and Fig.~\ref{fig:logit_sparse}, respectively.
From Fig.~\ref{fig:linear_sparse}, we see that the naive method suffers from undercoverage since the effects induced by the unmeasured confounders are not taken into account.  Our proposed method has coverage at approximately 0.95 and is similar to that of the oracle method. The proposed method also has similar length of the confidence intervals as that of the oracle method.
As a comparison, the doubly debiased lasso method suffers from undercoverage when $p$ is relatively small ($p < 100$ and $n = 500$): our finding is consistent with that of \cite{guo2021doubly}, where they indicate that when $p$ is relatively small to $n$, bias from the hidden confounding effects can be relatively large, which leads to undercoverage of their method.
As $p$ increases, we see that coverage for the doubly debiased lasso method approaches to 0.95, but its confidence interval lengths are larger than that of our proposed method, indicating our method is preferred in this case.

Under logistic regression model, our proposed method performs similarly to the oracle method in terms of both coverage and length of the confidence intervals. 
Results for doubly debiased lasso are not reported as it is not directly applicable to generalized linear model. 

In the remaining subsection, we show that when the working identifiability condition fails, our proposed method can still perform well. More discussions about the working identifiability condition are in Appendix B. Specifically, we set $W_{jk}^*{\sim} \text{ Unif}[0, 1]$ for $j = 1,\dots,p$ and $k = 1,2,$ and $3$. This loading matrix implies that each covariate vector $\bX_i$ is related to all the confounders. The identifiability condition fails in such case as $p^{-1}\bW^*(\bSigma_e^*)^{-1}(\bW^*)^\T$ is not a diagonal matrix. We continue with the parameter setting at the start of this section and the data generation process is unchanged except for replacing the original loading matrix with the new loading matrix where elements are uniformly distributed. 

\begin{figure}[h!]
	\centering
	\includegraphics[scale=0.7]{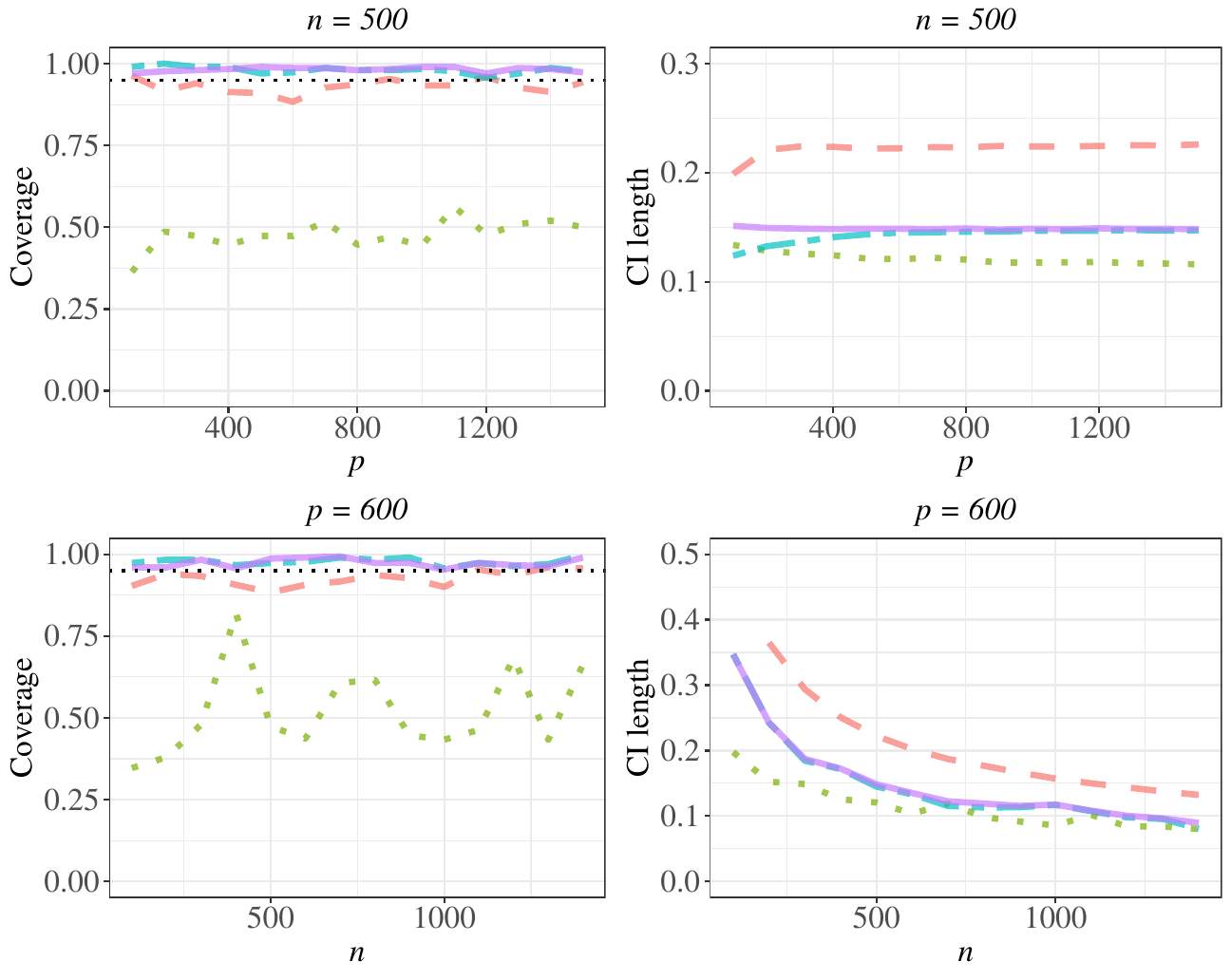}
	\caption{
	Coverage and length of the confidence interval under linear regression with new loading matrix, averaged over 300 replications, with varying $p$ and fixed $n = 500$ (top), and with varying $n$ and fixed $p= 600$ (bottom). Black dashlines indicate the 0.95 level. Purple solid lines (\protect\includegraphics[height=0.5em]{Figures/purple_solid.png}) are our proposed method. 
		Blue two-dashed lines (\protect\includegraphics[height=0.5em]{Figures/blue_twodahsed.png})  represent the oracle case. 
		Green dotted lines (\protect\includegraphics[height=0.5em]{Figures/green_dotted.png}) indicate the naive method. 
		Orange dashed lines(\protect\includegraphics[height=0.5em]{Figures/orange_dashed.png}) represent the doubly debiased lasso method.}
	\label{fig:linear_dense}
\end{figure}

\begin{figure}[h!]
	\centering
	\includegraphics[scale=0.7]{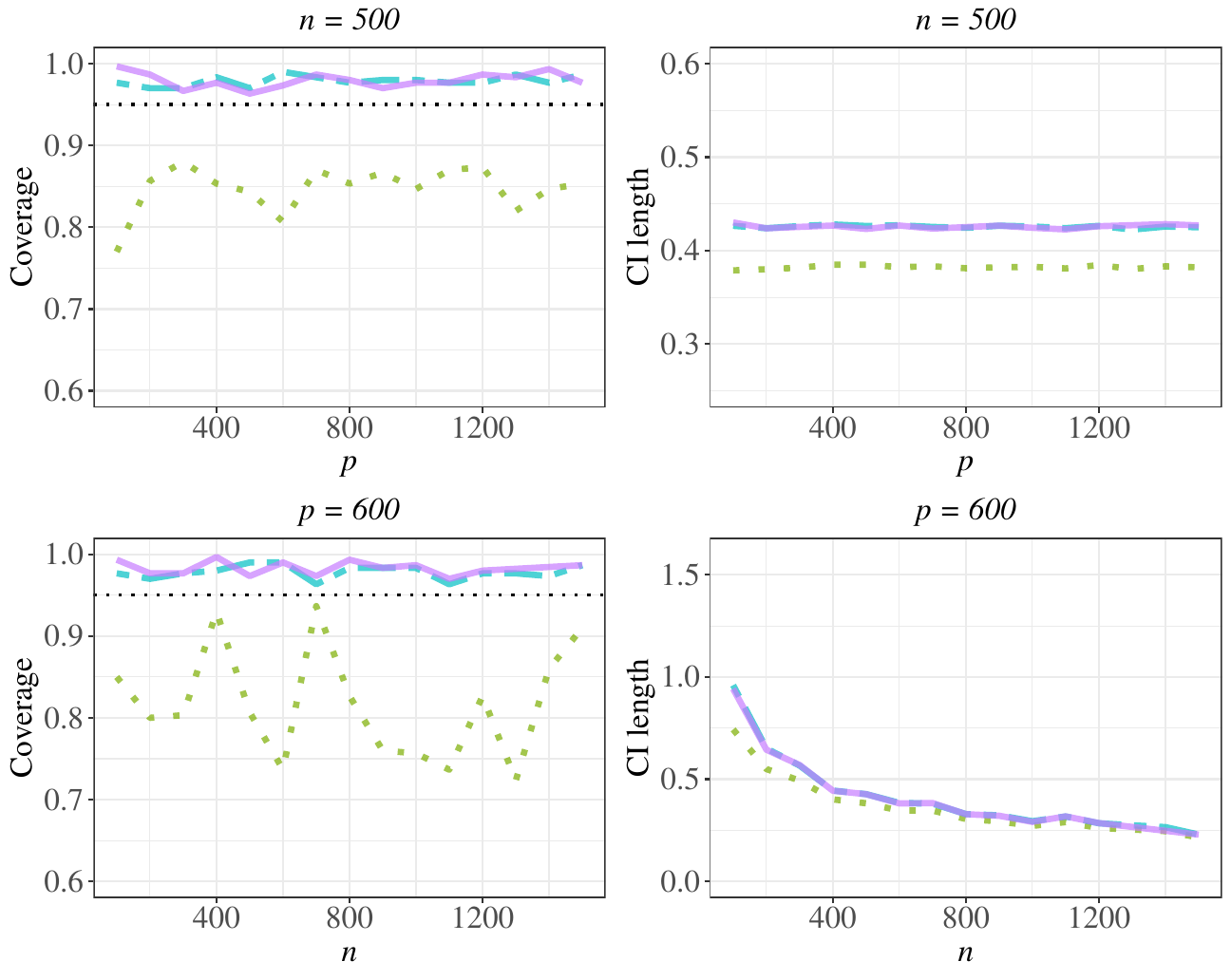}
	\caption{Coverage and length of confidence interval under logistic regression new loading matrix, averaged over 300 replications, with varying $p$ and fixed $n = 500$ (top), and with varying $n$ and fixed $p= 600$ (bottom).
	Black dashlines indicate the 0.95 level. 
	Purple solid lines (\protect\includegraphics[height=0.5em]{Figures/purple_solid.png}) are our proposed method. 
	Blue two-dashed lines (\protect\includegraphics[height=0.5em]{Figures/blue_twodahsed.png})  represent the oracle case. 
	Green dotted lines (\protect\includegraphics[height=0.5em]{Figures/green_dotted.png}) indicate the naive method. }
		\label{fig:logit_dense}
		\end{figure}

 At varying dimensions $n$ and $p$, we construct the confidence intervals using each method over 300 simulations, and then compute the coverage probabilities of the 95\% confidence intervals on the true parameter and  the average confidence interval lengths. Under the new loading matrix, the results for linear regression model are shown in Fig.~\ref{fig:linear_dense} and the results for logistic regression model are shown in Fig.~\ref{fig:logit_dense}.
We observe the coverage rates of our proposed method can still achieve the desirable 0.95 level and are close to the oracle case. The naive method performs worse than the proposed method, with coverage rates much less than the 0.95 level for most cases. For the linear regression results, although a little under coverage in some cases, the doubly debiased lasso method exhibits good performance in this setting with coverage rates approximating to the oracle case, while its average confidence interval length is larger than that of the proposed method.

\subsection{Data Application}
\label{Sec:real data application}
In this section, we apply the proposed method to a genetic data containing gene expression quantifications and stimulation statuses in mouse bone marrow derived dendritic cells. The data were also previously analyzed in \cite{Shalek2014nature} and \cite{cai2021glm}. Specifically,  \cite{cai2021glm} aimed to find significant gene expressions in response to three stimulations, which are (a) PAM, a synthetic mimic of bacterial lipopeptides; (b) PIC, a viralike ribonucleic acid; and (c) LPS, a component of bacteria.  However, there may be potential unmeasured confounders that may lead to spurious discoveries.  Our goal is to investigate the relationship between gene expressions levels and the different stimulations while accounting for possible unmeasured confounders. 

We start with pre-processing the data: we consider expression profiles after six hours of stimulations and perform genes filtering, expression level transformation, and normalization.  The details can be found in \citet{cai2021glm}. 
After the pre-processing steps, we have three groups of cells including  64 PAM stimulated cells, 96 PIC stimulated cells, and 96 LPS stimulated cells, respectively. Moreover, each of the three groups contains 96 control cells without any stimulation. 
 The gene expression quantifications are computed on 768 genes for PAM stimulation group, 697 genes on  PIC stimulation group, and 798 genes for LPS stimulation group.  
For each group of cells, the high-dimensional covariates are the gene expression levels, the stimulation status is a binary response variable recording whether a cell is stimulated.  
 
\begin{figure}[h!]
\centering
\includegraphics[scale = 0.7]{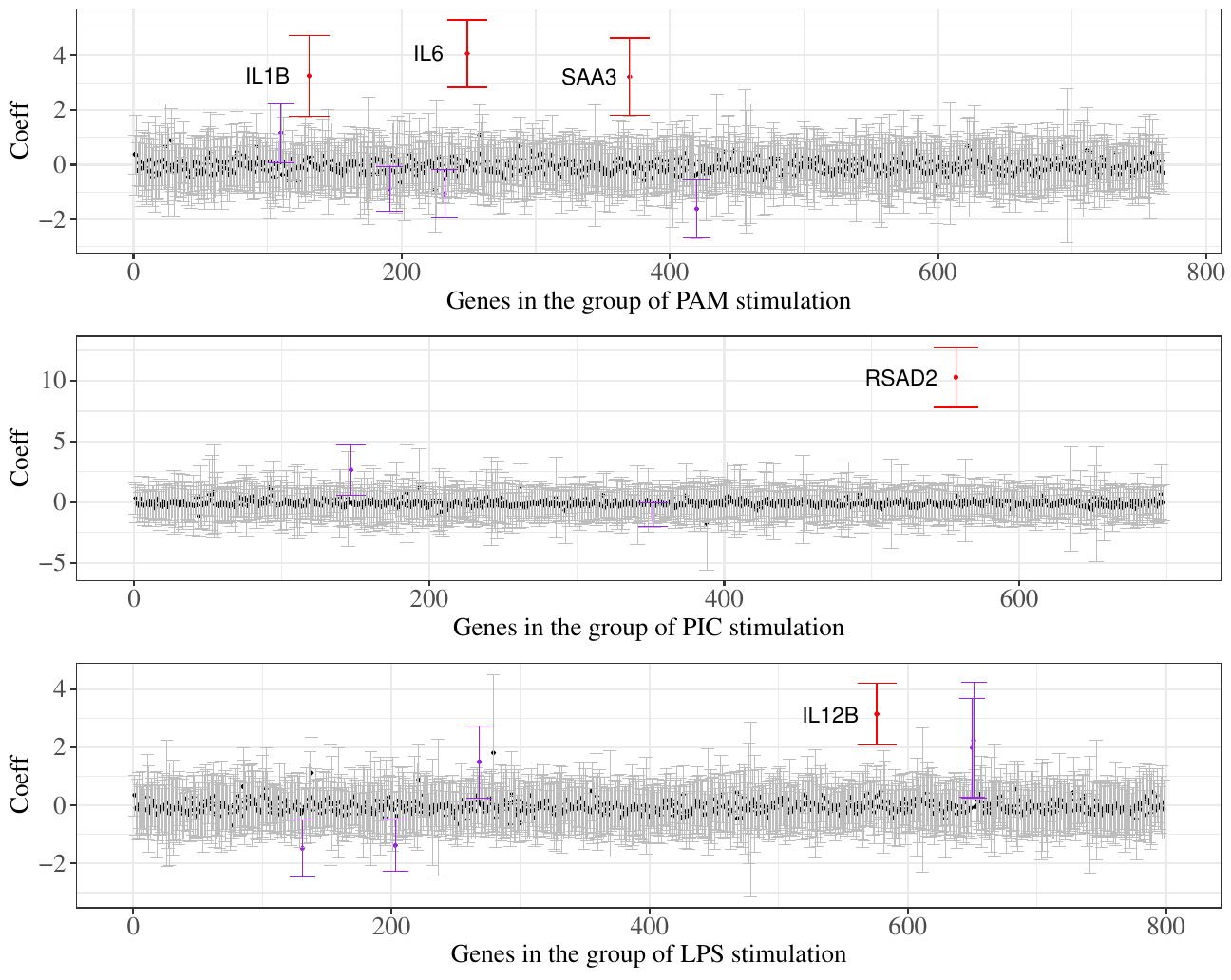}
	\caption{Confidence interval estimation using proposed method under logistic regression for the stimulation groups of PAM (top), PIC (middle) and LPS (bottom).
	Purple intervals indicate confidence intervals that do not cover zero. 
	Red intervals represent confidence intervals that are among purple intervals and significant after Bonferroni Correction.}
	\label{fig:CI genes}
\end{figure}

 We fit the data using our proposed method and the naive method in which we perform the debiasing approach without adjusting for unmeasured confounders. Applying each method, we construct 95\% confidence intervals for the regression coefficient for every gene.  In addition, we compute the $p$-value and effect size for each gene using each of two methods, respectively.  Our proposed method found that there are 
  7 possible confounders for the PAM and PIC stimulation groups, and 9 possible confounders for the LPS stimulation group. The 95\% confidence intervals constructed with our proposed method for each of the three groups are shown in Fig.~\ref{fig:CI genes}.

 We see from Fig.~\ref{fig:CI genes} that the confidence intervals for some genes do not contain zero, suggesting that they are possibly associated with their respective stimulations. 
For valid statistical inference, we perform a Bonferroni correction to adjust for multiple hypothesis testing. As a comparison, we perform similar procedures using the confidence intervals constructed with the naive method to identify the significant genes. Genes that are significantly associated with the stimulation after Bonferroni correction from the proposed method and naive method respectively are reported in Table~\ref{table:bf significant}.
The small $p$-values and large effect sizes indicate that the corresponding genes are strongly associated with their respective stimulations. 
Some gene codes in Table~\ref{table:bf significant} coincide with the findings in~\cite{cai2021glm}, and their functional consequences to the stimulation are also supported in existing literature. Specifically, \cite{cai2021glm} suggested that there exist significant association for ``IL6'' with PAM stimulation, ``RSAD2'' with PIC stimulation, and ``CXCL10'' and ``IL12B'' with LPS stimulation. 
Applying our proposed method in which the effects of unmeasured confounders are considered, we identify that ``IL1B'', ``IL6'', and ``SAA3'' are significant genes for PAM stimulated cells, ``RSAD2'' is significant for PIC stimulated cells, and ``IL12B'' is significant for LPS stimulated cells. On the other hand, when we apply the naive method which does not adjust for unmeasured confounders, we identify significant genes ``IL1B'', ``IL6'' for PAM stimulated cells, ``RSAD2'' for PIC stimulated cells and ``IL12B'' for LPS stimulated cells. Comparing the proposed method with the naive method, our method that adjusts for unmeasured confounders identifies an additional gene ``SAA3".  In the genetics literature, experimental studies support the finding that ``SAA3" plays important roles in immune reactions~\citep{SAA3}. For instance, mice lacking the gene ``SAA3'' develop metabolic dysfunction along with defects in innate immune development~\citep{SAA3}. This comparison suggests that our proposed method can identify significant genes/variables that are not captured by existing debiasing procedures without for adjusting unmeasured confounders.
\begin{table}[h!]
\centering
    \begin{tabular}{c|c|ccc}
        \hline
 Method &  Stimulation      & Gene code   & $p$-value & Effect size  \\
    \hline
 \multirow{5}{*}{Proposed method}  & \multirow{3}{*}{PAM}  & IL1B & 1$\cdot$743 $\times 10^{-5}$ & 4$\cdot$295  \\
   &       &          IL6    &   8$\cdot$940$\times 10^{-11}$ & 6$\cdot$484   \\
    &                &          SAA3    &   8$\cdot$800$\times 10^{-6}$ & 4$\cdot$445   \\
    \cline{2-5}
   &    PIC         &  RSAD2     &  4$\cdot$441$\times 10^{-16}$  &  8$\cdot$144  \\
    \cline{2-5}
  &     LPS       &  IL12B     &  5$\cdot$180$\times 10^{-9}$  &  5$\cdot$841  \\
           \hline
    \multirow{4}{*}{Naive method}  &   \multirow{2}{*}{PAM}  & IL1B &  6$\cdot$908$\times 10^{-7}$ & 4$\cdot$963  \\
             &       &          IL6    & 2$\cdot$745$\times 10^{-12}$   &  6$\cdot$990  \\
             \cline{2-5}
              &    PIC         &  RSAD2     &  $<1\times 10^{-16}$  & 8$\cdot$604  \\
    \cline{2-5}
  &     LPS       &   IL12B     &  4$\cdot$187$\times 10^{-8}$    &  5$\cdot$482   \\
  \hline
    \end{tabular}%
    \caption{Significant gene expression associated with stimulations after Bonferroni correction estimated using the proposed method and naive method, respectively.}
  \label{table:bf significant}%
\end{table}%

 Moreover, many genetic studies support that in addition to the ``SAA3", genes discovered by our proposed method all play important roles in the immune response.
The gene ``IL1B'' encodes protein in the family of interleukin 1 cytokine, which is an important mediator of the inflammatory response and its induction can contribute to inflammatory pain hypersensitivity~\citep{IL1B}. The genes ``IL6'' and ``IL12B'' are also known to encode cytokines that play key roles in hosting defense through stimulation and immune reactions~\citep{IL12B, IL6}. The expression of the gene ``RSAD2'' is important in antiviral innate immune responses and ``RSAD2'' is also a powerful stimulator of adaptive immune response mediated via mature dendritic cells~\citep{rsad2}.

\section{Discussion}
\label{Sec:discussion}

This manuscript studies statistical inference problem for the high-dimensional generalized linear model under the presence of hidden confounding bias. 
We propose a debiasing approach to construct a consistent estimator of the individual coefficient of interest and its corresponding confidence intervals, which generalizes the existing debiasing approach to account for the effects induced by the unmeasured confounders.
Theoretical properties were also established for the proposed procedure.

Our goal of this paper is to conduct statistical inference for individual coefficients $\theta$. The purpose of adjusting for the unmeasured confounders in the proposed method is to use the latent information for the better inference of covariates coefficient~$\theta$, while the inference for the coefficient of unmeasured confounders~$\bbeta$ is not the focus of the current work. This problem setting has wide scientific applications. For instance, as introduced in our data application, biologists are interested in the significance of the genes under the simulations with the unmeasured confounders adjusted. Moreover, in many practices with finite samples, consistent estimation of the number of factors may be problematic; our method only requires that the estimated factors can well approximate the confounders and the interpretation of the factors is of less interest in this setting. Here to investigate how the accuracy of the estimation of the dimension of unmeasured confounders $K$ affects the asymptotic distribution of debiased estimator, we conducted a simulation study in Appendix C and the results suggest that in practice, the overestimation of $K$ may not affect the asymptotic normality results, which may be because an abundant amount of unmeasured confounders can still fully capture the covariate information and would not incur bias in the point and interval estimation.

Nevertheless, the inference of unmeasured confounders is also of great interest in many econometrics and psychometrics applications. For instance, \cite{fan2022latent} recently proposed an ANOVA-type testing procedure for testing the existence of unmeasured confounders or not. It would be interesting to develop similar inference testing procedures under the generalized linear models in applications when the inference of the unmeasured confounders of scientific interest.

Another interesting related problem is regarding the inference on group-wise covariate coefficients. We next briefly describe how to generalize our method to obtain the group-wise asymptotic distribution of $\max_{j \in G}|\gamma_j|$ via a bootstrap-assisted procedure for $\bgamma = (\theta, \bv^{\T}) \in \RR^p$ and any subset $G \subseteq \{1, \dots, p\}$. The procedures follow simultaneous inference results from~\cite{zhang2017simultaneous} and are shown as follows.

\begin{itemize}
    \item Step 1: For each $\gamma_j$, we construct a debiased estimator $\tilde{\gamma}_j$ using our proposed method for $j \in G$. We generate a sequence of i.i.d. standard normal random variables and denote them as $\{\varpi_i\}_{i = 1,\dots, n}$. 
    \item Step 2: 
Then, under the null hypothesis that $H_{0, G}: \gamma_j^* = \gamma_j^0 $ for $j \in G$,  let ${W}_G = \max_{j \in G} $ $ |\sum_{i=1}^n\hat{I}_{\gamma_j\mid \bm{\eta}_{-j}}^{-1} \hat{S}(\gamma_j^0, \hat{\bm{\eta}}_{-j}) \varpi_i/\sqrt{n}|$,
where $\hat{S}(\gamma_j^0, \hat{\bm{\eta}}_{-j})$ can be similarly calculated as $\hat{S}(\theta^0, \hat{\bzeta})$ and  $\hat{I}_{\gamma_j\mid \bm{\eta}_{-j}}$ can be similarly calculated as $\hat{I}_{\theta\mid \bzeta}$ in Section~\ref{Sec:methodology} by treating $\gamma_j$ as the coefficient of interest and $\bm{\eta}_{-j}$ as the other coefficient for nuisance covariates and unmeasured confounders. 
\item Step 3: 
Next, we calculate the critical value $C_G(\alpha)$ at $(1-\alpha)$ significance level by
$C_G(\alpha) = \inf [t \in \RR: P\{ W_G \leq t \mid (y_i, \bX_i)_{i = 1, \dots, n} \} \geqslant 1-\alpha ]$.
\end{itemize}
With the debiased estimators and the critical value, we expect a similar result as Theorem 4.1 in \cite{zhang2017simultaneous} that the asymptotic distribution of $\max _{j \in G} \sqrt{n}|\tilde{\gamma}_j-\gamma_j^*|$ satisfy
$
    \sup_{\alpha \in(0,1)}|P\{\max _{j \in G} \sqrt{n}|\tilde{\gamma}_j-\gamma_j^*|>C_G(\alpha)\}-\alpha|=o(1)$. While the inference on group-wise maximum coefficients is an interesting problem, it is not the focus of this work and we leave the related theoretical proof for future research.

 Besides the aforementioned extensions, there are several related problems worth investigating in the future. For instance, as discussed in Section~\ref{sec:related models}, there are many models in existing literature related to our problem, in particular, the factor-adjusted method can be extended to the situation involving hidden confounding~\citep{fan2020}. In addition, in our theoretical analysis, we assume that the dimension of unmeasured confounders, $K$, is fixed, which has also been imposed in the existing literature  \citep[e.g.,][]{wang2017asymptotics,fan2022latent}. 
 One possible extension is to allow $K$ to grow as $n$ and $p$ increase, and it involves generalizations of the theoretical results on the maximum likelihood estimation for the unmeasured confounders. 
 It would also be interesting to generalize the factor model to a nonlinear structure and investigate the theoretical properties of the debiased estimator under generalized factor model \citep{chen2020structured,liu2021GFA}.
Besides generalized linear models, the high-dimensional debiasing technique is also popularly studied in a variety of models such as Gaussian graphical models \citep{zhao2015, zhu2020} and
 additive hazards models \citep{lin1994semiparametric, lin2013}. For instance, consider additive hazards models, which are popularly used in survival analysis and assume the conditional hazard function at time $t$ as $\lambda( t \mid D,\ \bQ,\ \bU) = \lambda_0(t) + \theta D + \bv^{\T} \bQ + \bbeta^{\T} \bU$ with $D$ and $\bQ$   covariates and $\bU$  unmeasured confounders. The relationship between covariates $\{D, \bQ\}$ and unmeasured confounders $\bU$ are also modeled by a linear factor model. Our method can be used to perform inference on $\theta$ under the aforementioned model setup based on the quadratic loss function. Generalizing our proposed approach to these models to adjust for possible unmeasured confounders is also an interesting direction to investigate.

\newpage

\appendix
\renewcommand{\theequation}{A\arabic{equation}}
\section*{Appendix A. Preliminaries}
\label{preliminaries}
We start with introducing some notations and definitions. For a vector $\br = (r_1, \dots, r_l)^{\T}$, we let $\mathcal{S}_{r} = \{j: {r}_j \neq 0\}$,  $\|\br \|_{\infty}= \max_{j = 1, \ldots, l} |r_j|$, $\|\br\|_q = (\sum_{j=1}^l |r_j|^q)^{1/q}$ for $q \geq 1$ and $s_r = \text{card}(\mathcal{S}_{r})$. 
For a matrix $\bA = (a_{ij})_{n\times l}$, let $\|\bA\|_{\infty,1} = \max_{j=1,\ldots, l} \sum_{i = 1}^n |a_{ij} |$ to be the maximum absolute column sum, $\| \bA\|_{1,\infty} = \max_{i=1,\ldots, n} \sum_{j=1}^l |a_{ij}| $ to be the maximum of the absolute row sum, $\| \bA\|_{\max} = \max_{i,j} |a_{ij}|$ to be the maximum of the matrix entry,  $\lambda_{\min}(\bA)$ and $\lambda_{\max}(\bA)$ to be the smallest and largest eigenvalues of $\bA$ and $\| \bA\|_{F} = (\sum_{i=1}^n \sum_{j=1}^l |a_{ij}|^2)^{1/2}$ to be the Frobenius norm of $\bA$. 
For sequences $\{a_n\}$ and $\{b_n\}$, we write $a_n \lesssim b_n $ if there exists a constant $C >0$ such that $a_n \leq C b_n$ for all $n$, and $a_n \asymp b_n$ if $a_n \lesssim b_n $ and $b_n \lesssim a_n$. 
For any sub-exponential random variable $Y_1$, we define the sub-exponential norm as $\|Y_1\|_{\varphi_1}=\inf [s>0 : E \{\exp(Y_1/s)\} \leq 2]$.
For any sub-Gaussian random variable $Y_2$, we define the sub-Gaussian norm as $\|Y_2\|_{\varphi_2}=\inf [s>0 : E \{\exp{(Y_2^2/s^2)}\} \leq 2]$.

 Next, we give a review of our model framework. We assume  the response $y$ given the covariate of interest $D$, the nuisance covariates $\bQ$ and the unmeasured confounders $\bU$ follows the generalized linear model with the probability density (mass) function to be  
 \begin{equation}
 \label{eq:full y model_appendix}
 	f(y) = \exp \left[ \{y ( \theta {D}  +\bv^\T \bQ + \bbeta^\T\bU )-b( \theta {D}  +\bv^\T \bQ + \bbeta^\T\bU )\}/{a(\phi)} +c(y, \phi)\right],
 \end{equation}
 and the relationship between $\bX$ and $\bU$ is  
 \begin{equation}
\label{eq:XUmodel_appendix}
 \bX=  \bW^{\T}\bU + \bE. 
\end{equation}
We let $\bZ = (D, \bQ^\T, \bU^\T)^\T$ to be a vector that includes all of the covariates and the unmeasured confounders, and let $\bm{\eta}^\T = (\theta, \bv^\T, \bbeta^\T)$ to be the corresponding parameters. For notational convenience, we also let $\bM =(\bQ^\T, \bU^\T)^\T$ and its coefficient $\bzeta = (\bv^{\T}, \bbeta^{\T})^\T$, so $\bm{\eta} = (\theta, \bzeta^\T)^\T$.  We denote $\bgamma = (\theta, \bv^\T)^\T$ as the coefficient for covariates $\bX$, so $\bm{\eta} = (\bgamma^\T, \bbeta^\T)^\T$. 

Throughout the appendix, we use an asterisk on the upper subscript to indicate  the population parameters. We assume that the observed data $\{y_i, \bX_i\}_{i=1,\dots, n}$ and the unmeasured confounders $\{\bU_i\}_{i= 1,\dots, n}$ are realizations of~\eqref{eq:full y model_appendix} and~\eqref{eq:XUmodel_appendix}. The noise for the factor model are denoted as $\{\bE_i\}_{i=1, \ldots, n}$ with $\bE_i = (E_{ij}: j=1, \ldots, p)$. 

In the proposed inferential procedure, we first obtain the maximum likelihood estimator for unmeasured confounders $\hat{\bU}_i$.
With the unmeasured confounder estimators, we let $\dot{\bZ}_i = (D_i, \bQ_i^\T, \hat{\bU}_i^\T)^\T$, $\dot{\bM}_i = (\bX_i^\T, \hat{\bU}_i^\T)^\T$ and get an initial lasso estimator
\begin{equation}
\hat{\bm{\eta}} = \underset{\eta \in \RR^{(p+K)}}{\operatorname{argmin}}~ l(\bm{\eta})+\lambda\|\bgamma\|_1, \label{eq:initial estimator}
\end{equation}
where the loss function is the negative log-likelihood function defined as
\begin{eqnarray}
l( \bm{\eta})=  -\frac{1}{n} \sum_{i=1}^{n}\{ y_i \bm{\eta}^{\T} \dot{\bZ}_{i}- b(\bm{\eta}^{\T} \dot{\bZ}_{i} ) \}. \nonumber
\end{eqnarray}
The gradient $\nabla l(\bm{\eta})$ and Hessian   $\nabla^{2} l(\bm{\eta})$ of the loss function are commonly used in our subsequent proofs, which are expressed as
\begin{eqnarray}
	\nabla l (\bm{\eta}) =  - \frac{1}{ n} \sum_{i=1}^{n}\{y_i- b^{\prime}(\bm{\eta}^{\T} \dot{\bZ}_{i} )\} \dot{\bZ}_i; \quad
	\nabla^{2} l(\bm{\eta}) =  \frac{1}{ n } \sum_{i=1}^{n}b^{\prime\prime}(\bm{\eta}^{\T} \dot{\bZ}_{i} )\dot{\bZ}_i \dot{\bZ}_i^\T. \nonumber
\end{eqnarray}

In constructing the debiased estimator, we define $\bw^* =\bI_{\theta \bzeta}^*(\bI_{\bzeta \bzeta}^{*})^{-1}$ and $\btau^* = (1, - \bw^*)^\T$, where $\bI_{\theta \bzeta}^* = E[ b^{\prime\prime}\{(\bm{\eta}^{*})^\T {\bZ}_{i}\} D_i\bM_i^\T ]$ and $\bI_{\bzeta \bzeta}^* = E[ b^{\prime\prime}\{(\bm{\eta}^{*})^\T {\bZ}_{i}\} \bM_i\bM_i^\T ]$. We denote two sub-vectors of $\bw^*$ as $\bw_q^* =(w_2^*, \ldots, w_q^*)^\T$ and $\bw_u^* =(w_{p+1}^*, \ldots, w_{p+K}^*)^\T$. We define the estimator for the sparse vector $\bw$ as
\begin{equation}
	\hat{\bw} = \underset{\bw \in \mathbb{R}^{(p+K-1)}}{\operatorname{argmin}}~\frac{1}{2n} \sum_{i=1}^{n} \{\bw^\T \nabla_{\bzeta\bzeta} l_{i}(\hat{\theta}, \hat{\bzeta}) \bw - 2 \bw^{\T} \nabla_{\bzeta\theta} l_{i}(\hat{\theta}, \hat{\bzeta})\}+\lambda^{\prime} \|\bw\|_1, \label{eq:estimator w}
\end{equation}
where $l_i(\theta, \bzeta) = -y_i (\theta D_i + \bzeta^T \dot{\bM}_i) + b(\theta D_i + \bzeta^T \dot{\bM}_i)$ is equivalent to $l_i(\bm{\eta})$, the $i$th component of the loss function. Equivalently, the estimator $\hat{\bw}$ is obtained by
\begin{equation}
\hat{\bw} = \underset{\bw \in \mathbb{R}^{(p+K-1)}}{\operatorname{argmin}}~\frac{1}{2n} \sum_{i=1}^{n}b^{\prime\prime} (\hat{\theta}D_i + \hat{\bzeta}^{\T} \dot{\bM}_i)(D_i -  \bw^{\T}\dot{\bM}_i )^2+\lambda^{\prime} \|\bw\|_1. \nonumber
\end{equation}The estimator $\hat{\bw}$ is used in estimating the generalized decorrelated score function and the partial Fisher information matrix. Under null hypothesis $H_0: \theta^* = \theta^0$, the two estimators are 
\begin{eqnarray}
	\hat{S}(\theta^0, \hat{\bzeta}) &=& -\frac{1}{n}\sum_{i=1}^n\{y_i - b^{\prime}(\theta^0 D_i + \hat{\bv}^\T \bQ_i + \hat{\bbeta}^\T\hat{\bU}_i)\}(D_i - \hat{\bw}^\T\dot{\bM}_i); \nonumber \\
		\hat{{I}}_{\theta\mid \bzeta} &=& \frac{1}{n} \sum_{i=1}^n b^{\prime\prime}(\hat{\theta} D_{i}+\hat{\bv}^\T \bQ_i + \hat{\bbeta}^\T\hat{\bU}_i) D_i(D_i - \hat{\bw}^\T\dot{\bM_i}). \nonumber
\end{eqnarray}

Our theoretical results are established under the asymptotic regime with  $n, p \rightarrow \infty$. Regarding the factor model, as stated in Assumption~\ref{assumption1} in Section~\ref{Sec:theoretical results} of the main text, we do not assume the random errors $\bE_i$ to be identically distributed nor does the model assumption require the covariance of $\bE_i$ to be diagonal. Specifically, we assume for some large constant $C > 0$: (a) $\EE(E_{ij}) = 0$, $\EE({E}_{ij}^{8}) \leq C$; (b) $\EE(E_{ih}E_{ij}) = \tau_{i, hj}$ with $| \tau_{i, hj}| \leq  \tau_{hj}$ for some $\tau_{hj} > 0$ and all $i = 1, \dots, n$, and $\sum_{h=1}^p\tau_{hj}\leq C$ for all $j=1,\dots,p$.
 (c) $\EE(E_{ij}E_{sj}) = \rho_{is, j}$ with $| \rho_{is, j}| \leq \rho_{is}$ for some $\rho_{is} >0$ and all $j=1,\dots, p$, and $n^{-1}\sum_{i=1}^n \sum_{s=1}^n \rho_{is} $ $  \leq C$. (d) For all $j,q = 1, \dots, p$,
 \begin{equation}
     \EE \left\{\left|\frac{1}{\sqrt{n}} \sum_{i=1}^n [E_{ij} E_{iq} - \EE (E_{ij} E_{iq})]\right|^4  \right\} \leq C. \nonumber
 \end{equation}
 Regarding the loading matrix, we assume
$\|\bW_{j}^*\|_2 \leq C$. There exist positive definite matrices $\bGamma^*$ and $\Upsilon^*$ such that  $\lim _{p \rightarrow \infty} p^{-1} \bW^* (\bSigma_{e}^*)^{-1} (\bW^*)^\T=\bGamma^*$ and $\lim _{p \rightarrow \infty}p^{-1} \sum_{j=1}^{p} ({\sigma}_{j}^*)^{-4}$ $\{(\bW_{j}^*)^\T \otimes(\bW_{j}^*)^\T \}(\bW_{j}^* \otimes \bW_{j}^*)= \Upsilon^*$. 
 In addition, we also assume a working identifiability condition that $\bS_u = \bI_K$ and $p^{-1}\bW^*(\bSigma_e^*)^{-1}(\bW^*)^\T$ is a diagonal matrix with distinct entries.

For the assumptions regarding the generalized linear model relating ${y}$ and $(\bX, \bU)$, as mentioned in Assumption~\ref{glm assumption} in Section~\ref{Sec:theoretical results} of the main text, we let $\lambda_{\min}(\bI^*) \geq \kappa$, where $\bI^* = E[b^{\prime\prime}\{(\bm{\eta}^*)^\T \bZ_{i}\}{\bZ}_i{\bZ}_i^\T ]$ and $\kappa$ is some constant. 
The unmeasured confounders, the covariates and coefficients parameters are assumed to be bounded, that is, $\|\bU_i\|_{\infty} \leq M$, $\|\bX_i\|_{\infty} \leq M$, $\| \bm{\eta}^*\|_{\infty} \leq M$, and $|(\bw_q^{*})^\T\bQ_i| \leq M$ for some constant $M>0$.
We let $|y_i - b^{\prime}\{(\bm{\eta}^*)^\T \bZ_{i}\} |$ to be sub-exponential with $\|y_i - b^{\prime}\{(\bm{\eta}^*)^\T \bZ_{i}\} \|_{\varphi_1} \leq M$.
In addition, we assume $a_1 \leq (\bm{\eta}^*)^{\T}{\bZ}_i \leq a_2$, $0 \leq |b^{\prime}(t)| \leq B$ with $|b^{\prime}(t_{1})-b^{\prime}(t)| \leq B |(t_1 - t)b^{\prime}(t)| $ and $0 \leq b^{\prime\prime}(t) \leq B$ with $|b^{\prime\prime}(t_{1})-b^{\prime\prime}(t)| \leq B |t_1 - t| b^{\prime\prime}(t)$ for constants $a_1$, $a_2$ and $B$, where $t \in [a_1 - \epsilon, a_2 + \epsilon]$ for $\epsilon > 0$ and sequence $t_1$ satisfying $|t_1 - t| = o(1)$. 

 With the notations and assumptions revisited, we next present the discussion for the working identifiability condition and then the proofs of Theorems~\ref{Initial estimator consistency} and~\ref{normality theorem}.


\section*{Appendix B. Discussion on Working Identifiability Condition}
\label{Detailed Clarification on Remark}

In this section, we make a detailed clarification on the working identifiability condition in the main text. 
Specifically, we show that when the working identifiability condition does not hold, we can transform the model into an identifiable model and the transformed identifiable model has identical parameter of interest $\theta$ compared to that of the pre-transformed model.

 When the identifiability condition is not satisfied, the estimators are not fully identifiable in the sense that, for any invertible matrix $\bO$, we have $\tilde{\bW} = \bO^\T\hat{\bW}$ and $\tilde{\bS}_u = (\bO^{-1})^\T \hat{\bS}_u\bO^{-1} $ to be valid maximum likelihood estimators. At $\tilde{\bW} = \bO^\T \hat{\bW}$, we can substitute the expression for $\tilde{\bW}$ into~\eqref{Ui hat} to get the relationship between $\tilde{\bU}_i$ and $\hat{\bU}_i$ as follows,
\begin{eqnarray}
	\tilde{\bU}_{i}&=&(\tilde{\bW} \hat{\bSigma}_{e}^{-1} \tilde{\bW}^\T)^{-1} \tilde{\bW} \hat{\bSigma}_{e}^{-1}(\bX_{i}-\bar{\bX}) \nonumber \\
	&=&(\bO^\T\hat{\bW} \hat{\bSigma}_{e}^{-1} \hat{\bW}^\T \bO)^{-1} \bO^\T\hat{\bW}\hat{\bSigma}_{e}^{-1}(\bX_{i}-\bar{\bX}) \nonumber \\
	&=& \bO^{-1}(\hat{\bW} \hat{\bSigma}_{e}^{-1} \hat{\bW}^\T )^{-1}(\bO^{-1})^\T\bO^\T \hat{\bW}\hat{\bSigma}_{e}^{-1}(\bX_{i}-\bar{\bX}) \nonumber \\
	&=& \bO^{-1}\hat{\bU}_i. \nonumber
\end{eqnarray}
 Suppose $\hat{\bW}$ and $\hat{\bU}_i$ are the asymptotically unbiased estimators for $\bW^*$ and $\bU_i^*$ respectively. Here we want to clarify that we use $\bS_u^*$ and $\bU_i^*$ to differentiate the unmeasured confounders before the transformation from that of transformed model but this does not imply that $\bS_u^*$ and $\bU_i^*$ are population parameters.
 we have $\tilde{\bW}$ and $\tilde{\bU}_i$ to be the asymptotically unbiased estimators for $\bO^\T \bW^*$ and $\bO^{-1} \bU_i^*$.

When the working identifiability condition does not hold, that is, we have $\bS_u^* \neq \Ib_K$, and/or $p^{-1}\bW^*\bSigma_e^{-1}(\bW^*)^\T$ is not diagonal matrix with distinct entries, we can find an invertible matrix $\bO$ to transform the true parameters to satisfy the assumption. Then we build a correspondence between the true model and the model corresponding to the transformed parameters, showing that the two models have parameter of interest $\theta$ to be identical.

The transformed parameters are $\bW^r = \bO_2^{-1}\bO_1^\T\bW^*$ and $\bU_i^r = \bO_2^\T(\bO_1^{-1})\bU_i^*$, which are constructed in two steps.  At first step, for any $\bS_u^* \neq \bI_K$, we can find a matrix $\bO_1$ such that 
\begin{equation}
	 n^{-1}\sum_{i=1}^n\bO_1^{-1}(\bU_i^* - \bar{\bU}^*)(\bU_i^* - \bar{\bU}^*)^\T(\bO_1^{-1})^\T = \bO_1^{-1}\bS_u^*(\bO_1^{-1})^\T = \bI_K.  \nonumber 
\end{equation}
At next step, given the matrix $p^{-1}\bW^*\bSigma_e^{-1}(\bW^*)^\T$ is symmetrical, there exists an orthogonal matrix $\bO_2$ whose columns correspond to the eigenvectors of $p^{-1}\bO_1^\T\bW^*\bSigma_e^{-1}(\bO_1^\T\bW^*)^\T$ as we could decompose this symmetric matrix into $p^{-1}\bO_1^\T\bW^*\bSigma_e^{-1}(\bO_1^\T\bW^*)^\T $ $ = \bO_2 {\bLambda} \bO_2^\T$ where ${\bLambda}$ has distinct eigenvalues in the diagonal. 

We verify the transformed parameters satisfy the identifiability condition as follows. At $\bW^r = \bO_2^{-1}\bO_1^\T\bW^*$ and $\bU_i^r = \bO_2^\T(\bO_1^{-1})\bU_i^*$, we have
\begin{equation}
	\bS_u^r = n^{-1}\sum_{i=1}^n\bO_2^\T \bO_1^{-1}(\bU_i^* - \bar{\bU}^*)(\bU_i^* - \bar{\bU}^*)^\T(\bO_1^{-1})^\T \bO_2= \bO_2^\T\bI_K\bO_2 = \bI_K,  \nonumber
\end{equation}
and 
\begin{equation}
	p^{-1}\bW^r\bSigma_e^{-1}(\bW^r)^\T = p^{-1}\bO_2^{-1}\bO_1^\T\bW^*\bSigma_e^{-1}(\bW^*)^\T \bO_1(\bO_2^{-1})^\T = \bO_2^{-1} \bO_2 {\bLambda} \bO_2^\T (\bO_2^{-1})^\T={\bLambda}, \nonumber
\end{equation}
is a diagonal matrix with distinct entries.

We let $\bO= \bO_2^\T(\bO_1^{-1})$, so accordingly the parameters for the transformed factor model are $\bU^r = \bO\bU^*$ and $(\bW^r)^\T = (\bW^*)^\T \bO^{-1}$. The relationship between the true model and the transformed model is as follows. The factor model structure corresponding to the  true parameter is the same as the model corresponding to the transformed parameters as  
\begin{equation}
\label{eq:XUmodel_matrixform}
 \bX = (\bW^r)^\T \bU^r + \bE = (\bW^*)^\T\bO^{-1} \bO \bU^*  + \bE  =  (\bW^*)^\T \bU^*+ \bE. \nonumber
\end{equation}
The generalized linear framework according to the rotated true parameter is 
 \begin{equation}
 	f(y) = \exp \left( [y \{ \theta^r {D}  + (\bv^r)^\T \bQ + (\bbeta^r)^\T\bU \}-b\{ \theta^r {D}  + (\bv^r)^\T \bQ + (\bbeta^r)^\T\bU ]/{a(\phi)} +c(y, \phi)\right), \nonumber
 \end{equation}
 whereas the framework according to true parameters is
 \begin{equation}
 	f(y) = \exp \left([y \{ \theta^* {D}  + (\bv^*)^\T \bQ + (\bbeta^*)^\T\bU \}-b\{ \theta^* {D}  + (\bv^*)^\T \bQ + (\bbeta^*)^\T\bU \}]/{a(\phi)} +c(y, \phi)\right). \nonumber
 \end{equation}
At $\bbeta^r = \bO^{-1}\bbeta^*$, $\theta^r = \theta^*$ and $\bv^r = \bv^*$, the two frameworks are identical. That is, when the confounders are not identifiable, only the coefficient $\bbeta$ will be affected accordingly; the parameter of interest will not change and thus the theoretical results on $\tilde{\theta}$ are not affected.

\section*{Appendix C. Estimation of the Dimension of Unmeasured Confounders}
\label{appendix:estimation K}
In the numerical studies of this paper, we use parallel analysis~\citep{horn1965} to estimate the dimension of unmeasured confounders. Because in factor analysis, parallel analysis is a popular approach to selecting the number of factors as it is accurate and easy to use~\citep{hayton2004factor, costello2005best, brown2015confirmatory}. 
In the existing literature, several authors have conducted extensive simulation studies to assess the performance of parallel analysis relative to other existing approaches \citep{zwick1986comparison,pedro2005stopping}. 
They have shown that parallel analysis has better numerical performance in terms of selecting $K$ than many existing approaches \citep{zwick1986comparison,pedro2005stopping}.
Furthermore, parallel analysis is also a commonly used statistical tool for dimension reduction~\citep{lin2016simultaneous}, multiple testing dependence~\citep{leek2008general}, and finds wide applications in other scientific disciplines including virology~\citep{quadeer2014statistical} and genetic studies~\citep{leek2007capturing}.

The implementation of parallel analysis is as follows. With our given matrix $\bX_{n\times p} = (D, \bQ^{\T})^{\T}$, we denote $p$ columns in the design matrix as $\bX_1, \dots, \bX_p$, respectively. Then we repeatedly generate matrices $\bX_{\pi}$'s where each matrix is generated by randomly permuting every column $\bX_j$ for $j = 1,\dots, p$. Next, we select the first factor when the top singular value of $\bX$ is larger than a certain percentile of the top singular value of the permuted matrices $\bX_{\pi}$'s. If the first factor is selected, we repeat this procedure to determine whether the second factor can be selected. The process is repeated until no more factor is selected. The main intuition of this approach is that the factor model is considered a summation of the signal (factors) and noise (random error). The permutation destroys the original signal structure and turns it into a matrix of random noise.  Thus, identifying factors based on large singular values of $\bX$ can be interpreted as selecting factors that are above the noise level.

Besides parallel analysis, there are various methods to estimate the dimension of unmeasured confounders such as scree plot~\citep{cattell1966scree}, which empirically chooses the elbow point in the plot of descending eigenvalues of factors; method based on cross validation~\citep{owen2016bicv}, which uses random held-out matrices of data to choose the number of factors; method based on information criteria including AIC and BIC to select the number of factors in high-dimensional factor model~\citep{bai2002determining}; the eigenvalue ratio method~\citep{lam2012factor, ahn2013eigenvalue}, which chooses $\hat{K}$ by $\hat{K} = \arg\max_{K \leq \mathcal{K}} \lambda_{k} (\bX \bX^{\T}) / \lambda_{k+1} (\bX \bX^{\T})$ where $\lambda_{k}(\bX \bX^{\T})$ denotes the $k$-th eigenvalue of $\bX \bX^{\T}$ and $\mathcal{K}$ is a prespecified threshold which is often set to be $\mathcal{K} = p/2$ in practice. 
Among these methods, the information-criteria-based method and eigenvalue ratio method have the theoretical guarantees to be consistent under similar assumptions as Assumption~\ref{assumption1} in our paper. These assumptions follow the common conditions in theoretical analysis for the approximate factor model that allows weak correlation among the random error. Besides the method to determine the number of factors for linear factor models, \cite{chen2021determining} proposed a method based on joint-likelihood-based information criterion to determine the number of factors for generalized linear factor models. All these selection methods are well-established tools and can possibly be used to select $K$. Nonetheless, our theoretical results hold as long as the dimension of unmeasured confounders is consistently estimated. In our manuscript, we use the parallel analysis for good empirical performance, as illustrated in \citet{pedro2005stopping}.

To further investigate how the accuracy of the estimation of $K$ affects the asymptotic distribution of the debiased estimation, we conduct simulation studies where we keep all the estimation settings to be the same as in Section~5.1 in the paper except manually replacing the $\hat{K}$ estimated by parallel analysis with specified values: 2, 4, 5, and 10. Recall that we set the true $K = 3$, so this simulation provides us with some insights on how the overestimation $(\hat{K} = 4, 5$ and $10)$ and underestimation $(\hat{K} = 2)$ of unmeasured confounder dimension affect the inference results.

\begin{figure}[h!]
    \centering
    \includegraphics[scale = 0.7]{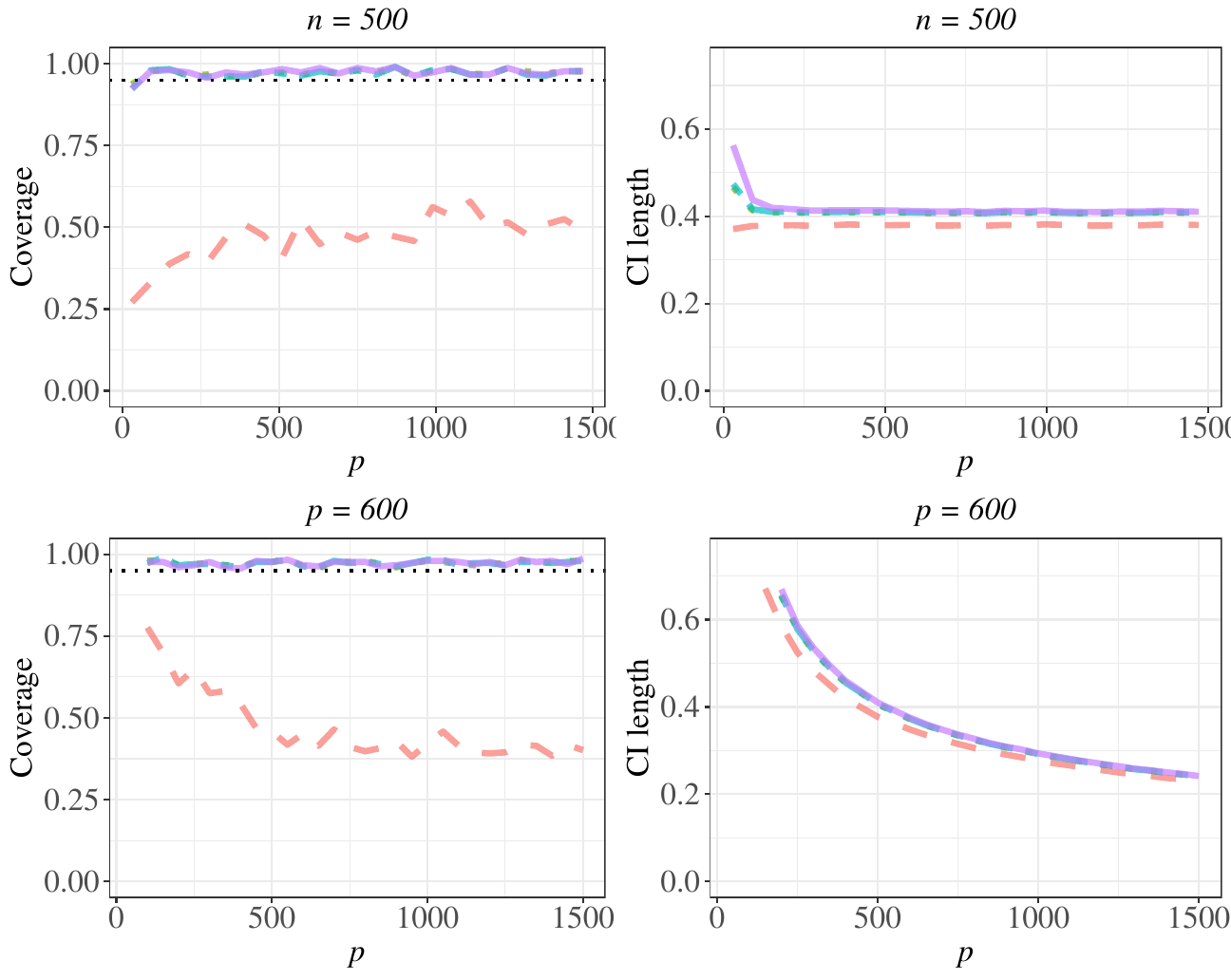}
    \caption{ Coverage and length of the confidence interval under logistic regression with prespecified $\hat{K} = 2, 4, 5$ and 10, averaged over 300 replications, with varying $p$ and fixed $n = 500$ (top), and with varying $n$ and fixed $p= 600$ (bottom). Black dashlines indicate the 0.95 level.
    Orange dashed lines
		(\protect\includegraphics[height=0.5em]{Figures/orange_dashed.png}) represent the results at $\hat{K} = 2$.
  Green dotted lines (\protect\includegraphics[height=0.5em]{Figures/green_dotted.png}) represent the results at $\hat{K} = 4$. 
  Blue two-dashed lines (\protect\includegraphics[height=0.5em]{Figures/blue_twodahsed.png})  represent the results at $\hat{K} = 5$.  
  Purple solid lines (\protect\includegraphics[height=0.5em]{Figures/purple_solid.png}) represent the results at $\hat{K} = 10$.}
    \label{fig:mis-esimate K_appendix}
\end{figure}

\begin{figure}[h!]
    \centering
    \includegraphics[scale = 0.7]{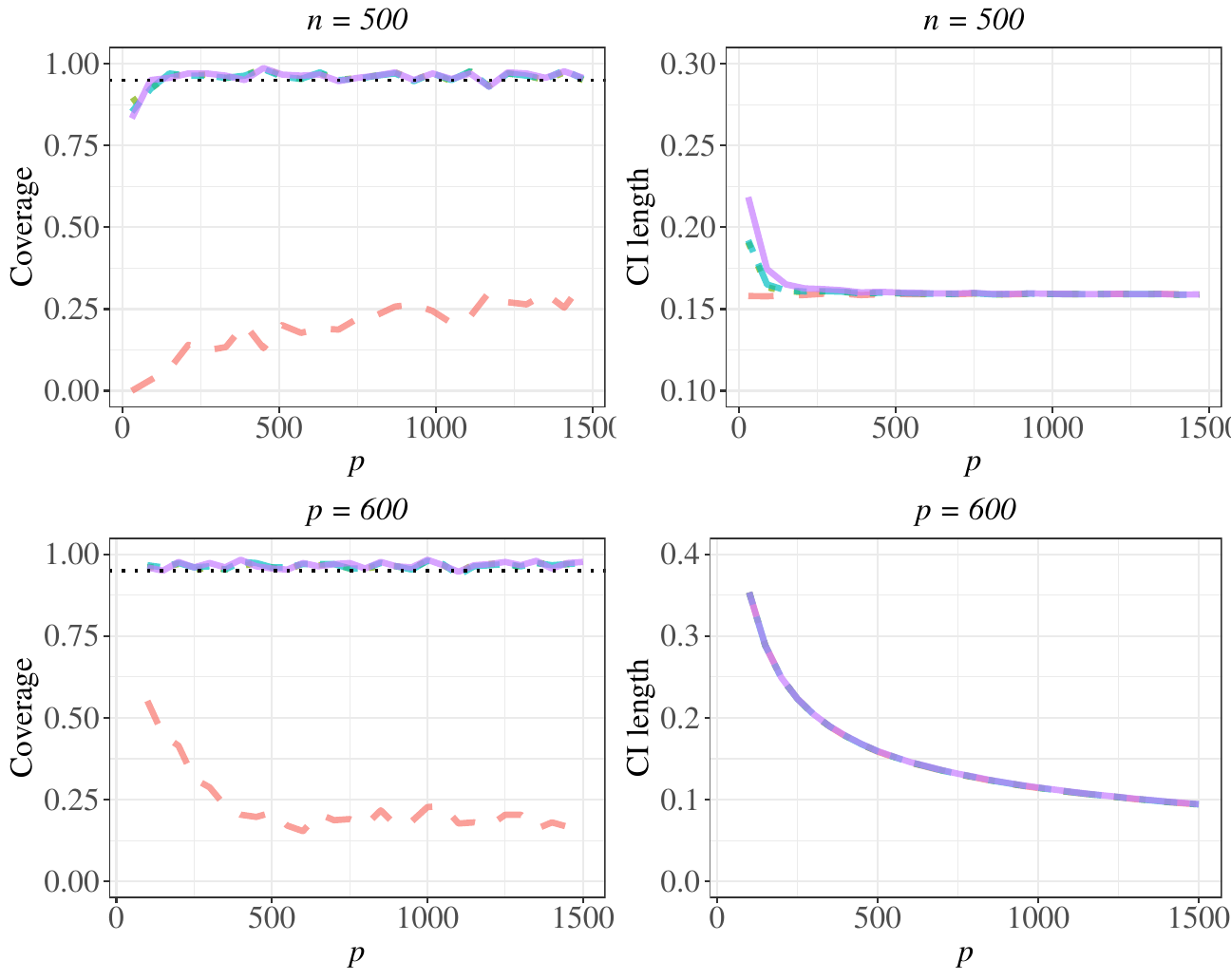}
    \caption{ Coverage and length of the confidence interval under linear model with prespecified $\hat{K} = 2, 4, 5$ and 10, averaged over 300 replications, with varying $p$ and fixed $n = 500$ (top), and with varying $n$ and fixed $p= 600$ (bottom). Black dashlines indicate the 0.95 level.
    Orange dashed lines
		(\protect\includegraphics[height=0.5em]{Figures/orange_dashed.png}) represent the results at $\hat{K} = 2$.
  Green dotted lines (\protect\includegraphics[height=0.5em]{Figures/green_dotted.png}) represent the results at $\hat{K} = 4$. 
  Blue two-dashed lines (\protect\includegraphics[height=0.5em]{Figures/blue_twodahsed.png})  represent the results at $\hat{K} = 5$.  
  Purple solid lines (\protect\includegraphics[height=0.5em]{Figures/purple_solid.png}) represent the results at $\hat{K} = 10$.}
    \label{fig:mis-esimate K_appendix linear}
\end{figure}

The results under the logistic regression model at (a) $n=500$ and $p$ vary from 30 to 1500; (b) $p = 600$ and $n$ varies from 100 to 1500 are presented in Fig.~\ref{fig:mis-esimate K_appendix}. The results under the linear model at the same regime of $n$ and $p$ are presented in Fig.~\ref{fig:mis-esimate K_appendix linear}.
From the results, we find that  the overestimation of $K$ appears not to affect the asymptotic normality results but the underestimation of $K$ can influence the asymptotic distribution of debiased estimation and further affects confidence interval estimation. Intuitively, as long as the corresponding linear combinations of the true underlying factors $\bU$ in the considered models can be well approximated by those of the estimated $\hat \bU$, the developed inference results for $\theta^*$ would still hold.

\section*{Appendix D. Proof of Theorem~\ref{Initial estimator consistency}}
\label{sec:proof of thm1}
\subsection*{D.1 Estimation Consistency of $\hat{\bm{\eta}}$}
\label{estimation consistency of eta}

Theorem~\ref{Initial estimator consistency} shows that the estimators $\hat{\bm{\eta}}$ and $\hat{\bw}$ can be consistently estimated. Before proving these results, we first introduce some lemmas that will be used in the proofs.

\begin{lemma}[Concentration of the Gradient and Hessian]
\label{concentration of gradient and Hessian}
 Under Assumptions~\ref{assumption1}--\ref{glm assumption} and the scaling condition $n, p \rightarrow \infty$, we have 
 
 $(i)$ $\|\nabla l(\bm{\eta}^{*})\|_{\infty}= 	\| n^{-1} \sum_{i=1}^{n}[y_i- b^{\prime}\{(\bm{\eta}^*)^{\T} \dot{\bZ}_{i} \}] \dot{\bZ}_i\|_{\infty}= O_{p}\{ n^{-1/2}(\log p)^{1/2} + p^{-1/2}(\log n)^{1/2} \}$;

$(ii)$ $\|(\btau^*)^\T \nabla^{2} l(\bm{\eta}^{*})-E_{\bm{\eta}^{*}}\{(\btau^*)^\T \nabla^{2} l(\bm{\eta}^{*})\}\|_{\infty} \\
= \|n^{-1} \sum_{i=1}^{n}(\btau^{*})^\T b^{\prime\prime}\{(\bm{\eta}^{*})^\T \dot{\bZ}_{i} \}\dot{\bZ}_i \dot{\bZ}_i^\T -E_{\bm{\eta}^{*}}[(\btau^{*})^\T b^{\prime\prime}\{(\bm{\eta}^{*})^\T \dot{\bZ}_{i} \} \dot{\bZ}_i \dot{\bZ}_i^\T ] \|_{\infty} \\
=  O_{p}\{ n^{-1/2}(\log p)^{1/2} + p^{-1/2}(\log n)^{1/2} \}$.
\end{lemma}

Lemma~\ref{concentration of gradient and Hessian} shows that there exist sub-exponential type of bounds on the gradient and linear combination of the Hessian of the loss functions.  This is motivated by Assumption 3.2 in~\cite{ning2017}. They construct the loss function based on the observed covariates whereas our results are established upon the fact that the gradient and the hessian of the loss function involve not only observed covariates but also the estimated unmeasured confounders.

\begin{remark}
\label{remark:estimate U} 
As the decomposition techniques commonly used in linear models may not be applicable in generalized linear model settings, it is necessary to establish stronger and more general intermediate results as the foundation of our theoretical analysis. For instance, different from linear models, where average estimation consistency on unmeasured confounders suffices, stronger uniform estimation consistency is necessary for the generalized linear framework. Specifically, in~\cite{fan2022latent}, the linear model form and projection-based techniques enable the reduction of the gradient max-norm into $\|\hat{\bE}^{\T}(\tilde{\by} - \hat{\bE}\bgamma^*)\|_{\infty}$, where $\bgamma^* = (\theta^*, (\bv^*)^{\T})^{\T}$ and $\tilde{\by} = (\bI_n - n^{-1}\hat{\bU}\hat{\bU}^{\T}) \by$ is the residual of the response $\by$ after projecting in onto space of $\hat{\bU}$. To show the concentration of the gradient, a key step is to upper bound $\| (\bU^*)^{\T}\hat{\bE} \bphi\|_{\infty}$ with $\|\bphi\|_2=1$ by $\|(\hat{\bE} - \bE) ^{\T}(\bU \bH^{\T} - \hat{\bU}) \bH^{-\T} \bphi\|_{\infty}$ and $\|\bE^{\T} (\bU \bH^{\T} - \hat{\bU}) \bH^{-\T} \bphi\|_{\infty}$, where they apply Cauchy-Schwartz inequality and use the Frobenius norm of estimated unmeasured confounders $\|(\bU \bH^{\T} - \hat{\bU})\|_{F}$. Here $\bH_{K \times K}$ is some transformation matrix and with a slight abuse of notation, we use $\bE \in \RR^{n \times p}$, $\bU \in \RR^{n \times K}$ and $\by \in \RR^n$ to denote the matrix (vector) form of random error, unmeasured confounders, and responses. However in the generalized linear model, to derive the bound for gradient max-norm to be $\| n^{-1} \sum_{i=1}^{n}[y_i- b^{\prime}\{(\bm{\eta}^*)^{\T} \dot{\bZ}_{i} \}] \dot{\bZ}_i\|_{\infty}$, we instead apply Bernstein inequality which requires the uniform estimation bound of unmeasured confounders, that is, $\max_i\|\hat{\bU}_i - \bU_i\|_{\infty}$. We leave the detailed proof of Lemma~\ref{concentration of gradient and Hessian} in Appendix~G.1. 
\end{remark}

We next obtain the upper bound for the estimation error of $\hat{\bm{\eta}}$.  
To this end, we define the first-order approximation to the loss difference as 
$D(\hat{\bm{\eta}}, \bm{\eta}^{*}) = (\hat{\bm{\eta}} - \bm{\eta}^{*})^\T \{\nabla l({\hat{\bm{\eta}}}) - \nabla l(\bm{\eta}^{*})\}$.
We will obtain upper and lower bounds of $D(\hat{\bm{\eta}}, \bm{\eta}^{*})$. As we will show, the $\ell_2$ norm $\|\hat{\bm{\eta}} - \bm{\eta}^{*}\|_2$ is involved in both upper and lower bounds, combining which will give an inequality of $\|\hat{\bm{\eta}}-\bm{\eta}^{*}\|_2$ and thus result in a bounded estimation error of $\hat{\bm{\eta}}$.

 {\em Upper Bound for $D(\hat{\bm{\eta}}, \bm{\eta}^{*})$}: recall that $\hat{\bm{\eta}}^\T = (\hat{\bgamma}^\T, \hat{\bbeta}^\T)$ is a solution obtained from solving the convex optimization problem in~\eqref{eq:initial estimator}.  And thus, we have the following optimality conditions $ -\nabla l(\hat{\bgamma})= \lambda\brho$ and $\nabla l(\hat{\bbeta}) = 0$, where $\bgamma$ is a $p \times 1$ vector with entries
\begin{eqnarray}
\brho_{j} = \left\{\begin{array}{ll}
\operatorname{sign}(\hat{{\gamma}}_{j}), &\ \hat{{\gamma}}_{j} \neq 0 \\
{[-1,1]}, &\  \hat{{\gamma}}_{j}=0
\end{array}, \quad (j=1, \ldots p) .\right. \nonumber
\end{eqnarray}
This implies $\|\nabla l({\hat{\bm{\eta}}})\|_{\infty} =  \| n^{-1} \sum_{i=1}^{n}\dot{\bZ}_{i}\{y_i- b^{\prime}(\hat{\bm{\eta}}^{\T} \dot{\bZ}_{i} )\} \|_{\infty}\leq \lambda$.

Recall that we denote the support set for $\bm{\eta}^*$ as $\cS_{\eta} = \{ j: {\eta}_j^* \neq 0\}$. 
We denote the difference between the estimator $\hat{\bm{\eta}}$ and the true parameter $\bm{\eta}^*$ as $\hat{\Delta}=\hat{\bm{\eta}}-\bm{\eta}^{*}$, and its two sub-vectors are $\hat{\Delta}_{\cS} = (\hat{\eta}_j - \eta_j^* : j \in \cS_{\eta} )$ and  $\hat{\Delta}_{\bar{\cS}} = (\hat{\eta}_j - \eta_j^* : j \notin \cS_{\eta} )$. 
Similarly, we denote the sub-vectors of $\dot{\bZ}_i$ corresponding to non-zero entries as $\dot{\bZ}_{i,\cS} = \{\dot{\bZ}_{ij}: j \in \mathcal{S}_{\eta}\}$ and that corresponding to zero entries as $\dot{\bZ}_{i,\bar{\cS}} = \{\dot{\bZ}_{ij}: j \notin \mathcal{S}_{\eta}\}$. 
 With the notations introduced, the quadratic difference $D(\hat{\bm{\eta}}, \bm{\eta}^{*})$ can then be written as 
\begin{eqnarray}
D(\hat{\bm{\eta}}, \bm{\eta}^{*}) &=& (\hat{\bm{\eta}}-\bm{\eta}^{*})^\T \{\nabla l({\hat{\bm{\eta}}}) - \nabla l(\bm{\eta}^{*}) \}	\nonumber \\
&=&	 -\frac{1}{n} \sum_{i=1}^{n} \hat{{\Delta}}_{\cS}^{\T} \dot{\bZ}_{i,\cS}\{y_i- b^{\prime}(\hat{\bm{\eta}}^{\T} \dot{\bZ}_{i} )\}-\frac{1}{n} \sum_{i=1}^{n} \widehat{{\Delta}}_{\bar{\cS}}^{\T} \dot{\bZ}_{i,\bar{\cS}}\{y_i- b^{\prime}(\hat{\bm{\eta}}^{\T} \dot{\bZ}_{i} )\}-\widehat{{\Delta}}^{\T} \nabla l({\bm{\eta}}^{*}) \nonumber \\
& \leq & \lambda\|\hat{\Delta}_{S}\|_{1} - \lambda\|\hat{\Delta}_{\bar{S}}\|_{1} + \|\hat{\Delta}\|_{1}\|\nabla l(\bm{\eta}^{*}) \|_{\infty} \nonumber 
\end{eqnarray}
where the last inequality is by H{\"o}lder's inequality.

From Lemma~\ref{concentration of gradient and Hessian}, we have $\|\nabla l(\bm{\eta}^{*})\|_{\infty} \lesssim  {n^{-1/2}(\log p)^{1/2}} + p^{-1/2} (\log n)^{1/2}$. Since $\|\hat{\Delta}\|_{1} \leq \|\hat{\Delta}_{\cS}\|_{1} + \|\hat{\Delta}_{\bar{\cS}}\|_{1}$ and by taking $\lambda = 2c \{n^{-1/2}(\log p)^{1/2} + p^{-1/2} (\log n)^{1/2}\}$ where $c >0$ is some constant, we further have
\begin{eqnarray}
D(\hat{\bm{\eta}}, \bm{\eta}^{*})  &\leq & c  \left( \sqrt{\frac{\log p}{n}} + \sqrt{\frac{\log n}{p}} \right)(3\|\hat{\Delta}_S\|_{1}-\|\hat{\Delta}_{\bar{S}}\|_{1}) \nonumber \\
&\leq&  3 cs_{\eta}^{1/2} \left( \sqrt{\frac{\log p}{n}} + \sqrt{\frac{\log n}{p}} \right)\|\hat{\Delta}\|_{2}, \label{eq:3dS 1dS_bar} 
\end{eqnarray}
where the last inequality is because  $\| \hat{\Delta}_{\cS}\|_1 \leq s_{\eta}^{1/2} \|\hat{\Delta}_{\cS}\|_2 \leq s_{\eta}^{1/2} \|\hat{\Delta}\|_2$.

 {\em Lower Bound for $D(\hat{\bm{\eta}}, \bm{\eta}^{*})$}: after establishing the upper bound for the quadratic difference, we next obtain a lower bound for it. 
Because $ l(\bm{\eta}^{*})$ is convex function, $D(\hat{\bm{\eta}}, \bm{\eta}^{*}) \geq 0$. We have $\|\hat{\Delta}_{\bar{S}}\|_{1} \leq 3\|\hat{\Delta}_{S}\|_{1}$ from~\eqref{eq:3dS 1dS_bar}, and based on this result and the restricted strong
convexity condition for generalized linear model in Proposition 1 of \cite{loh2013}, 
\begin{equation}
    D(\hat{\bm{\eta}}, \bm{\eta}^{*})  \geq \kappa_2 \|\hat{\Delta}\|_{2}^{2}, \label{eq:lb for D}
\end{equation}
for some constant $\kappa_2 > 0$.

Then combining the upper bound~\eqref{eq:3dS 1dS_bar} and the lower bound~\eqref{eq:lb for D} for $D(\hat{\bm{\eta}}, \bm{\eta}^{*})$, we have 
\begin{equation}
\|\hat{\Delta}\|_{2} \leq \frac{3cs_{\eta}^{1/2}}{\kappa_2}  \left( \sqrt{\frac{\log p}{n}} + \sqrt{\frac{\log n}{p}} \right), \label{eq:Delta norm2}
\end{equation}
and so
\begin{eqnarray}
\|\hat{\Delta}\|_{1} &\leq & \|\hat{\Delta}_{\cS}\|_{1} + \|\hat{\Delta}_{\bar{\cS}}\|_{1}  \leq 4\|\hat{\Delta}_{S}\|_{1} \leq 4 s_{\eta}^{1/2}\|\hat{\Delta}_S\|_{2} \leq 4 s_{\eta}^{1/2}\| \hat{\Delta} \|_{2} \nonumber \\
&\leq & \frac{12 c s_{\eta}}{\kappa_2} \left( \sqrt{\frac{\log p}{n}} + \sqrt{\frac{\log n}{p}} \right), \nonumber \end{eqnarray}
 which completes the first part proof.
\subsection*{D.2 Estimation Consistency of $\hat{\bw}$}
\label{pf:estimation consistency of w}

Before we present the proof for the estimation consistency of the estimator $\hat{\bw}$, we introduce  additional results established based on the estimation consistency of $\hat{\bm{\eta}}$ and used in proving the estimation consistency of $\hat{\bw}$. 
\begin{lemma}
\label{glm complementary of consistency results}
 Under Assumptions~\ref{assumption1}--\ref{glm assumption}, with $\lambda \asymp \lambda^{\prime} \asymp  {n^{-1/2}(\log p)^{1/2}} + p^{-1/2}(\log n)^{1/2} $ and $(s_{\eta} \vee s_w)({n^{-1/2}(\log p)^{1/2}} + p^{-1/2}(\log n)^{1/2} ) = o_p(1)$, then we have
	\begin{eqnarray}
		&& \frac{1}{n} \sum_{i=1}^{n}(\hat{\bm{\eta}}-\bm{\eta}^{*})^{\T} {\dot{\bZ}}_{i} b^{\prime \prime}\{(\bm{\eta}^{*})^\T \dot{\bZ}_{i}\} \dot{\bZ}_{i}^{\T}(\hat{\bm{\eta}}-\bm{\eta}^{*})= O_p\left\{ s_\eta\left( \frac{\log p}{n} + \frac{\log n}{p} \right)\right\} , \label{eta bprime z} \\
		&& \frac{1}{n} \sum_{i=1}^{n}(\hat{\bw}-\bw^{*})^{\T} \dot{\bM}_{i} b^{\prime \prime}\{(\bm{\eta}^{*})^\T \dot{\bZ}_{i}\} \dot{\bM}_{i}^{\T}(\hat{\bw}-\bw^{*}) 
		=  O_p\left\{ (s_{\eta} \vee s_{w})\left( \frac{\log p}{n} + \frac{\log n}{p} \right)\right\}, \nonumber \\
  && \label{w bprime m} \\
		&& \frac{1}{n} \sum_{i=1}^{n}(\hat{\bw}-\bw^{*})^{\T} \dot{\bM}_{i} b^{\prime \prime}\{\hat{\bm{\eta}}^\T \dot{\bZ}_{i}\} \dot{\bM}_{i}^{\T}(\hat{\bw}-\bw^{*}) 
		=  O_p\left\{ (s_{\eta} \vee s_{w})\left( \frac{\log p}{n} + \frac{\log n}{p} \right)\right\}.  \nonumber \\
  && \label{w bprimehat m}
	\end{eqnarray}
\end{lemma}
\begin{remark}
\label{rmk:discuss lemma 4}
For Lemma~\ref{glm complementary of consistency results}, we prove the results~\eqref{w bprime m} and~\eqref{w bprimehat m} along the way in proving our main result of the estimation consistency of $\hat{\bw}$ as the two inequalities are direct consequences of certain intermediate steps. 
We leave the proof for~\eqref{eta bprime z} in Appendix G.2.
\end{remark}

To prove the consistency of $\hat{\bw}$, we denote  $\hat{\bdelta}=\hat{\bw}-\bw^{*}$ and establish the estimation error bound for $\|\hat{\bdelta}\|_1$.  Recall that $\hat{\bw}$ is a solution to~\eqref{eq:estimator w}, that is 
\begin{equation}
		\hat{\bw} = \underset{\bw}{\operatorname{argmin}}~\frac{1}{2 n} \sum_{i=1}^{n}\{ \bw^\T b^{\prime\prime}(\hat{\bm{\eta}}^\T\dot{\bZ}_i) \dot{\bM}_i \dot{\bM}_i^T\bw -2\bw^{\T} b^{\prime\prime}(\hat{\bm{\eta}}^\T\dot{\bZ}_i)D_i  \dot{\bM}_i\}+\lambda^{\prime} \|\bw \|_1. \nonumber
\end{equation}

By definition, we have
\begin{eqnarray}
&&\frac{1}{2 n} \sum_{i=1}^{n}\{ \hat{\bw}^\T b^{\prime\prime}(\hat{\bm{\eta}}^\T\dot{\bZ}_i) \dot{\bM}_i \dot{\bM}_i^T\hat{\bw} -2\hat{\bw}^{\T} b^{\prime\prime}(\hat{\bm{\eta}}^\T\dot{\bZ}_i)D_i  \dot{\bM}_i\}+\lambda^{\prime} \|\hat{\bw} \|_1 \nonumber \\
&&\leq  \frac{1}{2 n} \sum_{i=1}^{n}\{ (\bw^*)^\T b^{\prime\prime}(\hat{\bm{\eta}}^\T\dot{\bZ}_i) \dot{\bM}_i \dot{\bM}_i^T\bw^*-2(\bw^*)^\T b^{\prime\prime}(\hat{\bm{\eta}}^\T\dot{\bZ}_i)D_i  \dot{\bM}_i\}+\lambda^{\prime} \|\bw^* \|_1. \nonumber
\end{eqnarray}
Write $(\hat{\bdelta}^\T\dot{\bM}_i)^2 =(\hat{\bw}^\T \dot{\bM}_i)^2 - \{(\bw^*)^\T \dot{\bM}_i\}^2 - 2\hat{\bw}^\T \dot{\bM}_i \dot{\bM}_i^\T \bw^* + 2\{(\bw^*)^\T \dot{\bM}_i\}^2 $, then the above inequality can be rearranged as
\begin{eqnarray}
\frac{1}{2 n} \sum_{i=1}^{n} b^{\prime\prime}(\hat{\bm{\eta}}^\T\dot{\bZ}_i)  (\hat{\bdelta}^\T\dot{\bM}_i)^2 \leq \frac{1}{n} \sum_{i=1}^{n} b^{\prime\prime}(\hat{\bm{\eta}}^\T\dot{\bZ}_i)\{D_i - (\bw^*)^\T\dot{\bM}_i\} \dot{\bM}_i^\T \hat{\bdelta} +\lambda^{\prime} \|\bw^* \|_1 - \lambda^{\prime} \|\hat{\bw} \|_1. \nonumber \\ \label{eq:w norm inequality} 
\end{eqnarray}

The proof techniques are mostly motivated by~\cite{ning2017}. To present our proof, we define two quadratic difference terms as $Q(\hat{\bw}, \bw^*)$ $= (\hat{\bw}-\bw^*)^{\T}\nabla^{2} l(\hat{\bm{\eta}})(\hat{\bw}-\bw^*) =  n^{-1}  \sum_{i=1}^{n} b^{\prime\prime}(\hat{\bm{\eta}}^\T\dot{\bZ}_i) (\dot{\bM}_{i}^{\T} \hat{\bdelta})^2$ and $Q^*(\hat{\bw}, \bw^*)  = (\hat{\bw}-\bw^*)^{\T} $ $\nabla^{2} l({\bm{\eta}}^*) (\hat{\bw}-\bw^*)= n^{-1}  \sum_{i=1}^{n} b^{\prime\prime}\{(\bm{\eta}^*)^\T\dot{\bZ}_i\}  (\dot{\bM}_{i}^{\T}\hat{\bdelta})^2$.
The left hand side of the above inequality is $Q(\hat{\bw}, \bw^*)$, and next we investigate the upper bound for $Q(\hat{\bw}, \bw^*)$ in details.

 For the right hand side of~\eqref{eq:w norm inequality}, we first consider $\lambda^{\prime} \|\bw^* \|_1 - \lambda^{\prime} \|\hat{\bw} \|_1$. Recall that we denote the support set for $\bw^*$ as $\cS_{w} = \{ j: {\bw}_j^* \neq 0\}$. 
We also denote $\bw^*_{\cS} = (\bw^*_j : j \in \cS_w)$, $\bw^*_{\bar{\cS}} = (\bw^*_j : j \notin \cS_w)$, $\hat{\bdelta}_{\cS} = (\hat{\bw}_j - \bw_j^* : j \in \cS_{w} )$ and  $\hat{\bdelta}_{\bar{\cS}} = (\hat{\bw}_j - \bw_j^* : j \notin \cS_{w} )$. So $\|\hat{\bdelta}_{\bar{\cS}}\|_1 = \|\hat{\bw}_{\bar{\cS}}\|_1$ and $\|\bw^*_{\bar{\cS}}\|_1= 0$. Therefore we have
\begin{eqnarray}
    \lambda^{\prime} \|\bw^* \|_1 - \lambda^{\prime} \|\hat{\bw} \|_1 &=& \lambda^{\prime} \|\bw_{\cS}^* \|_1  + \lambda^{\prime} \|\bw_{\bar{\cS}}^* \|_1 - \lambda^{\prime} \|\hat{\bw}_{\cS} \|_1 -\lambda^{\prime} \|\hat{\bw}_{\bar{\cS}} \|_1 \nonumber \\
    & = & \lambda^{\prime} \|\bw_{\cS}^* \|_1   - \lambda^{\prime} \|\hat{\bw}_{\cS} \|_1 -\lambda^{\prime} \|\hat{\bw}_{\bar{\cS}} \|_1 \nonumber \\
    & \leq & \lambda^{\prime} \|\hat{\bdelta}_{\cS} \|_1   - \lambda^{\prime} \|\hat{\bdelta}_{\bar{\cS}}\|_1. \label{eq: w l1 norm}
\end{eqnarray}
And according to Lagranian duality theory, an equivalent problem for~\eqref{eq:estimator w} is 
\begin{equation}
		\hat{\bw} = \underset{\bw}{\operatorname{argmin}}~\|\bw\|_1 ~\text{ s.t. }~ \frac{1}{2 n} \sum_{i=1}^{n}\{ \bw^\T \nabla_{\bzeta\bzeta} \ell_{i}(\hat{\bm{\eta}}) \bw -2\bw^{\T} \nabla_{\bzeta\theta} \ell_{i}(\hat{\bm{\eta}})\} \leq b^2, \nonumber
		\end{equation}
		for $b > 0$. This gives $ \|\hat{\bw}\|_1 \leq \|\bw^*\|_1$, which further results in $   \|\hat{\bdelta}_{\bar{\cS}}\|_1 \leq \|\hat{\bdelta}_{\cS} \|_1$ from~\eqref{eq: w l1 norm}.

We next consider the first term in~\eqref{eq:w norm inequality}. Denote $I_1 = n^{-1} \sum_{i=1}^{n} b^{\prime\prime}(\hat{\bm{\eta}}^\T\dot{\bZ}_i)\{D_i - (\bw^*)^\T\dot{\bM}_i\}$ $ \dot{\bM}_i^\T \hat{\bdelta}$, which is a summation of two terms $I_{11}$ and $I_{12}$ denoted as  
\begin{eqnarray}
I_1
 &=&	\frac{1}{n}\sum_{i=1}^{n}b^{\prime\prime}\{(\bm{\eta}^*)^\T\dot{\bZ}_i\} \{D_i - (\bw^*)^\T\dot{\bM}_i\}\dot{\bM}_{i}^{\T}\hat{\bdelta} \nonumber \\
 && + \frac{1}{n}\sum_{i=1}^{n} [b^{\prime\prime}(\hat{\bm{\eta}}^\T\dot{\bZ}_i) - b^{\prime\prime}\{(\bm{\eta}^*)^\T\dot{\bZ}_i\}]\{D_i - (\bw^*)^\T\dot{\bM}_i\}\dot{\bM}_{i}^{\T}\hat{\bdelta} \nonumber\\
  &=& I_{11} + I_{12}. \nonumber 
\end{eqnarray}

For $I_{11}$, using a similar argument as in the proof of Lemma \ref{concentration of gradient and Hessian}, we have $\|n^{-1}\sum_{i=1}^{n}$ $b^{\prime\prime}\{(\bm{\eta}^*)^\T\dot{\bZ}_i\}$ $ \{D_i - (\bw^*)^\T\dot{\bM}_i\}\dot{\bM}_{i}^{\T}\|_{\infty} \lesssim  n^{-1/2}(\log p)^{1/2} +  p^{-1/2}(\log n)^{1/2} $, hence
\begin{eqnarray}
|I_{11}|& \leq &	\|\hat{\bdelta}\|_1\|n^{-1}\sum_{i=1}^{n}b^{\prime\prime}\{(\bm{\eta}^*)^\T\dot{\bZ}_i\} \{D_i - (\bw^*)^\T\dot{\bM}_i\}\dot{\bM}_{i}^{\T}\|_{\infty} \nonumber \\
& \lesssim & \left( \sqrt{\frac{\log p}{n}} + \sqrt{\frac{\log n}{p}} \right)	\|\hat{\bdelta}\|_1. \label{I11 in w glm}
\end{eqnarray}

For $I_{12}$, by Assumption~\ref{glm assumption} that $|b^{\prime \prime}(t_{1})-b^{\prime \prime}(t)| \leq B|t_{1}-t| b^{\prime \prime}(t)$ with $t_1 = \hat{\eta}^{\T} \dot{\bZ}_{i}$ and $t =(\eta^{*})^\T \dot{\bZ}_{i}$ and applying Cauchy-Schwarz inequality, we have
\begin{eqnarray}
|I_{12}| & \leq & \left|n^{-1}\sum_{i=1}^{n} b^{\prime\prime}\{(\bm{\eta}^*)^\T\dot{\bZ}_i\}|\hat{\bm{\eta}}^{\T} \dot{\bZ}_{i}-(\bm{\eta}^{*})^\T \dot{\bZ}_{i}|\{D_i - (\bw^*)^\T\dot{\bM}_i\}\dot{\bM}_{i}^{\T}\hat{\bdelta} \right| \nonumber \\
& \leq & \left|n^{-1}\sum_{i=1}^{n} b^{\prime\prime}\{(\bm{\eta}^*)^\T\dot{\bZ}_i\} (\dot{\bM}_{i}^{\T}\hat{\bdelta})^2 \right|^{1/2}\left|n^{-1}\sum_{i=1}^{n} b^{\prime\prime}\{(\bm{\eta}^*)^\T\dot{\bZ}_i\}\{ (\hat{\bm{\eta}} - \bm{\eta}^*)^\T \dot{\bZ}_i \}^2\right|^{1/2} \nonumber \\
&\lesssim & |Q^*(\hat{\bw}, \bw^*) |^{1/2} s_{\eta}^{1/2}\left( \sqrt{\frac{\log p}{n}} + \sqrt{\frac{\log n}{p}} \right), \label{I12 in w glm}
\end{eqnarray}
where the last inequality is from~\eqref{eta bprime z} in Lemma \ref{glm complementary of consistency results}.

Combining the upper bounds in~\eqref{eq: w l1 norm}, ~\eqref{I11 in w glm} and~\eqref{I12 in w glm} into~\eqref{eq:w norm inequality}, we have
\begin{eqnarray}
	Q(\hat{\bw}, \bw^*)  &\lesssim &  \left( \sqrt{\frac{\log p}{n}} + \sqrt{\frac{\log n}{p}} \right)	\|\hat{\bdelta}\|_1 \nonumber \\
 &&+ |Q^*(\hat{\bw}, \bw^*) |^{1/2} s_{\eta}^{1/2}\left( \sqrt{\frac{\log p}{n}} + \sqrt{\frac{\log n}{p}} \right) \nonumber \\
 && + \lambda^{\prime} \|\hat{\bdelta}_{\cS} \|_1   - \lambda^{\prime} \|\hat{\bdelta}_{\bar{\cS}}\|_1. \nonumber \\
	\label{Q tau taus} 
\end{eqnarray}

 The above upper bound of $Q(\hat{\bw}, \bw^*)$ involves the expression of $Q^*(\hat{\bw}, \bw^*)$. We next investigate the relation between $Q(\hat{\bw}, \bw^*)$ and $Q^*(\hat{\bw}, \bw^*)$. 
We apply Assumption~\ref{glm assumption}  again on the difference between $Q(\hat{\bw}, \bw^*)$ and $Q^*(\hat{\bw}, \bw^*) $ as we do for bounding the term $I_{12}$, and have
\begin{eqnarray}
	|Q(\hat{\bw}, \bw^*)  - Q^*(\hat{\bw}, \bw^*)| &=& \left|\frac{1}{n} \sum_{i=1}^{n}[b^{\prime\prime}(\hat{\bm{\eta}}^\T\dot{\bZ}_i) -b^{\prime\prime}\{(\bm{\eta}^*)^\T\dot{\bZ}_i\}](\dot{\bM}_{i}^{\T}\hat{\bdelta} )^2 \right| \nonumber \\
	& \leq & \left|\frac{1}{n} \sum_{i=1}^{n}b^{\prime\prime}\{(\bm{\eta}^*)^\T\dot{\bZ}_i\}(\dot{\bM}_{i}^{\T}\hat{\bdelta} )^2 (\hat{\bm{\eta}} - \bm{\eta}^*)^\T \dot{\bZ}_i \right| \nonumber \\
	& \leq & Q^*(\hat{\bw}, \bw^*)   \|\hat{\bm{\eta}} - \bm{\eta}^*\|_1 \max_{i=1, \ldots, n}\|\dot{\bZ}_i\|_{\infty} \nonumber \\
	& \lesssim & Q^*(\hat{\bw}, \bw^*)  { s_{\eta}} \left(\sqrt{\frac{\log p}{n}} +\sqrt{\frac{\log n}{p}} \right) \left( M + \frac{1}{\sqrt{n}} +\sqrt{\frac{\log n}{p}} \right), \nonumber
 \end{eqnarray}
 where the last inequality is by the estimation consistency results of $\hat{\bm{\eta}}$ in Appendix D.1 and by Proposition~\ref{prop:uniform U} that $\|\dot{\bZ}_i\|_{\infty} = M + O_p(n^{-1/2} + p^{-1/2}(\log n)^{1/2})$ and $s_{\eta} ( n^{-1/2}(\log p)^{1/2} + p^{-1/2}(\log n)^{1/2}) = o_p(1)$. 
 From the above result, we further apply triangular inequality and have
 \begin{equation}
     Q^*(\hat{\bw}, \bw^*)  \left\{1- { s_{\eta}} \left(\sqrt{\frac{\log p}{n}} +\sqrt{\frac{\log n}{p}} \right) \left( M + \frac{1}{\sqrt{n}} +\sqrt{\frac{\log n}{p}} \right) \right\} \lesssim Q(\hat{\bw}, \bw^*). \label{eq:Qstar and Q}
 \end{equation}
 
 Combining the above result on $Q^*(\hat{\bw}, \bw^*)$ with~\eqref{Q tau taus} and taking $\lambda^{\prime} = 2\{n^{-1/2}(\log p)^{1/2} + p^{-1/2}(\log n)^{1/2}\}$, we get
\begin{eqnarray}
	Q(\hat{\bw}, \bw^*) 
&\lesssim &|Q(\hat{\bw}, \bw^*) |^{1/2} s_{\eta}^{1/2}\left( \sqrt{\frac{\log p}{n}} + \sqrt{\frac{\log n}{p}} \right) \nonumber \\
&&+ \left( \sqrt{\frac{\log p}{n}} + \sqrt{\frac{\log n}{p}} \right) (3\|\hat{\bdelta}_{\cS} \|_1   - \|\hat{\bdelta}_{\bar{\cS}}\|_1). 
\label{Q1 quadratic}
\end{eqnarray}

The above result will be used to prove~\eqref{w bprimehat m}, which will later be used to derive the error bound $\| \hat{\bdelta}\|_1$. Consider the following two cases.

 Case 1: $ |Q(\hat{\bw}, \bw^*)|^{1/2} \lesssim s_{\eta}^{1/2} (n^{-1/2} (\log p)^{1/2} + p^{-1/2} (\log n)^{1/2})$. We have~\eqref{w bprimehat m} naturally hold. 

  Case 2: $ |Q(\hat{\bw}, \bw^*)|^{1/2} \gtrsim s_{\eta}^{1/2} (n^{-1/2} (\log p)^{1/2} + p^{-1/2} (\log n)^{1/2})$. Then we can have 
  \begin{equation}
      Q(\hat{\bw}, \bw^*)  - |Q(\hat{\bw}, \bw^*) |^{1/2} s_{\eta}^{1/2}\left( \sqrt{\frac{\log p}{n}} + \sqrt{\frac{\log n}{p}} \right) >0. \nonumber
  \end{equation}
combining this result with~\eqref{Q1 quadratic}, we have $3\|\hat{\bdelta}_{\cS} \|_1 \geq \|\hat{\bdelta}_{\bar{\cS}}\|_1$.
We next use this cone condition to derive the lower bound for $Q(\hat{\bw}, \bw^*)$ for case 2.

Denote  $\tilde{\bU}_i^\T = ( \bm{0}_{p-1}, \hat{\bU}_i^\T - \bU_i^\T)_{p+K-1}$ as a vector including a vector of zeros and the differences between estimated unmeasured confounders and the true confounders. Hence we have $\dot{\bM}_i = \bM_i + \tilde{\bU}_i$. We  establish the lower bounds for $Q(\hat{\bw}, \bw^*)/\| \hat{\bdelta}\|_2^2$ first.

\begin{eqnarray}
\frac{Q(\hat{\bw}, \bw^*)}{\| \hat{\bdelta}\|_2^2} &=& \frac{\hat{\bdelta}^\T\{ n^{-1}  \sum_{i=1}^{n} b^{\prime\prime}(\hat{\bm{\eta}}^\T\dot{\bZ}_i) \dot{\bM}_{i} \dot{\bM}_{i}^{\T} \}\hat{\bdelta} }{\| \hat{\bdelta}\|_2^2}	\nonumber \\ 
&=& \frac{\hat{\bdelta}^\T\{ n^{-1}  \sum_{i=1}^{n} b^{\prime\prime}(\hat{\bm{\eta}}^\T\dot{\bZ}_i)  {\bM}_{i} {\bM}_{i}^{\T} \}\hat{\bdelta} }{\| \hat{\bdelta}\|_2^2}+\frac{2\hat{\bdelta}^\T\{ n^{-1}  \sum_{i=1}^{n} b^{\prime\prime}(\hat{\bm{\eta}}^\T\dot{\bZ}_i)  {\bM}_{i} \tilde{\bU}_{i}^{\T} \}\hat{\bdelta} }{\| \hat{\bdelta}\|_2^2} \nonumber\\
&& +\frac{\hat{\bdelta}^\T\{ n^{-1}  \sum_{i=1}^{n} b^{\prime\prime}(\hat{\bm{\eta}}^\T\dot{\bZ}_i)  \tilde{\bU}_{i} \tilde{\bU}_{i}^{\T} \}\hat{\bdelta} }{\| \hat{\bdelta}\|_2^2} \nonumber \\
&=& Q_1 + Q_2 + Q_3. \label{vzzv}
\end{eqnarray}

We  next derive the bounds for the three terms in~\eqref{vzzv} separately. For $Q_1$, we have 
\begin{equation}
	Q_1 =  \frac{\hat{\bdelta}^\T[ n^{-1}  \sum_{i=1}^{n} b^{\prime\prime}\{(\bm{\eta}^*)^\T{\bZ}_i\}  {\bM}_{i} {\bM}_{i}^{\T} ]\hat{\bdelta} }{\| \hat{\bdelta}\|_2^2} +\frac{1}{n} \sum_{i=1}^{n} \frac{(\bM_{i}^\T \hat{\bdelta})^{2}}{\| \hat{\bdelta}\|_2^2}[b^{\prime \prime}(\hat{\bm{\eta}}^\T \dot{\bZ}_{i})-b^{\prime \prime}\{(\bm{\eta}^{*})^\T \bZ_{i}\}]. \label{eq:Q1}
\end{equation}
For the second term in~\eqref{eq:Q1}, consider Assumption~\ref{glm assumption} that $|b^{\prime \prime}(t_{1})-b^{\prime \prime}(t)| \leq B|t_1 - t| b^{\prime \prime}(t)$ by letting $t_1 = \hat{\bm{\eta}}^{\T} \dot{\bZ}_{i}$ and $t =(\bm{\eta}^{*})^\T \dot{\bZ}_{i}$ and applying Cauchy-Schwarz inequality, we have
\begin{eqnarray}
	&&\max_{i=1, \ldots, n} | b^{\prime \prime}(\hat{\bm{\eta}}^\T \dot{\bZ}_{i})-b^{\prime \prime}\{(\bm{\eta}^{*})^\T \bZ_{i}\}| \nonumber \\
 &\leq & \max_{i=1, \ldots, n}B  b^{\prime \prime}\{(\bm{\eta}^{*})^\T \bZ_{i}\}|\hat{\bm{\eta}}^\T \dot{\bZ}_{i} -  (\bm{\eta}^{*})^\T \bZ_{i}| \nonumber \\
	&\leq & \max_{i=1, \ldots, n}B  b^{\prime \prime}\{(\bm{\eta}^{*})^\T \bZ_{i}\}|\hat{\bm{\eta}}^\T \dot{\bZ}_{i} -  (\bm{\eta}^{*})^\T \dot{\bZ}_{i}  + (\bm{\eta}^{*})^\T \dot{\bZ}_{i} - (\bm{\eta}^{*})^\T \bZ_{i}| \nonumber \\
	&\leq & \max_{i=1, \ldots, n}B  b^{\prime \prime} \{(\bm{\eta}^{*})^\T \bZ_{i}\}| (\hat{\bm{\eta}}  -  \bm{\eta}^{*} )^\T\dot{\bZ}_i + (\bm{\eta}^{*})^\T(\dot{\bZ}_i - \bZ_i)|. \nonumber \\
	&\leq &\max_{i=1, \ldots, n}B  b^{\prime \prime}\{(\bm{\eta}^{*})^\T \bZ_{i}\} (\|\hat{\bm{\eta}}  -  \bm{\eta}^{*}\|_1\|\dot{\bZ}_i\|_{\infty} + \|\bm{\eta}^{*}\|_{1} \|\dot{\bZ}_i - \bZ_i\|_{\infty}).
	\label{max bhat - bs 1}
\end{eqnarray}
From Assumption~\ref{glm assumption}, $b^{\prime \prime}\{(\bm{\eta}^{*})^\T \bZ_{i}\} \in [0,B]$. From the result for estimation consistency of $\hat{\bm{\eta}}$ in Appendix D.1, we have $ \|\hat{\bm{\eta}}  -  \bm{\eta}^{*}\|_1 \lesssim {s_{\eta}} (n^{-1/2} (\log p)^{1/2} + p^{-1/2} (\log n)^{1/2})$. In addition, we have $\|\dot{\bZ}_i\|_{\infty} =  M + O_p(n^{-1/2} + p^{-1/2}(\log n)^{1/2})$. Based on these results and by the scaling condition $n,p \rightarrow \infty$ and ${s_{\eta}} (n^{-1/2} (\log p)^{1/2} + p^{-1/2} (\log n)^{1/2}) = o_p(1)$, we have 
\begin{equation}
	\max_{i=1, \ldots, n} | b^{\prime \prime}(\hat{\bm{\eta}}^\T \dot{\bZ}_{i})-b^{\prime \prime}\{(\bm{\eta}^{*})^\T \bZ_{i}\}| = o_p(1). \label{eq:max bhat - bs 2}
\end{equation}
Hence we have
\begin{equation}
	Q_1 \geq  \frac{\hat{\bdelta}^\T[ n^{-1}  \sum_{i=1}^{n} b^{\prime\prime}\{(\bm{\eta}^*)^\T{\bZ}_i\}  {\bM}_{i} {\bM}_{i}^{\T} ]\hat{\bdelta} }{4\| \hat{\bdelta}\|_2^2} \label{eq:Q1 lb}
\end{equation}
with probability tending to 1 as the second term in~\eqref{eq:Q1} goes to 0. Recall that we denote $\bI_{\bzeta\bzeta}^* = E[b^{\prime\prime}\{(\bm{\eta}^*)^\T \bZ_{i}\}{\bM}_i{\bM}_i^\T ]$. Because
\begin{eqnarray}
&&  \frac{\hat{\bdelta}^\T[ n^{-1}  \sum_{i=1}^{n} b^{\prime\prime}\{(\bm{\eta}^*)^\T{\bZ}_i\}  {\bM}_{i} {\bM}_{i}^{\T} ]\hat{\bdelta} }{\| \hat{\bdelta}\|_2^2} \nonumber \\
	 &= & \frac{\hat{\bdelta}^\T\bI_{\bzeta\bzeta}^* \hat{\bdelta}}{\| \hat{\bdelta}\|_2^2} + \frac{\hat{\bdelta}^\T |n^{-1}  \sum_{i=1}^{n} b^{\prime\prime}\{(\bm{\eta}^*)^\T{\bZ}_i\}  {\bM}_{i} {\bM}_{i}^{\T} -\bI_{\bzeta\bzeta}^*|\hat{\bdelta} }{\| \hat{\bdelta}\|_2^2} \nonumber \\
  & \geq &\lambda_{\min} (\bI_{\bzeta\bzeta}^*) - \left| \frac{\hat{\bdelta}^\T [n^{-1}  \sum_{i=1}^{n} b^{\prime\prime}\{(\bm{\eta}^*)^\T{\bZ}_i\}  {\bM}_{i} {\bM}_{i}^{\T} -\bI_{\bzeta\bzeta}^*]\hat{\bdelta} }{\| \hat{\bdelta}\|_2^2}  \right| \nonumber \\
	  &\geq & \kappa - \frac{\| \hat{\bdelta}\|_1^2\| n^{-1}  \sum_{i=1}^{n} b^{\prime\prime}\{(\bm{\eta}^*)^\T{\bZ}_i\}  {\bM}_{i} {\bM}_{i}^{\T} -\bI_{\bzeta\bzeta}^* \|_{\max} }{\| \hat{\bdelta}\|_2^2} \nonumber 
\end{eqnarray}
where the last inequality is by the definition and properties of eigenvalue as well as matrix operations. Recall that we have obtained the cone condition $3\|\hat{\bdelta}_{\cS} \|_1 \geq \|\hat{\bdelta}_{\bar{\cS}}\|_1$, which results in $\|\hat{\bdelta} \|_1^2 \leq 16\|\hat{\bdelta}_{\cS} \|_1^2 \leq 16s_w \|\hat{\bdelta} \|_2^2$. Moreover, using Bernstein's inequality, it can be shown that $\| n^{-1}  \sum_{i=1}^{n} b^{\prime\prime}\{(\bm{\eta}^*)^\T{\bZ}_i\}  $ ${\bM}_{i} {\bM}_{i}^{\T} -\bI_{\bzeta\bzeta}^* \|_{\max} = O_p\{n^{-1/2} (\log p)^{1/2}  \}$. As  $(s_{\eta} \vee s_w)(n^{-1/2} (\log p)^{1/2} + p^{-1/2} (\log n)^{1/2}) = o_p(1)$, the term $\| \hat{\bdelta}\|_1^2\| n^{-1}  \sum_{i=1}^{n} b^{\prime\prime}\{(\bm{\eta}^*)^\T{\bZ}_i\}  {\bM}_{i} {\bM}_{i}^{\T} -\bI_{\bzeta\bzeta} \|_{\max}/\| \hat{\bdelta}\|_2^2 = o_p(1)$, which together with~\eqref{eq:Q1 lb} shows  $Q_1 \geq \kappa/4$ with probability tending to 1.

For $Q_2$, we can write it as
\begin{equation}
	Q_2 =  \frac{2\hat{\bdelta}^\T[ n^{-1}  \sum_{i=1}^{n} b^{\prime\prime}(\hat{\bm{\eta}}^\T\dot{\bZ}_i)  {\bM}_{i} \tilde{\bU}_{i}^{\T} ]\hat{\bdelta} }{\| \hat{\bdelta}\|_2^2} +\frac{2}{n} \sum_{i=1}^{n} \frac{\hat{\bdelta}^{\T}\bM_i \tilde{\bU}_i^{\T}\hat{\bdelta}  }{\| \hat{\bdelta}\|_2^2}[b^{\prime \prime}(\hat{\bm{\eta}}^\T \dot{\bZ}_{i})-b^{\prime \prime}\{(\bm{\eta}^{*})^\T \bZ_{i}\}]. \label{eq:Q2}
\end{equation}
Based on~\eqref{eq:max bhat - bs 2} and 
\begin{eqnarray}
    \frac{\hat{\bdelta}^{\T}\bM_i \tilde{\bU}_i^{\T}\hat{\bdelta}  }{\| \hat{\bdelta}\|_2^2} \leq   \frac{\|\hat{\bdelta}\|_1^2 \|\bM_i\|_{\infty} \|\tilde{\bU}_i\|_{\infty} }{\| \hat{\bdelta}\|_2^2} \lesssim s_w \left(\frac{1}{\sqrt{n}} + \sqrt{\frac{\log n}{p}} \right), \nonumber
\end{eqnarray}
and under the scaling condition $(s_{\eta} \vee s_w)(n^{-1/2} (\log p)^{1/2} + p^{-1/2} (\log n)^{1/2}) = o_p(1)$, we have the second term in~\eqref{eq:Q2} goes to 0 and thus
\begin{eqnarray}
Q_2 &\geq& \frac{\hat{\bdelta}^\T[ n^{-1}  \sum_{i=1}^{n} b^{\prime\prime}\{(\bm{\eta}^*)^\T{\bZ}_i\}  {\bM}_{i} \tilde{\bU}_{i}^{\T} ]\hat{\bdelta} }{2\| \hat{\bdelta}\|_2^2} \nonumber \\
	 & \geq &\frac{\lambda_{\min} (\bI_{\bzeta\bzeta}^*)}{2} - \left| \frac{\hat{\bdelta}^\T [n^{-1}  \sum_{i=1}^{n} b^{\prime\prime}\{(\bm{\eta}^*)^\T{\bZ}_i\}  {\bM}_{i} \tilde{\bU}_{i}^{\T} -\bI_{\bzeta\bzeta}^*]\hat{\bdelta} }{2\| \hat{\bdelta}\|_2^2}  \right| \nonumber \\
	  &\geq & \frac{\kappa}{2} - \frac{\| \hat{\bdelta}\|_1^2\| n^{-1}  \sum_{i=1}^{n} b^{\prime\prime}\{(\bm{\eta}^*)^\T{\bZ}_i\}  {\bM}_{i} \tilde{\bU}_{i}^{\T} -\bI_{\bzeta\bzeta}^* \|_{\max} }{\| \hat{\bdelta}\|_2^2}. \label{eq:Q2 lb1} 
\end{eqnarray}

Because we have
\begin{eqnarray}
   && \| n^{-1}  \sum_{i=1}^{n} b^{\prime\prime}\{(\bm{\eta}^*)^\T{\bZ}_i\}  {\bM}_{i} \tilde{\bU}_{i}^{\T} -\bI_{\bzeta\bzeta}^* \|_{\max} \nonumber \\
    &\leq & \| n^{-1}  \sum_{i=1}^{n} b^{\prime\prime}\{(\bm{\eta}^*)^\T{\bZ}_i\}  {\bM}_{i} \tilde{\bU}_{i}^{\T} - \EE [b^{\prime\prime}\{(\bm{\eta}^*)^\T{\bZ}_i\}  {\bM}_{i} \tilde{\bU}_{i}^{\T} ] \|_{\max}  \nonumber \\
    && + \|\EE [b^{\prime\prime}\{(\bm{\eta}^*)^\T{\bZ}_i\}  {\bM}_{i} \tilde{\bU}_{i}^{\T}]-\bI_{\bzeta\bzeta}^* \|_{\max}. \nonumber \\
    &\lesssim &  \sqrt{\frac{\log p}{n}} + \sqrt{\frac{\log n}{p}}, \label{eq:Q2 lb2}
\end{eqnarray}
where the last inequality is from techniques similar to the proof of Lemma~\ref{concentration of gradient and Hessian} Condition $(ii)$ that $\| n^{-1}  \sum_{i=1}^{n} b^{\prime\prime}\{(\bm{\eta}^*)^\T{\bZ}_i\}  {\bM}_{i} \tilde{\bU}_{i}^{\T} - \EE [b^{\prime\prime}\{(\bm{\eta}^*)^\T{\bZ}_i\}  {\bM}_{i} \tilde{\bU}_{i}^{\T} ] \|_{\max} = O_p\{n^{-1/2} (\log p)^{1/2} + p^{-1/2} (\log n)^{1/2}\}$
and from Proposition~\ref{prop:uniform U} that $\|\EE [b^{\prime\prime}\{(\bm{\eta}^*)^\T{\bZ}_i\}  {\bM}_{i} \tilde{\bU}_{i}^{\T}]-\bI_{\bzeta\bzeta}^* \|_{\max} = O_p\{ n^{-1/2} + p^{-1/2} (\log n)^{1/2}\}$.
 Because $\| \hat{\bdelta}\|_1^2 \leq 16 s_w \| \hat{\bdelta}\|_2^2$, under the scaling condition $(s_{\eta} \vee s_w)\{n^{-1/2}$ $ (\log p)^{1/2} + p^{-1/2} (\log n)^{1/2}\} = o_p(1)$, combining~\eqref{eq:Q2 lb1} and~\eqref{eq:Q2 lb2} gives $Q_2 \geq \kappa/2$ with probability tending to 1. 
Similarly, we have $Q_3\geq \kappa/4$.

Summarizing the above results, we have that with a high probability $Q_1 +Q_2 + Q_3 \geq \kappa$ as $n,p \rightarrow \infty$ and $(s_{\eta} \vee s_w)\{n^{-1/2}$ $ (\log p)^{1/2} + p^{-1/2} (\log n)^{1/2}\} = o_p(1)$. Substituting this result for the terms $Q_1$, $Q_2$ and $Q_3$ into~\eqref{vzzv}, and because $\| \hat{\bdelta}_{\cS}\|_1^2 \leq s_{w}\| \hat{\bdelta}_{\cS}\|_2^2  \leq s_{w}\| \hat{\bdelta}\|_2^2$,  we have 
\begin{equation}
	Q(\hat{\bw}, \bw^*) \geq {\kappa s_{w}^{-1}\|\hat{\bdelta}_S\|_1^2}. \label{Q1/2 amd delta}
\end{equation}

Recall we have proved~\eqref{Q1 quadratic}, which can be written as
\begin{eqnarray}
	&&Q(\hat{\bw}, \bw^*) ^{1/2} \left\{Q(\hat{\bw}, \bw^*) ^{1/2} - s_{\eta}^{1/2}\left(\frac{ \log p}{n} + \frac{ \log n}{p}\right)^{1/2} \right\} \nonumber \\
 &\lesssim &  \left(\frac{ \log p}{n} + \frac{ \log n}{p}\right)^{1/2}(3\|\hat{\bdelta}_{\cS}\|_1 -\|\hat{\bdelta}_{\bar{\cS}}\|_1) \nonumber \\
	&\lesssim & 3\left(\frac{ \log p}{n} + \frac{ \log n}{p}\right)^{1/2}\|\hat{\bdelta}_{\cS}\|_1 . \nonumber
\end{eqnarray}
From \eqref{Q1/2 amd delta}, we have $ \|\hat{\bdelta}_{\cS}\|_{1}^2  \leq s_{w} \kappa^{-1} Q(\hat{\bw}, \bw^*)$ with high probability tending to 1. Substitute this result into the above inequality, we have 
\begin{eqnarray}
	&&Q(\hat{\bw}, \bw^*) ^{1/2} \left\{Q(\hat{\bw}, \bw^*) ^{1/2} - s_{\eta}^{1/2}\left(\frac{ \log p}{n} + \frac{ \log n}{p}\right)^{1/2} \right\} \nonumber  \\
	&\lesssim & \frac{3 s_w^{1/2}}{\kappa^{1/2}}\left(\frac{ \log p}{n} + \frac{ \log n}{p}\right)^{1/2}Q(\hat{\bw}, \bw^*)^{1/2}. \nonumber
\end{eqnarray}
Cancelling the term $Q(\hat{\bw}, \bw^*)^{1/2}$, we have
\begin{equation}
	Q(\hat{\bw}, \bw^*) ^{1/2} - s_{\eta}^{1/2}\left(\frac{ \log p}{n} + \frac{ \log n}{p}\right)^{1/2}  \lesssim  \frac{3 s_w^{1/2}}{\kappa^{1/2}}\left(\frac{ \log p}{n} + \frac{ \log n}{p}\right)^{1/2},\nonumber
\end{equation}
which gives an upper bound for $Q(\hat{\bw}, \bw^*)^{1/2}$ that 
\begin{equation}
	Q(\hat{\bw}, \bw^*)^{1/2} \lesssim (s_{w} \vee s_{\eta})^{1/2} \left(\frac{ \log p}{n} + \frac{ \log n}{p}\right)^{1/2}. \label{eq:case 2 upb}
\end{equation}
which completes the proof of~\eqref{w bprimehat m} in case 2. Therefore, we prove~\eqref{w bprimehat m} holds in both cases. In both cases 1 and 2,
by replacing $Q(\hat{\bw}, \bw^*)$ with $Q^*(\hat{\bw}, \bw^*)$, we get~\eqref{w bprime m} as
\begin{equation}
      Q^*(\hat{\bw}, \bw^*) = O_p\left\{(s_{w} \vee s_{\eta})\left(\frac{ \log p}{n} + \frac{ \log n}{p}\right)\right\}. \nonumber
\end{equation}

To finally prove the estimation consistency of $\hat{\bw}$, or equivalently, to derive the error bound $\|\hat{\bdelta} \|_1$, we also consider two situations. If cone condition $6\|\hat{\bdelta}_{\cS} \|_1 \geq \|\hat{\bdelta}_{\bar{\cS}}\|_1$ holds, we then apply similar techniques as in case 2 to derive~\eqref{Q1/2 amd delta}. Therefore
\begin{eqnarray}
	\|\hat{\bdelta}\|_1 \leq \|\hat{\bdelta}_{\cS}\|_1 + \|\hat{\bdelta}_{\bar{\cS}}\|_1 \leq 7 \|\hat{\bdelta}_{\cS}\|_1 &\lesssim &  s_{w}^{1/2} Q(\hat{\bw}, \bw^*)^{1/2} \nonumber \\
 &\lesssim &  s_{w}^{1/2} (s_{\eta} \vee s_{w} )^{1/2}  \left(\frac{ \log p}{n} + \frac{ \log n}{p}\right)^{1/2}\nonumber \\
 &\lesssim &  (s_{\eta} \vee s_{w} ) \left(\frac{ \log p}{n} + \frac{ \log n}{p}\right)^{1/2}. \nonumber
\end{eqnarray}

Otherwise, we have $6\|\hat{\bdelta}_{\cS} \|_1 \leq \|\hat{\bdelta}_{\bar{\cS}}\|_1$. From~\eqref{Q1 quadratic}, we have
\begin{eqnarray}
	Q(\hat{\bw}, \bw^*) 
&\lesssim &|Q(\hat{\bw}, \bw^*) |^{1/2} s_{\eta}^{1/2}\left( \sqrt{\frac{\log p}{n}} + \sqrt{\frac{\log n}{p}} \right) \nonumber \\
&&+ \left( \sqrt{\frac{\log p}{n}} + \sqrt{\frac{\log n}{p}} \right) (3\|\hat{\bdelta}_{\cS} \|_1   - \|\hat{\bdelta}_{\bar{\cS}}\|_1). \nonumber \\
&\lesssim &|Q(\hat{\bw}, \bw^*) |^{1/2} s_{\eta}^{1/2}\left( \sqrt{\frac{\log p}{n}} + \sqrt{\frac{\log n}{p}} \right) - \frac{1}{2} \left( \sqrt{\frac{\log p}{n}} + \sqrt{\frac{\log n}{p}} \right) \|\hat{\bdelta}_{\bar{\cS}}\|_1, \nonumber
\end{eqnarray}
which together with $6\|\hat{\bdelta}_{\cS} \|_1 \leq \|\hat{\bdelta}_{\bar{\cS}}\|_1$ gives
\begin{eqnarray}
    \|\hat{\bdelta}\|_1 &\leq & \frac{7}{6} \|\hat{\bdelta}_{\bar{\cS}}\|_1 \nonumber \\
    &\leq & \frac{7}{3} \left( \sqrt{\frac{\log p}{n}} + \sqrt{\frac{\log n}{p}} \right)^{-1} \left\{ |Q(\hat{\bw}, \bw^*) |^{1/2} s_{\eta}^{1/2}\left( \sqrt{\frac{\log p}{n}} + \sqrt{\frac{\log n}{p}} \right) - Q(\hat{\bw}, \bw^*) \right\} \nonumber \\
    & \lesssim & (s_{\eta} \vee s_{w} ) \left(\frac{ \log p}{n} + \frac{ \log n}{p}\right)^{1/2}, \nonumber
\end{eqnarray}
   where the last inequality is from~\eqref{w bprimehat m}.

   In conclusion, under either $6\|\hat{\bdelta}_{\cS} \|_1 \geq \|\hat{\bdelta}_{\bar{\cS}}\|_1$ or $6\|\hat{\bdelta}_{\cS} \|_1 \leq \|\hat{\bdelta}_{\bar{\cS}}\|_1$, we prove 
\begin{equation}
    \|\hat{\bdelta}\|_1 = O_p\left\{(s_{w} \vee s_{\eta})\left(\frac{ \log p}{n} + \frac{ \log n}{p}\right)^{1/2}\right\}. \nonumber
\end{equation} 

\section*{Appendix E. Proof of Theorem~\ref{normality theorem}}
\label{proof of theorem 2}
Throughout the rest of appendix, we denote  
\begin{eqnarray}
	\hat{\bH} &=& (\hat{\bW} \hat{\bSigma}_{ e}^{-1}\hat{\bW}^{\T})^{-1}, \quad  \hat{\bH}_p = p \times \hat{\bH} = (p^{-1} \hat{\bW} \hat{\bSigma}_{e}^{-1}\hat{\bW}^{\T})^{-1}. \nonumber
\end{eqnarray}
From the Corollary S.1 in \cite{bai2016maximum}, we have $\hat{\bH}_p = O_p(1)$ and $\hat{\bH} = O_p(p^{-1})$. These results will play an important role in the following proofs.
We next introduce a few technical lemmas as tools in the proof of Theorem \ref{normality theorem}.

\begin{lemma}[Smoothness of loss function]
\label{smoothness}
 Suppose that Assumptions~\ref{assumption1}--\ref{glm assumption} hold. With $\lambda \asymp \lambda^{\prime} \asymp  {n^{-1/2}(\log p)^{1/2}} + p^{-1/2}(\log n)^{1/2} $ and $(s_{w} \vee s_{\eta})$ $(n^{-1/2}\log p + p^{-1}n^{1/2}\log n)= o_p(1)$, the following conditions hold.
 
 $(iii)$ $(\btau^{*})^\T\{\nabla l(\hat{\bm{\eta}})-\nabla l(\bm{\eta}^*)-\nabla^{2} l(\bm{\eta}^{*})(\hat{\bm{\eta}}-\bm{\eta}^{*})\}=o_p(n^{-1/2})$;
 
 $(iv)$ $(\hat{\btau} - \btau^*)^\T\{\nabla l(\hat{\bm{\eta}} ) - \nabla l(\bm{\eta}^* )\} = o_p(n^{-{1}/{2}})$.

\end{lemma}
Lemma \ref{smoothness}  shows the loss functions are smooth in a sense that they are second-order differentiable around the true parameter value. The conditions hold for quadratic loss functions as well as non-quadratic functions given the function $b(\cdot)$ is properly constrained. 
\begin{lemma}[Central limit theorem of score function]
\label{clm}
	Under Assumptions \ref{assumption1}--\ref{glm assumption}, with $\lambda \asymp \lambda^{\prime} \asymp  {n^{-1/2}(\log p)^{1/2}} + p^{-1/2}(\log n)^{1/2} $ and $(s_{w} \vee s_{\eta})$ $(n^{-1/2}\log p + p^{-1}n^{1/2}\log n)= o_p(1)$, it holds that 
	\begin{equation}
		n^{1/2}(\btau^{*})^\T \nabla l(\bm{\eta}^{*}) (I_{\theta \mid \bzeta}^{*})^{-1/2} \rightarrow_d N(0,1). \label{eq:clm lemma}
	\end{equation}
\end{lemma}
Lemma \ref{clm} implies that a linear combination of the gradient of loss function, in other words, the score function is asymptotically normal. 
\begin{lemma}[Partial information estimator consistency]
\label{Partial information estimator consistency} Suppose that Assumptions \ref{assumption1}--\ref{glm assumption} hold. With $\lambda \asymp \lambda^{\prime} \asymp  {n^{-1/2}(\log p)^{1/2}} + p^{-1/2}(\log n)^{1/2} $, if $n,p \rightarrow \infty$ and $(s_{w} \vee s_{\eta})$ $(n^{-1/2}\log p + p^{-1}n^{1/2}\log n)= o_p(1)$, then the estimator for the partial information $\hat{I}_{\theta \mid \bzeta}= \nabla_{\theta \theta}^{2} l(\hat{\bm{\eta}})-\hat{\bw} \nabla_{\bzeta \theta}^{2} l(\hat{\bm{\eta}})$ is consistent,
\begin{equation}
	\hat{I}_{\theta \mid \bzeta} - {I}_{\theta \mid \bzeta}^* = o_p(1).  \nonumber
\end{equation} 
\end{lemma}

With these result, we now prove the asymptotic normality of the debiased estimator, which generalizes the proof of  Theorem 3.2 in \cite{ning2017} to the setting with unmeasured confounders.

The goal is to show the debiased estimator $\tilde{\theta}=\hat{\theta}-{\hat{I}_{\theta \mid \bzeta}}^{-1}{\hat{S}(\hat{\bm{\eta}})}$ is asymptotically normal. First, by Lemma \ref{clm}, we have~\eqref{eq:clm lemma} hold.
 Therefore, it suffices to show that
\begin{equation}
n^{1/2}|(\tilde{\theta}-\theta^{*}) (I_{\theta \mid \bzeta}^{*})^{1/2} +(\btau^*)^\T \nabla l(\bm{\eta}^{*})(I_{\theta \mid \bzeta}^{*})^{-1/2}|=o_{p}(1). \nonumber
\end{equation}
which is equivalent to show
\begin{equation}
n^{1/2}|(\tilde{\theta}-\theta^{*}) I_{\theta \mid \bzeta}^{*} +(\btau^*)^\T \nabla l(\bm{\eta}^{*})  |=o_{p}(1). \nonumber
\end{equation}
since $I_{\theta \mid \bzeta}^{*}$ is constant. Note that we have $\hat{S}(\hat{\bm{\eta}})=  \hat{\btau}^{\T} \nabla l(\hat{\bm{\eta}})$ by the definition of estimated decorrelated score function, we next decompose the left hand side of the above expression and apply triangular inequality as \begin{eqnarray}
&&n^{1/2}|\{\hat{\theta}-{\hat{I}_{\theta \mid \bzeta}}^{-1}\hat{S}(\hat{\bm{\eta}})-\theta^{*}\} I_{\theta \mid \bzeta}^{*}+(\btau^{*})^\T \nabla l(\bm{\eta}^{*})| \nonumber \\
&=& n^{1/2} | (\hat{\theta} - \theta^{*})I_{\theta \mid \bzeta}^{*}-(\btau^{*})^\T \nabla l(\hat{\bm{\eta}})+(\btau^{*})^\T \nabla l(\hat{\bm{\eta}})-\hat{\btau}^{\T} \nabla l(\hat{\bm{\eta}})+\hat{\btau}^{\T} \nabla l (\hat{\bm{\eta}}) \nonumber \\
&&  -I_{\theta \mid \bzeta}^{*} \hat{I}_{\theta \mid \bzeta}^{-1} \hat{\btau}^{\T} \nabla l(\hat{\bm{\eta}})+(\btau^*)^\T \nabla l(\bm{\eta}^{*})| \nonumber \\
&\leq & n^{1/2} | (\hat{\theta}-\theta^{*})I_{\theta \mid \bzeta}^{*}-(\btau^*)^\T\{\nabla l (\hat{\bm{\eta}})-\nabla l(\bm{\eta}^{*})\} | \nonumber \\
&& +n^{1/2} |(\hat{\btau}-\btau^{*})^{\T} \nabla l(\hat{\bm{\eta}})|+n^{1/2}|(I_{\theta \mid \bzeta}^{*} \hat{I}_{\theta \mid \bzeta}^{-1}-1) \hat{\btau}^{\T} \nabla l (\hat{\bm{\eta}}) |  \nonumber \\
&=:&  I_1 + I_2 + I_3. \nonumber
\end{eqnarray}

By an application of Lemma \ref{smoothness} condition $(iii)$, we  write $I_1$ as 
\begin{eqnarray}
I_1 &=& n^{1/2}|(\hat{\theta}-\theta^{*}) I_{\theta \mid \bzeta}^{*}-(\btau^*)^\T\{\nabla l(\hat{\bm{\eta}})-\nabla l(\bm{\eta}^{*})\}| \nonumber \\
&\leq & n^{1/2}|(\hat{\theta}-\theta^{*}) I_{\theta \mid \bzeta}^{*}-(\btau^*)^\T \nabla^{2} l(\bm{\eta}^*)(\hat{\bm{\eta}}-\bm{\eta}^{*})|+o_p(1)  \nonumber \\
& \leq &  n^{1/2} |(\hat{\theta}-\theta^{*})I_{\theta \mid \bzeta}^{*}-\{\nabla_{\theta \theta}^{2} l(\bm{\eta}^{*})-(\bw^*)^\T \nabla_{\bzeta \theta}^{2} l(\bm{\eta}^{*})\}(\hat{\theta}-\theta^{*})  \nonumber \\
&& -\{\nabla_{\theta \bzeta}^{2} l(\bm{\eta}^{*})-\bw^\T \nabla_{\bzeta \bzeta}^{2} l(\bm{\eta}^{*})\}(\hat{\bzeta} - \bzeta^*)|+o_p(1) \nonumber \\
&\leq & n^{1/2}\|\hat{\bm{\eta}}-\bm{\eta}^{*}\|_1 \|T\|_{\infty}+o_{p}(1) \nonumber
\end{eqnarray}
where the last inequality is by H\"older inequality with $T=\{I_{\theta \mid \bzeta}^{*}-\nabla_{\theta \theta}^{2} l(\bm{\eta}^{*})+ (\bw^*)^\T \nabla_{\bzeta \theta}^{2} l(\bm{\eta}^{*})$, $\nabla_{\theta \bzeta}^{2} l(\bm{\eta}^{*})-(\bw^*)^\T \nabla_{\bzeta \bzeta}^{2} l(\bm{\eta}^{*})\}^\T$. 
As a consequence of Lemma~\ref{concentration of gradient and Hessian} condition $(ii)$, we have $\|T\|_{\infty}=O_p\{n^{-1/2}(\log p)^{1/2} + p^{-1/2}(\log n)^{1/2}\}$. In addition, Theorem \ref{Initial estimator consistency}  gives $\|\hat{\bm{\eta}}-\bm{\eta}^{*}\|_1  = O_p\{s_{\eta} (n^{-1/2}(\log p)^{1/2} + p^{-1/2}(\log n)^{1/2})\}$. Hence under the scaling condition of 
$(s_{w} \vee s_{\eta})$ $(n^{-1/2}\log p + p^{-1}n^{1/2}\log n)= o_p(1)$, we have 
\begin{equation}
	I_1 \lesssim \sqrt{n} s_{\eta} \left(\frac{\log p}{n} + \frac{\log n}{p} \right) + o_p(1) = o_p(1) \label{I1 op1}.
\end{equation}

By an application of Lemma~\ref{smoothness} condition $(iv)$, we write $I_2$ as
 \begin{eqnarray}
  I_2 & = & {n}^{1/2}|(\hat{\btau}-\btau^{*})^{\T} \nabla l(\hat{\bm{\eta}})| \nonumber \\
  & \leq & {n}^{1/2}|(\hat{\btau}-\btau^{*})^{\T} \nabla l(\bm{\eta}^*)| + o_p(1) \nonumber \\
  & \leq & {n}^{1/2}\|\hat{\btau}-\btau^{*}\|_1 \|\nabla l(\bm{\eta}^*)\|_{\infty} + o_p(1). \label{tau tau hat grad eta hat}
 \end{eqnarray}
From Theorem \ref{Initial estimator consistency}, we have $\|\hat{\btau}-\btau^{*}\|_1 = \|\hat{\bw}-\bw^{*}\|_1 = O_p\{(s_{w}\vee s_{\eta}) (n^{-1/2}(\log p)^{1/2} + p^{-1/2}(\log n)^{1/2})\}$. From Lemma \ref{concentration of gradient and Hessian} Condition (i), we have $\| \nabla l(\eta^*) \|_{\infty} =O_p\{n^{-1/2}(\log p)^{1/2} + p^{-1/2}(\log n)^{1/2}\}$. Hence under the condition of $(s_{w} \vee s_{\eta})$ $(n^{-1/2}\log p + p^{-1}n^{1/2}\log n)= o_p(1)$, we have
\begin{equation}
		I_2 \lesssim \sqrt{n} (s_w \vee s_{\eta}) \left(\frac{\log p}{n} + \frac{\log n}{p} \right) + o_p(1) = o_p(1) \label{I2 op1}.
\end{equation}

For $I_3$, since $\hat{I}_{\theta \mid \bzeta} - {I}_{\theta \mid \bzeta}^* = o_p(1)$ from Lemma \ref{Partial information estimator consistency}, we have $I_{\theta \mid \bzeta}^{*} \hat{I}_{\theta \mid \bzeta}^{-1}= O_p(1)$. From \eqref{tau tau hat grad eta hat} and $\btau^*$ being fixed, the term $I_2 = o_p(1)$ implies that   $ {n}^{1/2}|\hat{\btau}^{\T} \nabla l (\hat{\bm{\eta}})| = o_p(1)$. Hence we have
\begin{equation}
	I_3 = {n}^{1/2}|(I_{\theta \mid \bzeta}^{*} \hat{I}_{\theta \mid \bzeta}^{-1}-1) \hat{\btau}^{\T} \nabla l (\hat{\bm{\eta}})| = o_p(1). \label{I3 op1}
\end{equation}

Combining \eqref{I1 op1}, \eqref{I2 op1} and \eqref{I3 op1}, we obtain
\begin{equation}
	{n}^{1/2}|(\tilde{\theta}-\theta^{*}) I_{\theta \mid \bzeta}^{*} +(\btau^*)^\T \nabla l(\bm{\eta}^{*})| \leq I_1 + I_2 + I_3 = o_{p}(1), \nonumber
\end{equation}
which completes the proof.

\section*{Appendix F. Proof of Proposition~\ref{prop:uniform U}}

 To prove Proposition~\ref{prop:uniform U},  we show that the estimated unmeasured confounders are bounded with convergence guarantees. From the expression of the estimator for unmeasured confounders in~\eqref{Ui hat} of the main text and recall that $\hat{H} = (\hat{\bW} \hat{\bSigma}_{ e}^{-1}\hat{\bW}^{\T})^{-1}$ and $\hat{\bH}_p = p \times \hat{\bH} = (p^{-1} \hat{\bW} \hat{\bSigma}_{e}^{-1}\hat{\bW}^{\T})^{-1}$, we apply triangular inequality and have 

\begin{eqnarray}
\max_{i=1, \ldots, n} \| \hat{\bU}_i - \bU_i \|_{\infty}	&=& \max_{i=1, \ldots, n} \|\hat{\bH} \hat{\bW} \hat{\bSigma}_{e}^{-1}\{(\bW^* - \hat{\bW})^{\T} \bU_{i}+{\bE}_{i}\}\|_{\infty} \nonumber \\
& \leq &\max_{i=1, \ldots, n} \|\hat{\bH} \hat{\bW} \hat{\bSigma}_{e}^{-1} (\bW^* - \hat{\bW})^{\T} \bU_{i}\|_{\infty}+ \max_{i=1, \ldots, n} \|\hat{\bH} \hat{\bW} \hat{\bSigma}_{e}^{-1} \bE_{i}\|_{\infty}. \nonumber \\ \label{eq:max uhat}
\end{eqnarray}

For the first term in~\eqref{eq:max uhat},  we have
\begin{eqnarray}
	\max_{i=1, \ldots, n}\|\hat{\bH} \hat{\bW} \hat{\bSigma}_{e}^{-1} (\bW^* - \hat{\bW})^{\T} \bU_{i}\|_{\infty} \nonumber &\leq & \|\hat{\bH} \hat{\bW} \hat{\bSigma}_{e}^{-1} (\bW^* - \hat{\bW})^{\T} \|_{1,\infty}\max_{i=1, \ldots, n} \| \bU_i \|_{1} \nonumber \\
	&\leq & \sqrt{K} \|\hat{\bH} \hat{\bW} \hat{\bSigma}_{e}^{-1} (\bW^* - \hat{\bW})^{\T} \|_{F} \max_{i=1, \ldots, n} K \| \bU_i\|_{\infty}. \nonumber
\end{eqnarray}

From Assumption~\ref{glm assumption}, we have $\|\bU_i \|_{\infty} \leq M$ for all $i = 1, \dots, n$ and some constant $M > 0$.
For the norm $\|\hat{\bH} \hat{\bW} \hat{\bSigma}_{e}^{-1} (\bW^* - \hat{\bW})^{\T} \|_{F}$, we apply the Cauchy-Schwarz inequality to bound the matrix norm 
\begin{eqnarray}
   \left\| \hat{\bH} \hat{\bW} \hat{\bSigma}_{e}^{-1} (\bW^* - \hat{\bW})^{\T} \right\|_{F} &= & \left\| \hat{\bH}_p \frac{1}{p} \sum_{j=1}^p \frac{1}{\hat{\sigma}_j^2} \hat{\bW}_j(\bW_j^* -\hat{\bW}_j)^{\T} \right\|_{F} \nonumber \\
   & \leq & \| \hat{\bH}_p\|_{F} \left(\frac{1}{p} \sum_{j=1}^p \frac{1}{\hat{\sigma}_j^2} \| \hat{\bW}_j\|_2^2 \right)^{1/2}\left(\frac{1}{p} \sum_{j=1}^p \frac{1}{\hat{\sigma}_j^2} \| \bW_j^* - \hat{\bW}_j\|_2^2 \right)^{1/2} \nonumber \\
   & = & O_p\left(\frac{1}{\sqrt{n}} + \frac{1}{p}\right) \nonumber
\end{eqnarray}
where the last equality follows because $\| \hat{\bH}_p\|_{F} = O_p(1)$ from Corollary S.1(b), $(p^{-1} \sum_{j=1}^p $ $\hat{\sigma}_j^{-2} \| \hat{\bW}_j\|_2^2 )^{1/2} = O_p(1)$ from Corollary S.1(a) and $(p^{-1} \sum_{j=1}^p{\hat{\sigma}_j^{-2}}  \| \bW_j^* - \hat{\bW}_j\|_2^2 )^{1/2} =  O_p$ $(n^{-1/2} + p^{-1})$ from Proposition 1 in \cite{bai2016maximum}. Therefore \begin{equation}
\max_{i=1, \ldots, n}\|\hat{\bH} \hat{\bW} \hat{\bSigma}_{e}^{-1} (\bW^* - \hat{\bW})^{\T} \bU_{i}\|_{\infty}  = O_p\left(\frac{1}{\sqrt{n}} + \frac{1}{p}\right). \label{HWU bound}
\end{equation}

 For the second term in~\eqref{eq:max uhat}, we have
\begin{eqnarray} 
\max_{i=1, \ldots, n} \|\hat{\bH} \hat{\bW} \hat{\bSigma}_{e}^{-1} \bE_{i}\|_{\infty} &\leq & \|\hat{\bH}_p  \|_{1,\infty} \max_{i=1, \ldots, n} \|p^{-1} \hat{\bW} \hat{\bSigma}_{e}^{-1}\bE_i \|_{1} \nonumber \\
&\leq & \sqrt{K} \|\hat{\bH}_p \|_{F} \max_{i=1, \ldots, n} K \|p^{-1}  \hat{\bW} \hat{\bSigma}_{e}^{-1}  \bE_i \|_{\infty}. \label{eq:lemma1 HE bound} 
\end{eqnarray}
For $p^{-1}  \hat{\bW} \hat{\bSigma}_{e}^{-1}  \bE_i $, we express the $K$-dimensional vector as 
	\begin{eqnarray}
	 \frac{1}{p}	\sum_{j=1}^{p} {\hat{\sigma}_{j}^{-2}} \hat{\bW}_{j} E_{i j}&=&\frac{1}{p}\sum_{j=1}^{p} (\sigma_{j}^*)^{-2} \bW_{j}^* E_{i j}+\frac{1}{p}\sum_{j=1}^p \left\{\frac{1}{\hat{\sigma}_j^2} - \frac{1}{(\sigma_j^*)^2}\right\} \bW_j^*E_{ij} \nonumber \\
 && + \frac{1}{p}\sum_{j=1}^p \frac{1}{\hat{\sigma}_j^2} (\hat{\bW}_j - \bW_j)E_{ij} . \nonumber \\
 & = :& R_1 + R_2 + R_3. \label{eq:R1-3 decomp} 
	\end{eqnarray}
 We next investigate the bound for the term $R_1 = p^{-1}\sum_{j=1}^{p} (\sigma_{j}^*)^{-2} \bW_{j}^* E_{i j}$ and then show the other two terms $R_2$ and $R_3$ are of smaller order and dominated by $R_1$. From Assumption \ref{glm assumption},  we have $\|\bU_i \|_{\infty} \leq M$, $\|\bW_{j}^*\|_2 \leq C$ and $\|\bX_i \|_{\infty} \leq M$. We apply triangular inequality and have
\begin{eqnarray}
\|\bE_{i}\|_{\infty}&=&\|\bX_{i}-(\bW^*)^{\T} \bU_{i}\|_{\infty}	\nonumber \\
& \leq & \|\bX_{i}\|_{\infty}+\|(\bW^*)^{\T} \bU_{i}\|_{\infty} \nonumber \\
& \leq & M + \max_{j = 1,\ldots, p} \|\bW_j^* \|_{1} \| \bU_{i}\|_{1} \nonumber \\
& \leq & M + \max_{j = 1,\ldots, p} \sqrt{K} \|\bW_j^* \|_2 K \| \bU_{i}\|_{\infty} \nonumber \\
&\leq & M(1+ CK^{3/2}),\nonumber
\end{eqnarray}
where the last three inequalities follow from matrix operation and properties of norms.

We have $E_{ij}$ to be sub-exponential random variable since $E_{ij}$ is bounded with $|E_{ij}| \leq M(1+ CK^{3/2})$. Combining the bound for $E_{ij}$ with that $|W_{jk}^*| \leq \|\bW_{j}^*\|_{\infty} \leq \|\bW_{j}^*\|_2 \leq C$ and ${C^{-2}} \leq {(\sigma_{j}^*)^{-2}} \leq C^{2}$, we have $|{(\sigma_{j}^*)^{-2}} {W}_{jk}^* E_{i j}| \leq M	 C^{3}(1+C K^{{3}/{2}})$, for $j = 1,\dots, p$ and $k = 1, \dots, K$. So ${(\sigma_{j}^*)^{-2}} {W}_{jk}^* E_{i j}$ is sub-Gaussian random variable and thus is sub-exponential. As $E_{ij}$ has mean zero, $(\sigma_{j}^*)^{-2} {W}_{jk}^* E_{i j}$ has mean zero and by Bernstein inequality 
\begin{eqnarray}
&&P\left(  \left|\frac{1}{p}\sum_{j=1}^{p} (\sigma_{j}^*)^{-2} {W}_{jk}^* E_{i j}\right| \geq t \right)\nonumber\\
&& \leq  2 \exp \left[-C^{\prime\prime} \min \left\{\frac{t^{2}}{M^2C^{6} (1+C K^{{3}/{2}})^2}, \frac{t}{M	 C^{3} (1+C K^{{3}/{2}})}\right\} p\right]. \nonumber
\end{eqnarray}
Apply union bound inequality, we have
\begin{eqnarray}
&& P\left( \max_{i=1, \ldots, n} \left\|\frac{1}{p}\sum_{j=1}^{p} (\sigma_{j}^*)^{-2} {\bW}_{j}^* E_{i j}\right\|_{\infty} \geq t \right) \nonumber \\
&& \leq 2 nK\exp \left[-C^{\prime\prime} \min \left\{\frac{t^{2}}{M^2C^{6} (1+C K^{{3}/{2}})^2}, \frac{t}{M	 C^{3} (1+C K^{{3}/{2}})}\right\} p\right], \nonumber
\end{eqnarray}
where $C^{\prime\prime} > 0$ is a constant. At $t = M	 C^{3} (1+C K^{{3}/{2}}) p^{-1/2} (\log n)^{1/2}$, the inequality
 
 \begin{equation}
 	\max_{i=1, \ldots, n} \left\|\frac{1}{p}\sum_{j=1}^{p}(\sigma_{j}^*)^{-2} {W}_{j}^* E_{i j}\right\|_{\infty} \leq M	 C^{3} (1+C K^{{3}/{2}}) \left(\frac{\log n}{p}\right)^{1/2}, \label{eq:R1 bound}
 \end{equation}
 holds with probability at least $1 - n^{-1}$. 
 
 For the term $R_2$, we apply Cauchy-Schwarz inequality and have
 \begin{eqnarray}
  && \left\|  \frac{1}{p}\sum_{j=1}^p \left\{\frac{1}{\hat{\sigma}_j^2} - \frac{1}{(\sigma_j^*)^2}\right\} \bW_j^*E_{ij}\right\|_2 \nonumber \\
   \leq && \left[ \frac{1}{p}\sum_{j=1}^p \left\{\frac{1}{\hat{\sigma}_j^2} - \frac{1}{(\sigma_j^*)^2}\right\}^2\right]^{1/2} \left\{ \frac{1}{p}\sum_{j=1}^p E_{ij} (\bW_j^*)^{\intercal}\bW_j^*   E_{ij}\right\}^{1/2}.\nonumber
 \end{eqnarray}
 By Assumption~\ref{assumption1}(b), $\hat{\sigma}_j^{-2}$ and $(\sigma_j^*)^{-2}$ are bounded in $[C^{-2}, C^2]$ and as a result, we have 
\begin{equation}
    \left[ \frac{1}{p}\sum_{j=1}^p \left\{\frac{1}{\hat{\sigma}_j^2} - \frac{1}{(\sigma_j^*)^2}\right\}^2\right]^{1/2}  \leq 2C^2. \nonumber
\end{equation}
and
 \begin{eqnarray}
 \max_{i = 1,\dots, n} \left\| \frac{1}{p}\sum_{j=1}^p \left\{\frac{1}{\hat{\sigma}_j^2} - \frac{1}{(\sigma_j^*)^2}\right\} \bW_j^*E_{ij}\right\|_{\infty} &\leq & \max_{i = 1,\dots, n} \left\| \frac{1}{p}\sum_{j=1}^p \left\{\frac{1}{\hat{\sigma}_j^2} - \frac{1}{(\sigma_j^*)^2}\right\} \bW_j^*E_{ij}\right\|_{2} \nonumber \\
& \leq & 2C^2 \max_{i=1,\dots, n} \left\{ \frac{1}{p}\sum_{j=1}^p E_{ij} (\bW_j^*)^{\intercal}\bW_j^*   E_{ij}\right\}^{1/2}. \nonumber  \end{eqnarray}
Because $|{W}_{jk}^* {W}_{jk^{\prime}}^* E_{i j}^2| \leq M^2C^{2} (1+C K^{{3}/{2}})^2$ and by Berstein inequality,
 \begin{eqnarray}
&&P\left(  \left|\frac{1}{p}\sum_{j=1}^{p} {W}_{jk}^* {W}_{jk^{\prime}}^* E_{i j}^2\right| \geq t \right)\nonumber\\
&& \leq  2 \exp \left[-C^{\prime\prime} \min \left\{\frac{t^{2}}{M^4C^{4} (1+C K^{{3}/{2}})^4}, \frac{t}{M^2	 C^{2} (1+C K^{{3}/{2}})^2}\right\} p\right]. \nonumber
\end{eqnarray}
and then applying union bound gives
 \begin{eqnarray}
&&P\left( \max_{i=1,\dots,n} \left\{ \frac{1}{p}\sum_{j=1}^p E_{ij} (\bW_j^*)^{\intercal}\bW_j^*   E_{ij}\right\} \geq t \right)\nonumber\\
&& \leq  2nK^2 \exp \left[-C^{\prime\prime} \min \left\{\frac{t^{2}}{M^4C^{4} (1+C K^{{3}/{2}})^4}, \frac{t}{M^2	 C^{2} (1+C K^{{3}/{2}})^2}\right\} p\right]. \nonumber
\end{eqnarray}
From the above probability bound, we have $\max_{i=1,\dots,n} \left\{ p^{-1}\sum_{j=1}^p E_{ij} (\bW_j^*)^{\intercal}\bW_j^*   E_{ij}\right\} = O_p(p^{-1/2} (\log n)^{1/2})$ and
 \begin{equation}
 \max_{i = 1,\dots, n} \left\| \frac{1}{p}\sum_{j=1}^p \left\{\frac{1}{\hat{\sigma}_j^2} - \frac{1}{(\sigma_j^*)^2}\right\} \bW_j^*E_{ij}\right\|_{\infty} = O_p\left\{ \left(\frac{\log n}{p}\right)^{1/4}\right\} \label{eq:R2 bound}
 \end{equation}

For the term $R_3$, we apply Cauchy-Schwarz inequality and get
\begin{equation}
\left\|\frac{1}{p}\sum_{j=1}^p \frac{1}{\hat{\sigma}_j^2} (\hat{\bW}_j - \bW_j)E_{ij}\right\|_2 \leq \left(\frac{1}{p}\sum_{j=1}^p \frac{1}{\hat{\sigma}_j^2}\|\hat{\bW}_j - \bW_j\|_2^2\right)^{1/2} \left(\frac{1}{p}\sum_{j=1}^p\frac{1}{\hat{\sigma}_j^2}E_{ij}^2\right)^{1/2}.
\end{equation}
By Proposition 1 in \cite{bai2016maximum}, the first term
\begin{equation}
\left(\frac{1}{p}\sum_{j=1}^p \frac{1}{\hat{\sigma}_j^2}\|\hat{\bW}_j - \bW_j\|_2^2\right)^{1/2} = O_p\left(\frac{1}{\sqrt{n}} + \frac{1}{p}\right). \nonumber
\end{equation}
By a similar technique as in bounding $R_1$ and $R_2$ using the combination of Bernstein inequality and union bound, we can show $\max_{i=1,\dots, n}(p^{-1}\sum_{j=1}^p{\hat{\sigma}_j^{-2}}E_{ij}^2)^{1/2} = O_p (p^{-1/4} (\log n)^{1/4})$, and therefore
\begin{eqnarray}
&& \max_{i=1,\dots, n} \left\| \frac{1}{p}\sum_{j=1}^p \frac{1}{\hat{\sigma}_j^2} (\hat{\bW}_j - \bW_j)E_{ij} \right\|_{\infty} \nonumber \\
&\leq & \max_{i=1,\dots, n} \left\| \frac{1}{p}\sum_{j=1}^p \frac{1}{\hat{\sigma}_j^2} (\hat{\bW}_j - \bW_j)E_{ij} \right\|_{2} \nonumber \\
&\leq & \left(\frac{1}{p}\sum_{j=1}^p \frac{1}{\hat{\sigma}_j^2}\|\hat{\bW}_j - \bW_j\|_2^2\right)^{1/2} \max_{i=1,\dots , n}\left(\frac{1}{p}\sum_{j=1}^p\frac{1}{\hat{\sigma}_j^2}E_{ij}^2\right)^{1/2} \nonumber \\
&= & O_p \left\{\frac{(\log n)^{1/4}}{p^{1/4}} \left( \frac{1}{\sqrt{n}}+ \frac{1}{p}\right) \right\} \label{eq:R3 bound}
\end{eqnarray}

 Combining~\eqref{eq:R1-3 decomp}, \eqref{eq:R1 bound},~\eqref{eq:R2 bound} and~\eqref{eq:R3 bound}, we have
 \begin{eqnarray}
 	\max_{i=1, \ldots, n} \left\|\frac{1}{p}	\sum_{j=1}^{p} \hat{\sigma}_{j}^{-2} \hat{\bW}_{j} E_{i j}\right\|_{\infty}  = 
 	 O_p\left\{ \sqrt{\frac{\log n}{p}}\right\} . \nonumber
 \end{eqnarray}
 In addition, from Corollary S.1 (a) in \cite{bai2016maximum}, we have $\hat{\bH}_p = (\bGamma^{*})^{-1} + o_p(1)$ with $\|(\bGamma^{*})^{-1}\|_{\text{sp}} = \lambda_{\max}^{1/2}\{(\bGamma^*)^{-\T}(\bGamma^*)^{-1}\} = \lambda_{\max}\{(\bGamma^*)^{-1}\}$ being finite constant as $(\bGamma^*)^{-1}$ is symmetric and positive definite. Substituting these results into~\eqref{eq:lemma1 HE bound}, we have
\begin{equation}
 \max_{i=1, \ldots, n} \|	\hat{\bH} \hat{\bW} \hat{\bSigma}_{e}^{-1} \bE_{i} \|_{\infty} \lesssim  K^{{3}/{2}}\lambda_{\max}\{(\bGamma^*)^{-1}\} O_p\left\{ \sqrt{\frac{\log n}{p}}\right\}  \label{HE bound}
\end{equation}

Combining~\eqref{eq:max uhat} and~\eqref{HE bound},  we show 
\begin{equation}
    \max_{i=1, \ldots, n} \| \hat{\bU}_i - \bU_i \|_{\infty} = O_p \left( \frac{1}{\sqrt{n}} + \sqrt{\frac{\log n}{p}} \right). \nonumber
\end{equation}

\section*{Appendix G. Proofs of Lemmas}
\label{Logistic Model}

\subsection*{G.1 Proof of Lemma~\ref{concentration of gradient and Hessian}}
\label{pf: concentration of gradient and hessian}
{\em Proof of Condition (i)}. To prove the Condition (i) in Lemma~\ref{concentration of gradient and Hessian}, we need to   show that
	\begin{equation}
		\left\| - \frac{1}{ n} \sum_{i=1}^{n}[y_i- b^{\prime}\{(\bm{\eta}^*)^{\T} \dot{\bZ}_{i} \}] \dot{\bZ}_i\right\|_{\infty}= O_{p}\left\{\left({\frac{\log p}{n}}\right)^{1/2} + \left({\frac{\log n}{p}}\right)^{1/2}\right\}.\nonumber
	\end{equation}
	
	 From Assumption~\ref{glm assumption}, we have $y_i -b^{\prime}\{(\bm{\eta}^*)^\T{\bZ}_i\}$ to be sub-exponential variable with mean 0 and $\|y_i -b^{\prime}\{(\bm{\eta}^*)^\T{\bZ}_i\}\|_{\varphi_{1}}$ $ \leq M$. In addition, it is assumed that $a_1 \leq (\bm{\eta}^*)^\T{\bZ}_i \leq a_2$ and $0 \leq b^{\prime}\{(\bm{\eta}^*)^\T{\bZ}_i\}  \leq B$. 
  Denote $\max_{i} \| \hat{\bU}_i - \bU_i^* \|_{\infty} = \|\dot{\bZ}_i - \bZ_i \|_{\infty} = \epsilon$.
  From Proposition~\ref{prop:uniform U},  $\epsilon = O_p(n^{-1/2} + p^{-1/2}(\log n)^{1/2}) $. Since $\|\dot{\bZ}_i - \bZ_i \|_{\infty} = \epsilon$, it can be shown that
	 $|(\bm{\eta}^*)^{\T} (\bZ_i -  \dot{\bZ}_{i} )| \leq a_2 + s_{\eta}M \epsilon$. 

To prove condition (i), we focus on finding the appropriate sequence of $t$ such that the following probability tends to 0.
\begin{eqnarray}
  && P\left(\left\|\frac{1}{n} \sum_{i=1}^{n}[y_i- b^{\prime}\{(\bm{\eta}^*)^{\T} \dot{\bZ}_{i} \}] \dot{\bZ}_{i} \right\|_{\infty} \geq  t \right)  \nonumber \\
 &&  \leq  P\left(\left\|\frac{1}{n} \sum_{i=1}^{n}[y_i- b^{\prime}\{(\bm{\eta}^*)^{\T} \dot{\bZ}_{i} \}] \hat{\bU}_{i} \right\|_{\infty} \geq  t \right) +  P\left(\left\|\frac{1}{n} \sum_{i=1}^{n}[y_i- b^{\prime}\{(\bm{\eta}^*)^{\T} \dot{\bZ}_{i} \}] {\bX}_{i} \right\|_{\infty} \geq  t \right) \nonumber \\
 &&  \leq  P\left(\left\|\frac{1}{n} \sum_{i=1}^{n}[y_i- b^{\prime}\{(\bm{\eta}^*)^{\T} {\bZ}_{i} \}] {\bU}_{i} \right\|_{\infty} \geq  t_1 - \delta_1 \right) \nonumber \\
 &&  +  P\left(\left\|\frac{1}{n} \sum_{i=1}^{n}[b^{\prime}\{(\bm{\eta}^*)^{\T} {\bZ}_{i} \}- b^{\prime}\{(\bm{\eta}^*)^{\T} \dot{\bZ}_{i} \}] {\bU}_{i} \right\|_{\infty} \geq  \delta_1 \right) \nonumber  \\
 && +   P\left(\left\|\frac{1}{n} \sum_{i=1}^{n}[y_i- b^{\prime}\{(\bm{\eta}^*)^{\T} {\bZ}_{i} \}] (\hat{\bU}_{i} -\bU_i)\right\|_{\infty} \geq  t_2 - \delta_2 \right) \nonumber \\
 && +  P\left(\left\|\frac{1}{n} \sum_{i=1}^{n}[b^{\prime}\{(\bm{\eta}^*)^{\T} {\bZ}_{i} \}- b^{\prime}\{(\bm{\eta}^*)^{\T} \dot{\bZ}_{i} \}]  (\hat{\bU}_{i} -\bU_i) \right\|_{\infty} \geq  \delta_2 \right) \nonumber \\
 && +  P\left(\left\|\frac{1}{n} \sum_{i=1}^{n}[y_i- b^{\prime}\{(\bm{\eta}^*)^{\T} {\bZ}_{i} \}] {\bX}_{i} \right\|_{\infty} \geq  t_3 -\delta_3 \right)  \nonumber \\
 && +  P\left(\left\|\frac{1}{n} \sum_{i=1}^{n}[b^{\prime}\{(\bm{\eta}^*)^{\T} {\bZ}_{i} \}- b^{\prime}\{(\bm{\eta}^*)^{\T} \dot{\bZ}_{i} \}] {\bX}_{i} \right\|_{\infty} \geq  \delta_3\right)  \nonumber \\
 && =: P_1 + P_2 + P_3 + P_4 + P_5 + P_6, \label{lemma5: full pr}
\end{eqnarray}
where $t = \max\{t_1 + t_2, t_3\} $

We let $\delta_1= \delta_3 = BM(a_2 + s_{\eta} M \epsilon) $ and $\delta_2 = B\epsilon(a_2 + s_{\eta} M \epsilon)$, then it can be verified that the probability $P_2$, $P_4$ and $P_6$ tends to 0 under $\epsilon = O_p(n^{-1/2} + p^{-1/2}(\log n)^{1/2}) $. Next, we look at $P_1$, $P_3$ and $P_5$ to determine $t_1$, $t_2$ and $t_3$ separately.

For $P_1$, as $[y_i- b^{\prime}\{(\bm{\eta}^*)^{\T} {\bZ}_{i} \}] {U}_{ik}$ is independent mean 0 sub-exponential random variables and $\|y_i- b^{\prime}\{(\bm{\eta}^*)^{\T} {\bZ}_{i} \}] {U}_{ik}\|_{\varphi_1} \leq 2M^2$, by Bernstein inequality, we have
\begin{equation}
P\left(\frac{1}{n}\sum_{i=1}^{n}|y_i- b^{\prime}\{(\bm{\eta}^*)^{\T} {\bZ}_{i} \}] {U}_{ik}  | \geq t_1 - \delta_1\right) \leq 2 \exp \left\{-\tilde{M}^{\prime \prime} \min \left( \frac{( t_1 - \delta_1)^{2}}{ 4M^4}, \frac{ t_1 - \delta_1}{2M^2}\right) n\right\}. \nonumber
\end{equation}
Then applying union bound inequality, we have
\begin{eqnarray}
&&P\left(\left\|\frac{1}{n} \sum_{i=1}^{n}[y_i- b^{\prime}\{(\bm{\eta}^*)^{\T} {\bZ}_{i} \}] \bU_i \right\|_{\infty}  \geq t_1 - \delta_1 \right) \nonumber \\ \leq && 2 K \exp \left\{-\tilde{M}^{\prime \prime} \min \left( \frac{( t_1 - \delta_1)^{2}}{ 4M^4}, \frac{ t_1 - \delta_1}{2M^2}\right) n\right\}, \nonumber
\end{eqnarray}
so at $t_1 - \delta_1 = M^2 n^{-1/2}$, the probability $P_1$ tends to 0.
Using similar techniques, we can show that at $t_2 - \delta_2 = M\epsilon n^{-1/2}$, the probability $P_3$ tends to 0; at $t_3 - \delta_3 = M^2n^{-1/2} (\log p)^{1/2}$, the probability $P_5$ tends to 0. Hence at 
\begin{eqnarray}
    t &=& \max\{ t_1 + t_2, t_3\} \nonumber \\
   &=& \max\left\{\frac{M^2}{\sqrt{n}} + BM(a_2 + s_{\eta} M \epsilon) +\frac{M\epsilon}{\sqrt{n}} + B\epsilon(a_2 + s_{\eta} M \epsilon) , \right.\nonumber \\
   && \left.M^2 \sqrt{\frac{\log p}{n}} + BM(a_2 + s_{\eta} M \epsilon)  \right\}, \nonumber 
\end{eqnarray}
  the probability~\eqref{lemma5: full pr} tends to 0. As $t_3$ dominates $t_1 + t_2$ in the above expression, we eventually have
\begin{equation}
    \left\|\frac{1}{n} \sum_{i=1}^{n}[y_i- b^{\prime}\{(\bm{\eta}^*)^{\T} \dot{\bZ}_{i} \}] \dot{\bZ}_{i} \right\|_{\infty} = O_p \left(\sqrt{\frac{\log p}{n}} + \sqrt{\frac{\log n}{p}} \right), \nonumber
\end{equation}
  which completes the proof for condition (i).

{\em Proof of Condition $(ii)$}. To show condition $(ii)$ in Lemma~\ref{concentration of gradient and Hessian} holds, we use a similar technique as in proving condition (i) and decomposing the probability as follows. Recall that $\btau^{*} = (1, -(\bw^*)^{\T})^{\T}$ and the two sub-vectors of $\bw^*$ are denoted as $\bw_q^* =(w_2^*, \ldots, w_q^*)^\T$ and $\bw_u^* =(w_{p+1}^*, \ldots, w_{p+K}^*)^\T$. Denote $\btau_q^* = (1, -(\bw_q^*)^{\T})^{\T} = (1, -w_2^*, \dots, -w_q^*)^{\T}$.
	\begin{align}
	&	P \left(\left\| \frac{1}{ n } \sum_{i=1}^{n}(\btau^{*})^\T b^{\prime\prime}\{(\bm{\eta}^{*})^\T \dot{\bZ}_{i} \}\dot{\bZ}_i \dot{\bZ}_i^\T -E_{\eta^{*}}\left[\frac{1}{ n } \sum_{i=1}^{n}(\btau^{*})^\T b^{\prime\prime}\{(\bm{\eta}^{*})^\T \dot{\bZ}_{i} \}\dot{\bZ}_i \dot{\bZ}_i^\T \right] \right\|_{\infty} \geq t \right) \nonumber \\
	& \leq  P \left(\left\| \frac{1}{ n } \sum_{i=1}^{n}(\bw_u^{*})^\T b^{\prime\prime}\{(\bm{\eta}^{*})^\T \dot{\bZ}_{i} \}\hat{\bU}_i {\bX}_i^\T -E_{\eta^{*}}\left[\frac{1}{ n } \sum_{i=1}^{n}(\bw_u^{*})^\T b^{\prime\prime}\{(\bm{\eta}^{*})^\T \dot{\bZ}_{i} \}\hat{\bU}_i {\bX}_i^\T \right] \right\|_{\infty} \geq t \right)
   \nonumber
   \end{align}
   \begin{align}
   & + P \left(\left\| \frac{1}{ n } \sum_{i=1}^{n}(\bw_u^{*})^\T b^{\prime\prime}\{(\bm{\eta}^{*})^\T \dot{\bZ}_{i} \}\hat{\bU}_i \hat{\bU}_i^\T -E_{\eta^{*}}\left[\frac{1}{ n } \sum_{i=1}^{n}(\bw_u^{*})^\T b^{\prime\prime}\{(\bm{\eta}^{*})^\T \dot{\bZ}_{i} \}\hat{\bU}_i \hat{\bU}_i^\T \right] \right\|_{\infty} \geq t \right)
   \nonumber \\
   & + P \left(\left\| \frac{1}{ n } \sum_{i=1}^{n}(\btau_q^{*})^\T b^{\prime\prime}\{(\bm{\eta}^{*})^\T \dot{\bZ}_{i} \}{\bX}_i \hat{\bU}_i^\T -E_{\eta^{*}}\left[\frac{1}{ n } \sum_{i=1}^{n}(\btau_q^{*})^\T b^{\prime\prime}\{(\bm{\eta}^{*})^\T \dot{\bZ}_{i} \}{\bX}_i \hat{\bU}_i^\T \right] \right\|_{\infty} \geq t \right) \nonumber \\
    & + P \left(\left\| \frac{1}{ n } \sum_{i=1}^{n}(\btau_q^{*})^\T b^{\prime\prime}\{(\bm{\eta}^{*})^\T \dot{\bZ}_{i} \}{\bX}_i {\bX}_i^\T -E_{\eta^{*}}\left[\frac{1}{ n } \sum_{i=1}^{n}(\btau_q^{*})^\T b^{\prime\prime}\{(\bm{\eta}^{*})^\T \dot{\bZ}_{i} \}{\bX}_i {\bX}_i^\T \right] \right\|_{\infty} \geq t \right) \nonumber \\
  & \leq  P \left(\left\| \frac{1}{ n } \sum_{i=1}^{n}(\bw_u^{*})^\T b^{\prime\prime}\{(\bm{\eta}^{*})^\T {\bZ}_{i} \}\hat{\bU}_i {\bX}_i^\T -E_{\eta^{*}}\left[(\bw_u^{*})^\T b^{\prime\prime}\{(\bm{\eta}^{*})^\T {\bZ}_{i} \}\hat{\bU}_i {\bX}_i^\T \right] \right\|_{\infty} \geq t_4 - \delta_4 \right)
   \nonumber \\
   & + P \left(\left\| \frac{1}{ n } \sum_{i=1}^{n}(\bw_u^{*})^\T [b^{\prime\prime}\{(\bm{\eta}^{*})^\T \dot{\bZ}_{i} \} - b^{\prime\prime}\{(\bm{\eta}^{*})^\T {\bZ}_{i} \}]\hat{\bU}_i {\bX}_i^\T  \right.\right. \nonumber \\
   & \left.\left. - E_{\eta^{*}}\left[(\bw_u^{*})^\T [b^{\prime\prime}\{(\bm{\eta}^{*})^\T \dot{\bZ}_{i} \} - b^{\prime\prime}\{(\bm{\eta}^{*})^\T {\bZ}_{i} \}]\hat{\bU}_i {\bX}_i^\T \right] \right\|_{\infty} \geq \delta_4 \right)
   \nonumber \\
   & + P \left(\left\| \frac{1}{ n } \sum_{i=1}^{n}(\bw_u^{*})^\T b^{\prime\prime}\{(\bm{\eta}^{*})^\T {\bZ}_{i} \}\hat{\bU}_i \hat{\bU}_i^\T -E_{\eta^{*}}\left[(\bw_u^{*})^\T b^{\prime\prime}\{(\bm{\eta}^{*})^\T {\bZ}_{i} \}\hat{\bU}_i \hat{\bU}_i^\T \right] \right\|_{\infty} \geq t_5 - \delta_5 \right)
   \nonumber \\
   & + P \left(\left\| \frac{1}{ n } \sum_{i=1}^{n}(\bw_u^{*})^\T [b^{\prime\prime}\{(\bm{\eta}^{*})^\T \dot{\bZ}_{i} \} - b^{\prime\prime}\{(\bm{\eta}^{*})^\T {\bZ}_{i} \}] \hat{\bU}_i \hat{\bU}_i^\T  \right. \right. \nonumber \\ 
& \left.\left. - E_{\eta^{*}}\left[(\bw_u^{*})^\T [b^{\prime\prime}\{(\bm{\eta}^{*})^\T \dot{\bZ}_{i} \} - b^{\prime\prime}\{(\bm{\eta}^{*})^\T {\bZ}_{i} \}] \hat{\bU}_i \hat{\bU}_i^\T \right] \right\|_{\infty} \geq  \delta_5 \right)
   \nonumber \\
   & + P \left(\left\| \frac{1}{ n } \sum_{i=1}^{n}(\btau_q^{*})^\T b^{\prime\prime}\{(\bm{\eta}^{*})^\T {\bZ}_{i} \}{\bX}_i \hat{\bU}_i^\T -E_{\eta^{*}}\left[\frac{1}{ n } \sum_{i=1}^{n}(\btau_q^{*})^\T b^{\prime\prime}\{(\bm{\eta}^{*})^\T {\bZ}_{i} \}{\bX}_i \hat{\bU}_i^\T \right] \right\|_{\infty} \geq t_6 -\delta_6 \right) \nonumber \\
    & + P \left(\left\| \frac{1}{ n } \sum_{i=1}^{n}(\btau_q^{*})^\T [b^{\prime\prime}\{(\bm{\eta}^{*})^\T \dot{\bZ}_{i} \} - b^{\prime\prime}\{(\bm{\eta}^{*})^\T {\bZ}_{i} \}]{\bX}_i \hat{\bU}_i^\T \right.\right. \nonumber \\
    & \left.\left.-E_{\eta^{*}}\left[\frac{1}{ n } \sum_{i=1}^{n}(\btau_q^{*})^\T [b^{\prime\prime}\{(\bm{\eta}^{*})^\T \dot{\bZ}_{i} \} - b^{\prime\prime}\{(\bm{\eta}^{*})^\T {\bZ}_{i} \}]{\bX}_i \hat{\bU}_i^\T \right] \right\|_{\infty} \geq \delta_6 \right) \nonumber \\
    & + P \left(\left\| \frac{1}{ n } \sum_{i=1}^{n}(\btau_q^{*})^\T b^{\prime\prime}\{(\bm{\eta}^{*})^\T {\bZ}_{i} \}{\bX}_i {\bX}_i^\T -E_{\eta^{*}}\left[\frac{1}{ n } \sum_{i=1}^{n}(\btau_q^{*})^\T b^{\prime\prime}\{(\bm{\eta}^{*})^\T {\bZ}_{i} \}{\bX}_i {\bX}_i^\T \right] \right\|_{\infty} \geq t_7 - \delta_7 \right) \nonumber \\
    & + P \left(\left\| \frac{1}{ n } \sum_{i=1}^{n}(\btau_q^{*})^\T [b^{\prime\prime}\{(\bm{\eta}^{*})^\T \dot{\bZ}_{i} \} - b^{\prime\prime}\{(\bm{\eta}^{*})^\T {\bZ}_{i} \}]{\bX}_i {\bX}_i^\T \right.\right. \nonumber \\
    & \left.\left. -E_{\eta^{*}}\left[\frac{1}{ n } \sum_{i=1}^{n}(\btau_q^{*})^\T [b^{\prime\prime}\{(\bm{\eta}^{*})^\T \dot{\bZ}_{i} \} - b^{\prime\prime}\{(\bm{\eta}^{*})^\T {\bZ}_{i} \}]{\bX}_i {\bX}_i^\T \right] \right\|_{\infty} \geq \delta_7 \right) \nonumber \\
    &=: R_1 + R_2 + R_3 + R_4 + R_5 + R_6 + R_7 + R_8, \label{lemma5: full cond2 pr}
	\end{align}
	where $t = \max\{t_4, t_5, t_6, t_7\}$.

 Similarly as in proof of condition $(i)$, we let $\max_{i} \| \hat{\bU}_i - \bU_i^* \|_{\infty} = \|\dot{\bZ}_i - \bZ_i \|_{\infty} = \epsilon$ with $\epsilon = O_p(n^{-1/2} + p^{-1/2}(\log n)^{1/2}) $. From Assumption~\ref{glm assumption}, we have $b^{\prime\prime}\{(\bm{\eta}^*)^\T \bZ_i \} \in [0,B]$,  $|(\btau_q^*)^\T\bX_i| \leq 2M$ and as a result, $|(\bw_u^*)^\T\hat{\bU}_i| \leq K\|\bw^*\|_{\infty} (M+\epsilon)$. In addition, $|b^{\prime\prime}\{(\bm{\eta}^{*})^\T \dot{\bZ}_{i} \} - b^{\prime\prime}\{(\bm{\eta}^{*})^\T {\bZ}_{i} \}| \leq |b^{\prime\prime}\{(\bm{\eta}^{*})^\T {\bZ}_{i} \}|(\bm{\eta}^{*})^\T (\dot{\bZ}_i - \bZ_i) \leq B(a_2 + s_{\eta} M \epsilon)$.

We let 
\begin{align*}
& \delta_4 = 2BM(a_2 + s_{\eta} M \epsilon)\{K\|\bw^*\|_{\infty} (M+\epsilon)\};  \\
& \delta_5 = 2B(M+\epsilon)(a_2 + s_{\eta} M \epsilon)\{K\|\bw^*\|_{\infty} (M+\epsilon)\}; \\
& \delta_6 = 4BM(M+\epsilon) (a_2 + s_{\eta} M \epsilon);\\
& \delta_7 = 4BM^2(a_2 + s_{\eta} M \epsilon).
\end{align*}
It can be shown that $R_2$, $R_4$, $R_6$ and $R_8$ tends to 0 under $\epsilon = O_p(n^{-1/2} + p^{-1/2}(\log n)^{1/2})$. Then we determine $t_4$, $t_5$, $t_6$ and $t_7$ such that the probability~\eqref{lemma5: full cond2 pr} tends to 0.

For $R_1$, 
as $(\bw_u^{*})^\T b^{\prime\prime}\{(\bm{\eta}^{*})^\T {\bZ}_{i} \}\hat{\bU}_i {X}_{ij} -E_{\eta^{*}}[(\bw_u^{*})^\T b^{\prime\prime}\{(\bm{\eta}^{*})^\T {\bZ}_{i} \}\hat{\bU}_i {X}_{ij} ]$ is independent mean 0 sub-exponential random variables and 
\begin{eqnarray}
   && \|(\bw_u^{*})^\T b^{\prime\prime}\{(\bm{\eta}^{*})^\T {\bZ}_{i} \}\hat{\bU}_i {X}_{ij} -E_{\eta^{*}}[(\bw_u^{*})^\T b^{\prime\prime}\{(\bm{\eta}^{*})^\T {\bZ}_{i} \}\hat{\bU}_i {X}_{ij} ]\|_{\varphi_1} \nonumber \\
    && \leq 2BMK\|\bw^*\|_{\infty} (M+\epsilon) =:M_c (M+\epsilon), \nonumber
\end{eqnarray}
where we denote $M_c = 2BMK\|\bw^*\|_{\infty}$ for notational simplicity.
 By Bernstein inequality, we have
\begin{eqnarray}
&&P\left(\frac{1}{n}\sum_{i=1}^{n}\left|(\bw_u^{*})^\T b^{\prime\prime}\{(\bm{\eta}^{*})^\T {\bZ}_{i} \}\hat{\bU}_i {X}_{ij} -E_{\eta^{*}}[(\bw_u^{*})^\T b^{\prime\prime}\{(\bm{\eta}^{*})^\T {\bZ}_{i} \}\hat{\bU}_i {X}_{ij} ]  \right| \geq t_4 - \delta_4\right) \nonumber \\
&& \leq 2 \exp \left\{-\tilde{M}^{\prime \prime} \min \left( \frac{( t_4 - \delta_4)^{2}}{ M_c^2(M+\epsilon)^2}, \frac{ t_4 - \delta_4}{M_c^2(M+\epsilon)^2}\right) n\right\}. \nonumber
\end{eqnarray}
Then applying union bound inequality, we have
\begin{eqnarray}
&& P\left(\left\|\frac{1}{n} \sum_{i=1}^{n}(\bw_u^{*})^\T b^{\prime\prime}\{(\bm{\eta}^{*})^\T {\bZ}_{i} \}\hat{\bU}_i {\bX}_{i} -E_{\eta^{*}}[(\bw_u^{*})^\T b^{\prime\prime}\{(\bm{\eta}^{*})^\T {\bZ}_{i} \}\hat{\bU}_i {\bX}_{i} ]  \right\|_{\infty} \geq  t_4 - \delta_4 \right) \nonumber \\
&& \leq 2 p \exp \left\{-\tilde{M}^{\prime \prime} \min \left( \frac{( t_4 - \delta_4)^{2}}{ M_c^2(M+\epsilon)^2}, \frac{ t_4 - \delta_4}{M_c^2(M+\epsilon)^2}\right) n\right\}, \nonumber
\end{eqnarray}
so at $t_4 - \delta_4 = M_c(M+\epsilon) n^{-1/2}(\log p)^{1/2}$, the probability $R_1$ tends to 0.
Using similar techniques, we can show that at $t_5 - \delta_5 = 2BK \|\bw^*\|_{\infty} (M+\epsilon)^2n^{-1/2} $, the probability $R_3$ tends to 0; at $t_6 - \delta_6 = 4BM(M+\epsilon)n^{-1/2}$, the probability $R_5$ tends to 0; at $t_7 - \delta_7 = 4BM^2 n^{-1/2}(\log p)^{1/2}$, the probability $R_7$ tends to 0. Hence at
\begin{eqnarray}
    t &=& \max\{t_4, t_5, t_6, t_7\} \nonumber \\
    &=& \max \left\{ M_c(M+\epsilon) n^{-1/2}(\log p)^{1/2}+ 2BM(a_2 + s_{\eta} M \epsilon)\{K\|\bw^*\|_{\infty} (M+\epsilon)\}, \right. \nonumber\\
    &&2BK \|\bw^*\|_{\infty} (M+\epsilon)^2n^{-1/2} + 2B(M+\epsilon)(a_2 + s_{\eta} M \epsilon)\{K\|\bw^*\|_{\infty} (M+\epsilon)\} , \nonumber\\
    && 4BM(M+\epsilon)n^{-1/2} + 4BM(M+\epsilon) (a_2 + s_{\eta} M \epsilon) ,\nonumber \\
    && 4BM^2 n^{-1/2}(\log p)^{1/2} + 4BM^2(a_2 + s_{\eta} M \epsilon) \left. \right\} ,\nonumber
\end{eqnarray}
the probability~\eqref{lemma5: full cond2 pr} tends to 0. Since $t_7$ dominates $t_4$, $t_5$ and $t_6$, we have 
\begin{eqnarray}
    &&\left\| \frac{1}{ n } \sum_{i=1}^{n}(\btau^{*})^\T b^{\prime\prime}\{(\bm{\eta}^{*})^\T \dot{\bZ}_{i} \}\dot{\bZ}_i \dot{\bZ}_i^\T -E_{\eta^{*}}\left[\frac{1}{ n } \sum_{i=1}^{n}(\btau^{*})^\T b^{\prime\prime}\{(\bm{\eta}^{*})^\T \dot{\bZ}_{i} \}\dot{\bZ}_i \dot{\bZ}_i^\T \right] \right\|_{\infty} \nonumber \\
    && = O_p \left(\sqrt{\frac{\log p}{n}} + \sqrt{\frac{\log n}{p}} \right). \nonumber 
\end{eqnarray}
This completes the proof for Condition $(ii)$.

\subsection*{G.2 Proof of Lemma~\ref{glm complementary of consistency results}}
\label{sec: proof of lemma 4}
In this proof, we denote 
	\begin{eqnarray}
	H_{\eta} &=& \frac{1}{n} \sum_{i=1}^{n}(\hat{\bm{\eta}}-\bm{\eta}^{*})^{\T} \dot{\bZ}_{i} b^{\prime \prime}\{(\bm{\eta}^{*})^\T \dot{\bZ}_{i}\} \dot{\bZ}_{i}^{\T}(\hat{\bm{\eta}}-\bm{\eta}^{*}) \nonumber 
	 \end{eqnarray}
We continue to use the notations and results defined in Appendix D.1. Recall we define $D(\hat{\bm{\eta}}, \bm{\eta}^{*}) = ({\hat{\bm{\eta}}}-\bm{\eta}^{*})^\T \{\nabla l({\hat{\bm{\eta}}}) - \nabla l(\bm{\eta}^{*})\}$. To show \eqref{eta bprime z}, we consider the difference of $D(\hat{\bm{\eta}}, \bm{\eta}^{*})$ and $H_{\eta}$ and apply the mean value theorem with $\tilde{\bm{\eta}} = \xi \hat{\bm{\eta}}+(1-\xi) \bm{\eta}^{*}$ for $\xi \in [0,1]$ to get 
\begin{eqnarray}
&& | D({\hat{\bm{\eta}}}, \bm{\eta}^{*}) - H_{\eta} | \nonumber \\
&=& \left|(\hat{\bm{\eta}}-\bm{\eta}^{*})^{\T}\{\nabla l(\hat{\bm{\eta}})-\nabla l(\bm{\eta}^{*})\} - 	\frac{1}{n} \sum_{i=1}^{n}(\hat{\bm{\eta}}-\bm{\eta}^{*})^{\T} \dot{\bZ}_{i} b^{\prime \prime}\{(\bm{\eta}^{*})^\T \dot{\bZ}_{i}\} \dot{\bZ}_{i}^{\T}(\hat{\bm{\eta}}-\bm{\eta}^{*})\right| \nonumber \\
&=& \left|(\hat{\bm{\eta}}-\bm{\eta}^{*})^{\T} \left[\nabla^{2} l(\tilde{\bm{\eta}}) - \frac{1}{n} \sum_{i=1}^{n} b^{\prime \prime}\{(\bm{\eta}^{*})^\T \dot{\bZ}_{i} \} \dot{\bZ}_{i}\dot{\bZ}_{i}^\T \right](\hat{\bm{\eta}}-\bm{\eta}^{*})\right| \nonumber \\
&=& \left| \frac{1}{n} \sum_{i=1}^{n}[b^{\prime \prime}(\tilde{\bm{\eta}}^{\T} \dot{\bZ}_{i})-b^{\prime \prime}\{(\bm{\eta}^{*})^\T \dot{\bZ}_{i}\}] \{\dot{\bZ}_{i}^{\T}(\hat{\bm{\eta}}-\bm{\eta}^{*})\}^2 \right| \nonumber \\
&\leq & B\left|  \frac{1}{n} \sum_{i=1}^{n} b^{\prime \prime}\{(\bm{\eta}^{*})^\T \dot{\bZ}_{i}\} \{\dot{\bZ}_{i}^{\T}(\hat{\bm{\eta}}-\bm{\eta}^{*})\}^2 \right|\max _{i}\left| (\hat{\bm{\eta}}-\bm{\eta}^{*})^\T\dot{\bZ}_{i} \right| \label{assumption 6.3} \\
&\lesssim & B \left|  \frac{1}{n} \sum_{i=1}^{n} b^{\prime \prime}\{(\bm{\eta}^{*})^\T \dot{\bZ}_{i}\} \{\dot{\bZ}_{i}^{\T}(\hat{\bm{\eta}}-\bm{\eta}^{*})\}^2 \right|  \|\hat{\bm{\eta}}-\bm{\eta}^{*} \|_1 \max_{i=1, \ldots, n} \|\dot{\bZ}_i \|_{\infty} \nonumber \\
&\lesssim & H_{\eta}  { s_{\eta}} \left(\sqrt{\frac{\log p}{n}} +\sqrt{\frac{\log n}{p}} \right) \left( M + \frac{1}{\sqrt{n}} +\sqrt{\frac{\log n}{p}} \right) \label{estimation bound for eq11} 
\end{eqnarray}
where the inequality \eqref{assumption 6.3} is based on Assumption \ref{glm assumption}(3) that $|b^{\prime \prime}(t_{1})-b^{\prime \prime}(t)| \leq B|t_{1}-t| b^{\prime \prime}(t)$ with $t_1 = \tilde{\bm{\eta}}^{\T} \dot{\bZ}_{i}$ and $t =(\bm{\eta}^{*})^\T \dot{\bZ}_{i}$. The last inequality \eqref{estimation bound for eq11} is from the estimation error bound in the proof of Theorem \ref{Initial estimator consistency} that $\|\hat{\bm{\eta}}-\bm{\eta}^{*} \|_1  \lesssim s_{\eta} (n^{-1/2} (\log p)^{1/2} + p^{-1/2} (\log n)^{1/2})$ and from the result derived from Proposition~\ref{prop:uniform U} that $\max_i \|\dot{\bZ}_i\|_{\infty} = M + O_p(n^{-1/2} + p^{-1/2} (\log n)^{1/2})$. Under the assumption that $s_{\eta} (n^{-1/2} (\log p)^{1/2} + p^{-1/2} (\log n)^{1/2}) = o_p(1)$, we have
\begin{eqnarray}
	 H_{\eta} \left\{1-{ s_{\eta}} \left(\sqrt{\frac{\log p}{n}} +\sqrt{\frac{\log n}{p}} \right) \left( M + \frac{1}{\sqrt{n}} +\sqrt{\frac{\log n}{p}} \right)\right\}
& \lesssim &  D({\hat{\bm{\eta}}}, \bm{\eta}^{*})  \nonumber \\ & \lesssim & 9c^2s_{\eta} \left(\frac{\log p}{n} + \frac{\log n}{p} \right), \nonumber
\end{eqnarray}
where the last inequality is by combining~\eqref{eq:3dS 1dS_bar} and~\eqref{eq:Delta norm2}. Therefore 
\begin{equation}
	H_{\eta} = O_p\left\{ s_\eta\left( \frac{\log p}{n} + \frac{\log n}{p} \right)\right\}. \nonumber
\end{equation}
The two inequalities~\eqref{w bprime m} and~\eqref{w bprimehat m} are shown to hold in Appendix D.2. 
Hence the proof for Lemma~\ref{glm complementary of consistency results} is complete.

\subsection*{G.3 Proof of Lemma~\ref{smoothness}}
 As shown in the preliminaries, we have
\begin{equation}
\nabla l(\bm{\eta})=\frac{1}{n} \sum_{i=1}^{n}\{-y_{i}+b^{\prime}(\bm{\eta}^{\T} \dot{\bZ}_{i})\} \dot{\bZ}_{i}, \quad \nabla^{2} l(\bm{\eta})=\frac{1}{n} \sum_{i=1}^{n}\{-y_{i}+b^{\prime \prime}(\bm{\eta}^{\T} \dot{\bZ}_{i})\} \dot{\bZ}_{i} \dot{\bZ}_{i}^\T. \nonumber
\end{equation}

{\em Proof of Condition (iii)}. For condition $(iii)$, we apply mean value theorem with $\tilde{\bm{\eta}} = \xi \hat{\bm{\eta}} + (1 - \xi) \bm{\eta}^{*}$ for $\xi \in [0,1]$, the left hand side is equivalent to
\begin{eqnarray}
&& |(\btau^{*})^\T\{\nabla l(\hat{\bm{\eta}})-\nabla l(\bm{\eta}^{*})-\nabla^{2} l(\bm{\eta}^{*})(\hat{\bm{\eta}}-\bm{\eta}^{*})\}| \nonumber \\
&=& |(\btau^{*})^\T \{\nabla^{2} l(\tilde{\bm{\eta}})-\nabla^{2} l(\bm{\eta}^{*})\}(\hat{\bm{\eta}}-\bm{\eta}^{*}) | \nonumber \\
&=& \left| \frac{1}{n} \sum_{i=1}^{n}[b^{\prime \prime}(\tilde{\bm{\eta}}^{\T} \dot{\bZ}_{i})-b^{\prime \prime}\{(\bm{\eta}^{*})^\T \dot{\bZ}_{i}\}]\dot{\bZ}_{i}^{\T}(\hat{\bm{\eta}}-\bm{\eta}^{*})(\btau^{*})^\T \dot{\bZ}_{i} \right| \nonumber \\
&\leq & B\left|  \frac{1}{n} \sum_{i=1}^{n} \xi(\hat{\bm{\eta}}-\bm{\eta}^{*})^{\T} \dot{\bZ}_{i} b^{\prime \prime}\{(\bm{\eta}^{*})^\T \dot{\bZ}_{i}\} \dot{\bZ}_{i}^{\T}(\hat{\bm{\eta}}-\bm{\eta}^{*})\right|\max _{i}|(\btau^{*})^\T \dot{\bZ}_{i} | \label{ass4 con3 bound} \\
&\lesssim & s_\eta\left( \frac{\log p}{n} + \frac{\log n}{p} \right) \left\{M + O_p\left(\frac{1}{\sqrt{n}} + \sqrt{\frac{\log n}{p}} \right) \right\} \label{ass7 con3 bound}
\end{eqnarray}
where the inequality \eqref{ass4 con3 bound} is based on Assumption \ref{glm assumption}(4) that $|b^{\prime \prime}(t_{1})-b^{\prime \prime}(t)| \leq B|t_{1}-t| b^{\prime \prime}(t)$ with $t_1 = \tilde{\bm{\eta}}^{\T} \dot{\bZ}_{i}$ and $t =(\bm{\eta}^{*})^\T \dot{\bZ}_{i}$.  The last inequality \eqref{ass7 con3 bound} is from the \eqref{eta bprime z} of Lemma \ref{glm complementary of consistency results} and from the results of Proposition~\ref{prop:uniform U}.

As $n, p \rightarrow \infty$, under the scaling condition that $(s_{w} \vee s_{\eta})(n^{-1/2}\log p + p^{-1/2}n^{1/2}\log n)= o_p(1)$, we have condition $(iii)$ holds as
\begin{eqnarray}
	&&\sqrt{n} |(\btau^{*})^\T\{\nabla l(\hat{\bm{\eta}})-\nabla l(\bm{\eta}^{*})-\nabla^{2} l(\bm{\eta}^{*})(\hat{\bm{\eta}}-\bm{\eta}^{*})\}| \nonumber \\
	&\lesssim & \sqrt{n} s_\eta\left( \frac{\log p}{n} + \frac{\log n}{p} \right) \left\{M + O_p\left(\frac{1}{\sqrt{n}} + \sqrt{\frac{\log n}{p}} \right) \right\}\nonumber \\
	&= & o_p(1). \nonumber
\end{eqnarray}

{\em Proof of Condition $(iv)$}. We multiply ${n}^{1/2}$ to the left hand side of condition $(iv)$ and apply mean value theorem with $\tilde{\bm{\eta}} = \xi \hat{\bm{\eta}}+(1-\xi) \bm{\eta}^{*}$. Then we have
\begin{align}
& {n}^{1/2}|(\hat{\btau}-\btau^{*})^{\T}\{\nabla l(\hat{\bm{\eta}})-\nabla l(\bm{\eta}^{*})\}| \nonumber \\
 & =  {n}^{1/2}\left|(\hat{\btau}-\btau^{*})^{\T}\left[\frac{1}{n} \sum_{i=1}^{n}b^{\prime}(\hat{\bm{\eta}}^{\T} \dot{\bZ}_{i})-b^{\prime}\{(\bm{\eta}^*)^{\T} \dot{\bZ}_{i})\} \dot{\bZ}_{i}\right]\right| \nonumber \\
 & =  {n}^{1/2}\left|\frac{1}{n} \sum_{i=1}^{n} b^{\prime \prime}(\tilde{\bm{\eta}}^{\T} \dot{\bZ}_{i})(\hat{\btau}-\btau^{*})^{\T} \dot{\bZ}_{i}(\hat{\bm{\eta}}-\bm{\eta}^{*})^{\T} \dot{\bZ}_{i}\right| \nonumber \\
  & \leq  {n}^{1/2}\left|\frac{1}{n}  \sum_{i=1}^{n} b^{\prime \prime}\{(\bm{\eta}^{*})^\T \dot{\bZ}_{i}\} \{(\hat{\bw} - \bw^*)^\T \dot{\bM}_i\}^2 \right|^{1/2} \nonumber \\
 & \times \left|\frac{1}{n}  \sum_{i=1}^{n} b^{\prime \prime}\{(\bm{\eta}^{*})^\T \dot{\bZ}_{i}\} \{(\hat{\bm{\eta}} - \bm{\eta}^*)^\T \dot{\bZ}_i\}^2 \right|^{1/2} \label{eq:lemma 5.2 cauchy} \\
& \lesssim {n}^{1/2} \left\{ s_\eta\left( \frac{\log p}{n} + \frac{\log n}{p} \right)\right\}^{1/2} \left\{ (s_\eta \vee s_w) \left( \frac{\log p}{n} + \frac{\log n}{p} \right)\right\}^{1/2} \label{ass4 con4 bprime} \\
& \lesssim  n^{1/2} (s_{w} \vee s_{\eta}) \left( \frac{\log p}{n} + \frac{\log n}{p} \right)  =o_{p}(1), \label{eq:lemma 5.2 final result}
\end{align}
where~\eqref{eq:lemma 5.2 cauchy} is by applying Cauchy-Schwarz inequality,~\eqref{ass4 con4 bprime} is from the~\eqref{eta bprime z} and~\eqref{w bprime m} in Lemma~\ref{glm complementary of consistency results} and~\eqref{eq:lemma 5.2 final result} is from the condition $(s_{w} \vee s_{\eta})(n^{-1/2}\log p + p^{-1/2}n^{1/2}\log n)= o_p(1)$. This completes the proof of Lemma \ref{smoothness}.
\subsection*{G.4 Proof of Lemma~\ref{clm}}
According to the definitions in the preliminaries, we have 
	\begin{align}
	 	&	(\btau^*)^\T \nabla l(\bm{\eta}^*)  (I_{\theta \mid \bzeta}^{*})^{-1/2} \nonumber \\
    &=  \frac{1}{n} \sum_{i=1}^n (\btau^*)^\T \dot{\bZ}_i [-y_i +b^{\prime}\{(\bm{\eta}^{*})^\T\dot{\bZ}_i\}] (I_{\theta \mid \bzeta}^{*})^{-1/2}. 
   \nonumber \\
    & =  \frac{1}{n} \sum_{i=1}^n (\btau^*)^\T \dot{\bZ}_i [-y_i +b^{\prime}\{(\bm{\eta}^{*})^\T{\bZ}_i\}] (I_{\theta \mid \bzeta}^{*})^{-1/2} \nonumber \\
    & + \frac{1}{n} \sum_{i=1}^n (\btau^*)^\T \dot{\bZ}_i [b^{\prime}\{(\bm{\eta}^{*})^\T\dot{\bZ}_i\} - b^{\prime}\{(\bm{\eta}^{*})^\T{\bZ}_i\}] (I_{\theta \mid \bzeta}^{*})^{-1/2} \nonumber \\
    &=  \frac{1}{n} \sum_{i=1}^n (\btau^*)^\T {\bZ}_i [-y_i +b^{\prime}\{(\bm{\eta}^{*})^\T{\bZ}_i\}] (I_{\theta \mid \bzeta}^{*})^{-1/2} \nonumber \\
    & + \frac{1}{n} \sum_{i=1}^n (\bw_u^*)^\T (\hat{\bU}_i - \bU_i) [-y_i +b^{\prime}\{(\bm{\eta}^{*})^\T{\bZ}_i\}] (I_{\theta \mid \bzeta}^{*})^{-1/2} \nonumber  \end{align}
    \begin{align}
    & + \frac{1}{n} \sum_{i=1}^n (\btau^*)^\T {\bZ}_i [b^{\prime}\{(\bm{\eta}^{*})^\T\dot{\bZ}_i\} - b^{\prime}\{(\bm{\eta}^{*})^\T{\bZ}_i\}] (I_{\theta \mid \bzeta}^{*})^{-1/2} \nonumber \\
    & + \frac{1}{n} \sum_{i=1}^n (\bw_u^*)^\T (\hat{\bU}_i - \bU_i) [b^{\prime}\{(\bm{\eta}^{*})^\T\dot{\bZ}_i\} - b^{\prime}\{(\bm{\eta}^{*})^\T{\bZ}_i\}] (I_{\theta \mid \bzeta}^{*})^{-1/2} \nonumber \\
    & =:  G_1 + G_2 + G_3 + G_4 \nonumber 
	 \end{align}
  
Because $|(\btau^*)^\T {\bZ}_i|$ is bounded, $[-y_i +b^{\prime}\{(\bm{\eta}^{*})^\T{\bZ}_i\}]$ is sub-exponential from Assumption~\ref{glm assumption}, the term $(\btau^*)^\T {\bZ}_i [-y_i +b^{\prime}\{(\bm{\eta}^{*})^\T{\bZ}_i\}] (I_{\theta \mid \bzeta}^{*})^{-1/2}$ are independent and has finite moments. We apply Berry-Esseen Theorem and show that $G_1 \rightarrow_d N(0,1) $

For $G_2$, we apply similar techniques as in the proof of Lemma~\ref{concentration of gradient and Hessian} condition(i) by Bernstein's inequality and it can be verified that
\begin{eqnarray}
   G_2 =  \frac{1}{n} \sum_{i=1}^n (\bw_u^*)^\T (\hat{\bU}_i - \bU_i) [-y_i +b^{\prime}\{(\bm{\eta}^{*})^\T{\bZ}_i\}] (I_{\theta \mid \bzeta}^{*})^{-1/2}  = O_p\left(\frac{1}{\sqrt{n}} + \sqrt{\frac{\log n}{p}}  \right). \nonumber
\end{eqnarray}
As a result of Proposition~\ref{prop:uniform U}, we have
\begin{equation}
   G_3 =  \frac{1}{n} \sum_{i=1}^n (\btau^*)^\T {\bZ}_i [b^{\prime}\{(\bm{\eta}^{*})^\T\dot{\bZ}_i\} - b^{\prime}\{(\bm{\eta}^{*})^\T{\bZ}_i\}] (I_{\theta \mid \bzeta}^{*})^{-1/2} =O_p\left(\frac{1}{\sqrt{n}} + \sqrt{\frac{\log n}{p}}  \right), \nonumber
\end{equation}
and 
\begin{equation}
    G_4 = \frac{1}{n} \sum_{i=1}^n (\bw_u^*)^\T (\hat{\bU}_i - \bU_i) [b^{\prime}\{(\bm{\eta}^{*})^\T\dot{\bZ}_i\} - b^{\prime}\{(\bm{\eta}^{*})^\T{\bZ}_i\}] (I_{\theta \mid \bzeta}^{*})^{-1/2} = O_p\left(\frac{1}{{n}} + {\frac{\log n}{p}}  \right). \nonumber
\end{equation}

Under the scaling condition that $n, p \rightarrow \infty$ and $(s_{w} \vee s_{\eta}) (n^{-1/2}\log p + p^{-1}n^{1/2}\log n)= o_p(1)$, we show that $G_2 \rightarrow_p 0$, $G_3 \rightarrow_p 0$ and $G_4 \rightarrow_p 0$.
Applying Slutsky's Theorem, we have
\begin{equation}
	 		\frac{1}{n} \sum_{i=1}^n (\btau^*)^\T \dot{\bZ}_i [-y_i +b^{\prime}\{(\bm{\eta}^{*})^\T\dot{\bZ}_i\}](I_{\theta \mid \bzeta}^{*})^{-1/2}\rightarrow_d N(0,1) . \nonumber
	 \end{equation}
This completes the proof of central limit theorem for the score function.

\subsection*{G.5 Proof of Lemma~\ref{Partial information estimator consistency}}
	In this proof, we will show the partial information
	\begin{equation}
{I}_{\theta \mid \bzeta}^*=E[ b^{\prime\prime}\{(\bm{\eta}^{*})^\T {\bZ}_{i}\} D_{i}\{D_{i}-({\bw}^*)^{\T} {\bM}_{i}\} ], \nonumber
\end{equation}
	 is consistently estimated by
\begin{equation}
\hat{I}_{\theta \mid \bzeta}=\frac{1}{n} \sum_{i=1}^{n} b^{\prime\prime}(\hat{\bm{\eta}}^{\T} \dot{\bZ}_{i} ) D_{i}(D_{i}-\hat{\bw}^{\T} \dot{\bM}_{i}). \nonumber
\end{equation}

To see this, we write the difference between them as
\begin{eqnarray}
	\hat{I}_{\theta \mid \bzeta} - {I}_{\theta \mid \bzeta}^* &=&  \frac{1}{n} \sum_{i=1}^{n} b^{\prime\prime}(\hat{\bm{\eta}}^{\T} \dot{\bZ}_{i} ) D_{i}^2 - E[ b^{\prime\prime}\{(\bm{\eta}^{*})^\T {\bZ}_{i}\} D_{i}^2] \nonumber \\
	&& + \frac{1}{n} \sum_{i=1}^{n} b^{\prime\prime}(\hat{\bm{\eta}}^{\T} \dot{\bZ}_{i} ) \hat{\bw}^{\T} \dot{\bM}_{i}D_{i} - E[ b^{\prime\prime}\{(\bm{\eta}^{*})^\T {\bZ}_{i}\}(\bw^*)^\T {\bM}_{i} D_{i}] \nonumber \\
	& = & L_1 + L_2. \nonumber \end{eqnarray}

For $L_1$, we decompose it into
\begin{eqnarray}
L_1 & = & \frac{1}{n} \sum_{i=1}^{n} b^{\prime\prime}(\hat{\bm{\eta}}^{\T} \dot{\bZ}_{i} ) D_{i}^2 - E[ b^{\prime\prime}\{(\bm{\eta}^{*})^\T {\bZ}_{i}\} D_{i}^2] \nonumber \\
& = &  \frac{1}{n} \sum_{i=1}^{n}  b^{\prime\prime}\{(\bm{\eta}^*)^{\T} {\bZ}_{i} \} D_{i}^2 - E[ b^{\prime\prime}\{(\bm{\eta}^{*})^\T {\bZ}_{i}\} D_{i}^2] \nonumber \\
&& +  \frac{1}{n} \sum_{i=1}^{n}  [b^{\prime\prime}\{(\bm{\eta}^*)^{\T} \dot{\bZ}_{i} \} - b^{\prime\prime}\{(\bm{\eta}^*)^{\T} {\bZ}_{i} \}] D_{i}^2  \nonumber \\
&& +  \frac{1}{n} \sum_{i=1}^{n} [b^{\prime\prime}(\hat{\bm{\eta}}^{\T} \dot{\bZ}_{i} ) - b^{\prime\prime}\{(\bm{\eta}^*)^{\T} \dot{\bZ}_{i} \}] D_{i}^2 \nonumber \\
& = & L_{11} + L_{12} + L_{13}. \nonumber
\end{eqnarray}
Applying similar techniques as in the proof of Lemma~\ref{concentration of gradient and Hessian} condition (ii) by Bernstein inequality, we can show that 
\begin{equation}
    L_{11} =  \frac{1}{n} \sum_{i=1}^{n}  b^{\prime\prime}\{(\bm{\eta}^*)^{\T} {\bZ}_{i} \} D_{i}^2 - E[ b^{\prime\prime}\{(\bm{\eta}^{*})^\T {\bZ}_{i}\} D_{i}^2] = O_p\left(\frac{1}{\sqrt{n}} \right). \nonumber
\end{equation}
As a result of Proposition~\ref{prop:uniform U}, 
\begin{equation}
    L_{12} = \frac{1}{n} \sum_{i=1}^{n}  [b^{\prime\prime}\{(\bm{\eta}^*)^{\T} \dot{\bZ}_{i} \} - b^{\prime\prime}\{(\bm{\eta}^*)^{\T} {\bZ}_{i} \}] D_{i}^2 = O_p\left(\frac{1}{\sqrt{n}} + \sqrt{\frac{\log n}{p}}  \right). \nonumber
\end{equation}
Using the estimation consistency results in Theorem~\ref{Initial estimator consistency}, we have
\begin{eqnarray}
    &&L_{13} =  \frac{1}{n} \sum_{i=1}^{n} [b^{\prime\prime}(\hat{\bm{\eta}}^{\T} \dot{\bZ}_{i} ) - b^{\prime\prime}\{(\bm{\eta}^*)^{\T} \dot{\bZ}_{i} \}] D_{i}^2 \nonumber \\
    = && O_p\left\{ s_\eta\left( \sqrt{\frac{\log p}{n}} + \sqrt{\frac{\log n}{p}} \right)\left( M + \frac{1}{\sqrt{n}} + \sqrt{\frac{\log n}{p}}\right)  \right\} \nonumber
\end{eqnarray}

 Under the condition that $n, p \rightarrow \infty$ and $(s_{w} \vee s_{\eta})$ $(n^{-1/2}\log p + p^{-1}n^{1/2}\log n)= o_p(1)$, we have
 \begin{equation}
     L_1 = L_{11} + L_{12} + L_{13} = o_p(1). \label{lemma10: bound l1}
 \end{equation}

For $L_2$, we decompose it into
\begin{eqnarray}
L_2& =& 	\frac{1}{n} \sum_{i=1}^{n} b^{\prime\prime}(\hat{\bm{\eta}}^{\T} \dot{\bZ}_{i} ) \hat{\bw}^{\T} \dot{\bM}_{i}D_{i} - E[ b^{\prime\prime}\{(\bm{\eta}^{*})^\T {\bZ}_{i}\}(\bw^*)^\T {\bM}_{i} D_{i}] \nonumber \\
&= &	\frac{1}{n} \sum_{i=1}^{n} b^{\prime\prime}(\hat{\bm{\eta}}^{\T} \dot{\bZ}_{i} )( \hat{\bw} - \bw^*)^\T \dot{\bM}_{i}D_{i} \nonumber \\
&& + 	\frac{1}{n} \sum_{i=1}^{n}[ b^{\prime\prime}(\hat{\bm{\eta}}^{\T} \dot{\bZ}_{i} ) -b^{\prime\prime}\{(\bm{\eta}^{*})^\T \dot{\bZ}_{i}\}  ] (\bw^*)^\T \dot{\bM}_{i}D_{i} \nonumber \\
&& + 	\frac{1}{n} \sum_{i=1}^{n}b^{\prime\prime}\{(\bm{\eta}^{*})^\T \dot{\bZ}_{i}\}   (\bw^*)^\T \dot{\bM}_{i}D_{i} - E[ b^{\prime\prime}\{(\bm{\eta}^{*})^\T {\bZ}_{i}\}(\bw^*)^\T {\bM}_{i} D_{i}] \nonumber \\
&=& L_{21} + L_{22} + L_{23}. \nonumber
\end{eqnarray}

We apply H\"older's inequality on $L_{21}$ and get
\begin{eqnarray}
	L_{21} & = & \frac{1}{n} \sum_{i=1}^{n} b^{\prime\prime}(\hat{\bm{\eta}}^{\T} \dot{\bZ}_{i} )( \hat{\bw} - \bw^*)^\T \dot{\bM}_{i}D_{i}  \nonumber \\
	 &\leq & \left[ \frac{1}{n} \sum_{i=1}^{n} b^{\prime\prime}(\hat{\bm{\eta}}^{\T} \dot{\bZ}_{i} )\{( \hat{\bw} - \bw^*)^\T \dot{\bM}_{i}\}^2\right]^{1/2} \left\{ \frac{1}{n} \sum_{i=1}^{n} b^{\prime\prime}(\hat{\bm{\eta}}^{\T} \dot{\bZ}_{i} )D_i^2\right\}^{1/2} \nonumber \\
	& \lesssim & \left\{ (s_{\eta} \vee s_{w})\left( \frac{\log p}{n} + \frac{\log n}{p} \right)\right\}^{1/2}\left(M + \frac{1}{\sqrt{n}} + \sqrt{\frac{\log n}{p}}  \right)^{1/2}   \label{eq:lemma 7 bound l21} 
\end{eqnarray}
where the inequality~\eqref{eq:lemma 7 bound l21} is from the the equation \eqref{w bprimehat m} in Lemma \ref{glm complementary of consistency results} and similar arguments as in bounding $L_{12} + L_{13}$.

For $L_{22}$, we apply Assumption \ref{glm assumption}(3) that $|b^{\prime \prime}(t_{1})-b^{\prime \prime}(t)| \leq B|t_{1}-t| b^{\prime \prime}(t)$ with $t_1 = \hat{\bm{\eta}}^{\T} \dot{\bZ}_{i}$ and $t =(\bm{\eta}^{*})^\T \dot{\bZ}_{i}$. Then we apply H\"older's inequality and have
\begin{eqnarray}
L_{22} & = &		\frac{1}{n} \sum_{i=1}^{n} [b^{\prime\prime}(\hat{\bm{\eta}}^{\T} \dot{\bZ}_{i} ) -b^{\prime\prime}\{(\bm{\eta}^{*})^\T \dot{\bZ}_{i}\}] (\bw^*)^\T \dot{\bM}_{i}D_{i} \nonumber \\
& \leq & \frac{1}{n} \sum_{i=1}^{n} b^{\prime\prime}\{(\bm{\eta}^{*})^\T \dot{\bZ}_{i}\}(\hat{\bm{\eta}} - \bm{\eta}^*)^{\T} \dot{\bZ}_{i}  (\bw^*)^\T \dot{\bM}_{i}D_{i}  \nonumber \\
 &\leq & \left[ \frac{1}{n} \sum_{i=1}^{n} b^{\prime\prime}\{(\bm{\eta}^*)^\T \dot{\bZ}_{i} \}\{(\hat{\bm{\eta}} - \bm{\eta}^*)^{\T} \dot{\bZ}_{i} \}^2\right]^{1/2} \left[ \frac{1}{n} \sum_{i=1}^{n} b^{\prime\prime}\{(\bm{\eta}^*)^\T \dot{\bZ}_{i} \} \{(\bw^*)^\T \dot{\bM}_{i}\}^2D_i^2\right]^{1/2} \nonumber \\ 
 & \lesssim &  \left\{ s_\eta\left( \frac{\log p}{n} + \frac{\log n}{p} \right)\right\}^{1/2} \left(M + \frac{1}{\sqrt{n}} + \sqrt{\frac{\log n}{p}}  \right)^{3/2}\label{eq:lemma 7 bound l22}
\end{eqnarray}
where the inequality~\eqref{eq:lemma 7 bound l22} is from the results \eqref{eta bprime z} in Lemma \ref{glm complementary of consistency results} and from similar arguments as in bounding $L_{11}$.

For $L_{23}$, we follow the similar arguments in bounding $L_{11}$ and get 
\begin{eqnarray}
L_{23} &=&	\frac{1}{n} \sum_{i=1}^{n}b^{\prime\prime}\{(\bm{\eta}^{*})^\T \dot{\bZ}_{i}\}   (\bw^*)^\T \dot{\bM}_{i}D_{i} - E[ b^{\prime\prime}\{(\bm{\eta}^{*})^\T {\bZ}_{i}\}(\bw^*)^\T {\bM}_{i} D_{i}]  \nonumber \\
&\lesssim & \frac{1}{{n}} + {\frac{\log n}{p}}  .\label{L23} 
\end{eqnarray}

Under the condition that $n, p \rightarrow \infty$ and $(s_{w} \vee s_{\eta})$ $(n^{-1/2}\log p + p^{-1}n^{1/2}\log n)= o_p(1)$ and \eqref{eq:lemma 7 bound l21}, \eqref{eq:lemma 7 bound l22}, \eqref{L23}, we get
\begin{equation}
    L_2 = L_{21} + L_{22} + L_{23} = o_p(1) \label{lemma10: bound l2}
\end{equation}

Combining~\eqref{lemma10: bound l1} and~\eqref{lemma10: bound l2}, we have
\begin{equation}
	\hat{I}_{\theta \mid \bzeta} - {I}_{\theta \mid \bzeta}^*  =  o_p(1). \nonumber
\end{equation}
This completes the proof of Lemma~\ref{Partial information estimator consistency}.

\vskip 0.2in

\newpage
\bibliography{manuscript_final}

\end{document}